\documentclass[aps,eqsecnum,preprint,floats,epsf,epsfig,nofootinbib]{revtex4}
\usepackage{epsfig}
\usepackage{graphicx}% Include figure files
\usepackage{extarrows}
\usepackage{subfigure}

\begin{document}
\def\be{\begin{eqnarray}}
\def\en{\end{eqnarray}}
\def\non{\nonumber}
\def\la{\langle}
\def\ra{\rangle}
\def\A{{\cal A}}
\def\B{{\cal B}}
\def\c{{\cal C}}
\def\d{{\cal D}}
\def\e{{\cal E}}
\def\p{{\cal P}}
\def\t{{\cal T}}
\def\nc{N_c^{\rm eff}}
\def\CP{{\it CP}~}
\def\CPP{{\it CP}}
\def\acp{{\cal A}_{C\!P}}
\def\vp{\varepsilon}
\def\drho{\bar\rho}
\def\deta{\bar\eta}
\def\vma{{_{V-A}}}
\def\vpa{{_{V+A}}}
\def\J{{J/\psi}}
\def\ov{\overline}
\def\Lqcd{{\Lambda_{\rm QCD}}}
\def\pr{{ Phys. Rev.}~}
\def\prl{{ Phys. Rev. Lett.}~}
\def\pl{{ Phys. Lett.}~}
\def\np{{ Nucl. Phys.}~}
\def\zp{{ Z. Phys.}~}
\def\lsim{ {\ \lower-1.2pt\vbox{\hbox{\rlap{$<$}\lower5pt\vbox{\hbox{$\sim$}
}}}\ } }
\def\gsim{ {\ \lower-1.2pt\vbox{\hbox{\rlap{$>$}\lower5pt\vbox{\hbox{$\sim$}
}}}\ } }

%\font\el=cmbx10 scaled \magstep2{\obeylines \hfill October, 2020}

%\vskip 1.0 cm

\centerline{\large\bf Finite-Width Effects in Three-Body $B$ Decays}

\bigskip
\centerline{\bf Hai-Yang Cheng$^{1}$, Cheng-Wei Chiang$^{2}$, Chun-Khiang Chua$^{3}$}
\medskip
\centerline{$^1$ Institute of Physics, Academia Sinica}
\centerline{Taipei, Taiwan 115, Republic of China}
\medskip
\centerline{$^2$ Department of Physics, National Taiwan University}
\centerline{Taipei, Taiwan 106, Republic of China}
\medskip
\centerline{$^3$ Department of Physics and Center for High Energy Physics}
\centerline{Chung Yuan Christian University}
\centerline{Chung-Li, Taiwan 320, Republic of China}

\bigskip
\centerline{\bf Abstract}

\medskip

\small
It is customary to apply the so-called narrow width approximation
$\Gamma(B\to RP_3\to P_1P_2P_3)=\Gamma(B\to RP_3)\B(R\to P_1P_2)$
to extract the branching fraction of the quasi-two-body decay $B\to RP_3$, with $R$ and $P_3$ being an intermediate resonant state and a pseudoscalar meson, respectively. However, the above factorization is valid only in the zero width limit. We consider a correction parameter $\eta_R$ from finite width effects.
Our main results are:
(i) We present a general framework for computing $\eta_R$ and show that it can be expressed in terms of the normalized differential rate and determined by its value at the resonance.
(ii) We introduce a form factor $F(s_{12},m_R)$ for the strong coupling involved in the $R(m_{12})\to P_1P_2$ decay when $m_{12}$ is away from $m_R$.  We find that off-shell effects are small in vector meson productions, but prominent in the $K_2^*(1430)$, $\sigma/f_0(500)$ and $K_0^*(1430)$ resonances.
(iii) We evaluate $\eta_R$ in the theoretical framework of QCD factorization (QCDF) and in the experimental parameterization (EXPP) for three-body decay amplitudes.
In general, $\eta_R^{\rm QCDF}$ and $\eta_R^{\rm EXPP}$ are similar for vector mesons, but different for tensor and scalar resonances.  A study of the differential rates enables us to understand the origin of their differences.
(iv) Finite-width corrections to $\B(B^-\to RP)_{\rm NWA}$ obtained in the narrow width approximation are generally small, less than 10\%, but they are prominent in $B^-\to\sigma/f_0(500)\pi^-$ and $B^-\to \ov K_0^{*0}(1430)\pi^-$ decays.
The EXPP of the normalized differential rates should be contrasted with the theoretical predictions from QCDF calculation as the latter properly takes into account the energy dependence in weak decay amplitudes.
(v) It is common to use the Gounaris-Sakurai model to describe the line shape of the broad $\rho(770)$ resonance.  After including finite-width effects, the PDG value of $\B(B^-\to\rho\pi^-)=(8.3\pm1.2)\times 10^{-6}$  should be corrected to $(7.9\pm1.1)\times 10^{-6}$ in EXPP and $(7.7\pm1.1)\times 10^{-6}$ in QCDF.
(vi) For the very broad $\sigma/f_0(500)$ scalar resonance,  we use a simple pole model to describe its line shape and find a very large width effect: $\eta_\sigma^{\rm QCDF}\sim 2.15$ and $\eta_\sigma^{\rm EXPP}\sim 1.64$\,.  Consequently, $B^-\to \sigma\pi^-$ has a large branching fraction of order $10^{-5}$.
(vii) We employ the Breit-Wigner line shape to describe the production of $K_0^*(1430)$ in three-body $B$ decays and find large off-shell effects.  The smallness of $\eta^{\rm QCDF}_{K^*_0}$ relative to $\eta^{\rm EXPP}_{K^*_0}$ is ascribed to the differences in the normalized differential rates off the resonance.
(viii) In the approach of QCDF, the calculated \CP asymmetries of $B^-\to f_2(1270)\pi^-, \sigma/f_0(500)\pi^-, K^-\rho^0$ decays agree with the experimental observations. The non-observation of \CP asymmetry in $B^-\to \rho(770)\pi^-$ can also be accommodated in QCDF.

\pagebreak

\tableofcontents
\newpage

%%%%%%%%%%%%%%%%%%%%%
\section{Introduction}
%%%%%%%%%%%%%%%%%%%%%
In a three-body decay with resonance contributions,
it is a common practice to apply the factorization relation, also known as the narrow width approximation (NWA), to factorize the process as a quasi-two-body weak decay followed by another two-body strong decay.  Take a $B$ meson decay $B\to RP_3\to P_1P_2P_3$ as an example, where $R$ and $P_3$ are an intermediate resonant state and a pseudoscalar meson, respectively.  One then uses
\be \label{eq:fact}
\Gamma(B\to RP_3\to P_1P_2P_3)=\Gamma(B\to RP_3)\B(R\to P_1P_2),
\en
to extract the branching fraction of the quasi-two-body decay, $\B(B\to RP_3)$, which is then compared with theoretical predictions.  However, such an approach is valid only in the narrow width limit, $\Gamma_R\to 0$.  In other words, one should have instead
\be \label{eq:NWA}
\Gamma(B\to RP_3\to P_1P_2P_3)_{\Gamma_R\to 0}  = \Gamma(B\to RP_3)\B(R\to P_1P_2),
\en
where we have assumed that both $\Gamma(B\to RP_3)$ and $\B(R\to P_1P_2)$ are not affected by the NWA. In other words, while taking the $\Gamma(R\to P_1P_2) \to 0$ limit, the branching fraction of $R\to P_1P_2$ is assumed to remain intact.
For the case when $R$ has a finite-width, Eq.~(\ref{eq:fact}) does not hold.  Moreover, theoretical predictions of $\B(B\to RP_3)$ are normally calculated under the assumption that the both final-state particles are stable (i.e., $\Gamma_R , \Gamma_{P_3} \to 0$).  Therefore, the question is how one should extract $\B(B\to RP_3)$ from the experimental measurement of the partial rate of $B\to RP_3\to P_1P_2P_3$ and make a meaningful comparison with its theoretical predictions.

Let us define a quantity~\footnote{For later convenience, our definition of $\eta_R$ here is inverse to the one defined in~\cite{Cheng:2002mk}. A similar (but inversely) quantity ${\cal W}^{(\ell)}_R=\Gamma^{(\ell)}_R/\Gamma^{(\ell)}_{\rm R,NWL}$ was also considered in~\cite{Huber:2020pqb}, where $\Gamma_R^{(\ell)}$ is the partial-wave decay rate integrated in a region around a resonance and $\Gamma^{(\ell)}_{\rm R,NWL}$ denotes $\Gamma_R^{(\ell)}$ in the narrow width limit. }
\be \label{eq:eta}
\eta_{_R}\equiv \frac{\Gamma(B\to RP_3\to P_1P_2P_3)_{\Gamma_R\to 0}}{\Gamma(B\to RP_3\to P_1P_2P_3)}=\frac{\Gamma(B\to RP_3)\B(R\to P_1P_2)}{\Gamma(B\to RP_3\to P_1P_2P_3)}=1+\delta,
\en
so that the deviation of $\eta_{_R}$ from unity measures the degree of departure from the NWA when the width is finite. It is naively expected that the correction $\delta$ will be of order $\Gamma_R/m_R$. The quantity $\eta_R$ extrapolates the three-body decay from the physical width to the zero width. It is calculable theoretically but  depends on the line shape of the resonance and the approach of describing weak hadronic decays such as QCD factorization (QCDF), perturbative QCD and soft collinear effective theory. After taking into account the finite-width effect $\eta_R$ from the resonance, the branching fraction of the quasi-two-body decay reads
\be \label{eq:BRofRP}
\B(B\to RP_3)=\eta_{_R}{\B(B\to RP_3\to P_1P_2P_3)_{\rm expt} \over \B(R\to P_1P_2)_{\rm expt}}.
\en
Note that $\B(B\to RP_3)$ on the left-hand side of the above formula is the branching fraction under the assumption that both $R$ and $P_3$ are stable and thus have zero decay width.  Therefore, it is suitable for a comparison with theoretical calculations.

In the literature, such as the Particle Data Group~\cite{PDG}, the branching fraction of the quasi-two-body decay is often inferred from Eq.~(\ref{eq:BRofRP}) by setting $\eta_R$ equal to unity. While this is justified for narrow-width resonances, it is not for the broad ones. For example, $\Gamma_\rho/m_\rho=0.192$ for the $\rho$ vector meson, $\Gamma_{f_2}/m_{f_2}=0.146$ for the $f_2(1270)$ tensor meson, $\Gamma_\sigma/m_\sigma\sim {\cal O}(1)$ for the $\sigma/f_0(500)$ scalar meson, and $\Gamma_{K_2^*}/m_{K_2^*}\approx 0.189$ for the $K_2^*(1430)$ tensor meson. For these resonances, finite-width effects seem to be important and cannot be neglected. We shall see in this work that the deviation of $\eta_R$ from unity does not always follow the guideline from the magnitude of $\Gamma_R/m_R$.

It is worth mentioning that the finite-width effects play an essential role in charmed meson decays~\cite{Cheng:2002mk,Cheng:2003bn}. There exist some modes, {\it e.g.}, $D^0\to \rho(1700)^+K^-$,
$D^0\to K^*(1410)^-K^+$ which are not allowed kinematically can proceed through the finite-width effects.

In this work, we will calculate the parameter $\eta_R$ within the framework of QCDF for various resonances and use these examples to highlight the importance of finite-width effects.  First, we need to check the NWA relation Eq.~(\ref{eq:NWA}) both analytically and numerically. Once this is done, it is straightforward to compute $\eta_R$.

In the experimental analysis of $B\to RP_3\to P_1P_2P_3$ decays, it is customary to parameterize the amplitude as
$A(m_{12}, m_{23})=c\,F(m_{12}, m_{23})$,
where the strong dynamics is described by the function $F$ that parameterizes the intermediate resonant processes, while the information of weak interactions is encoded in the complex coefficient $c$ which is obtained by fitting to the measured Dalitz plot. The function $F$ can be further parameterized in terms of a resonance line shape, an angular dependence and Blatt-Weisskopf barrier factors.  Using the experimental parameterization of $F(m_{12}, m_{23})$, we can also compute the ratio of the three-body decay rate without and with the finite-width effects of the resonance, which we shall refer to as $\eta_R^{\rm EXPP}$. Obviously, $\eta_R^{\rm EXPP}$ is independent of $c$.
On the contrary, the weak decay amplitude of $B\to R(m_{12})P_3$ generally has some dependence on $m_{12}$ in QCDF calculations. Hence, $\eta_R^{\rm QCDF}$ is different from $\eta_R^{\rm EXPP}$ in general. It will be instructive to compare them to gain more insight to the underlying mechanism.

Although it is straightforward to estimate the parameter $\eta_R$ in a theoretical framework by computing the decay rates of the quasi-two-body decay and the corresponding three-body decay, we shall develop a general framework for the study of $\eta_R$. We will show that $\eta_R$ can be expressed in terms of a normalized differential decay rate. It turns out that $\eta_R$ is nothing but the value of the normalized differential decay rate evaluated at the contributing resonance. Not only is the calculation significantly simplified, the underlying physics also becomes more transparent.  Finally, we note in passing that while we focus on three-body $B$ meson decays in this paper to elucidate our point and explain the cause, our finding generally applies to all quasi-two-body decays.

The layout of the present paper is as follows. In Sec.~II, we present a general framework for the study of the parameter $\eta_R$ and show that it can be obtained from the normalized differential decay rate. The experimental analysis of $B\to RP_3\to P_1P_2P_3$ decays relies on a parameterization of the involved strong dynamics. This is discussed in detail in Sec.~III. We then proceed to evaluate $\eta_R^{\rm QCDF}$ within the framework of QCDF in Sec.~IV for some selected processes mediated by tensor, vector and scalar resonances, and compare them with $\eta_R^{\rm EXPP}$ determined from the experimental parameterization. We discuss our findings in Sec.~V. Sec.~VI comes to our conclusions. A more concise version of this work has been presented in~\cite{Cheng:2020mna}.

%%%%%%%%%%%%%%%%%%%%%
\section{General Framework }
%%%%%%%%%%%%%%%%%%%%%

In this section, we discuss how $\eta_R$ can be determined from a normalized differential decay rate.  We start by considering the simpler case where the mediating resonance is a scalar meson, and show that the result reduces to the usual one in the NWA.  We then generalize our discussions to resonances of arbitrary spin, and derive an important relation between $\eta_R$ and the normalized differential decay rate evaluated at the resonance mass.  Two examples of the $\rho(770)$ and $\sigma/f_0(500)$ resonances are presented at the end of the section.

\subsection{Scalar intermediate states}

We first consider the case that $R$ is a scalar resonance for simplicity.
The three-body $B\to R P_3\to P_1 P_2 P_3$ decay amplitude has the following form:
\be
A(m_{12}, m_{23})
=
\frac{{\cal M}[B\to R(m_{12}) P_3] {\cal M}[R(m_{12})\to P_1 P_2]}
{(m^2_{12}-m_R^2)+i m_R \Gamma_R},
\label{eq: A scalar}
\en
where ${\cal M}[B\to R(m_{12}) P_3] $ and ${\cal M}[R(m_{12})\to P_1 P_2]$ are weak and strong decay amplitudes of $B\to R(m_{12}) P_3$ and $R(m_{12})\to P_1 P_2$ decays, respectively, and $m^2_{ij}\equiv p^2_{ij}\equiv (p_i+p_j)^2$.
Note that at the resonance, we have
\be
i\sqrt{\pi m_R \Gamma_R} \, A(m_R,m_{23})={\cal M}[B\to R(m_R) P_3] \frac{{\cal M}[R(m_R)\to P_1 P_2]}{\sqrt{m_R \Gamma_R/\pi}},
\label{eq: A at mR}
\en
which contains the critical information of the physical $B\to R P_3$ and $R\to P_1 P_2$ decay amplitudes.

Using the standard formulas~\cite{PDG}, the three-body differential decay rate at the resonance is given by
\be
\frac{D \, \Gamma(m^2_R)}{d m^2_{12}}
=\frac{1}{(2\pi)^3}\frac{1}{32 m_B^3}\int |A(m_R,m_{23})|^2 \, dm_{23}^2,
\label{eq: formular 1}
\en
or, equivalently,
\be
\frac{D \, \Gamma(m^2_R)}{d m^2_{12}}
&=&\frac{1}{(2\pi)^5}\frac{1}{32 m_R m_B^2}
\int |A(m_R,m_{23})|^2  |\vec p_1| |\vec p_3| \, d\Omega_1\, d\Omega_3,
 \label{eq: formular 2}
\en
where $|\vec p_1|$ and $\Omega_1$ are evaluated in the $R$ rest frame.
With the help of Eq.~(\ref{eq: A at mR}), the above equation can be rewritten as
\be
\pi m_R \Gamma_R \frac{D \, \Gamma(m^2_R)}{d m^2_{12}}
&=&\frac{1}{32\pi^2}\int \left|{\cal M}[B\to R(m_R) P_3]\right|^2\frac{|\vec p_3|}{m_B^2} d\Omega_3
\non\\
&&\times
\bigg(\frac{1}{32\pi^2}\int \left|{\cal M}[R(m_R)\to P_1 P_2]\right|^2\frac{|\vec p_1|}{m_R^2} d\Omega_1\bigg) \bigg/ \Gamma_R,
\non\\
&=&\Gamma(B\to R P_3)\B(R\to P_1 P_2).
\label{eq: RP3 RPP}
\en
Hence,
we obtain
\be
\Gamma(B\to R P_3)\B(R\to P_1 P_2)
 &=&\pi m_R \Gamma_R \frac{D \, \Gamma(m^2_R)}{d m^2_{12}}
 \non\\
&=&\frac{\pi m_R \Gamma_R}{(2\pi)^3}\frac{1}{32 m_B^3}
\int_{(m_{23}^2)_{\rm min.}(m_R)}^{(m_{23}^2)_{\rm max.}(m_R)}
 |A(m_R,m_{23})|^2 dm_{23}^2.
\label{eq: Gamma B}
\en
Consequently, Eqs.~(\ref{eq: Gamma B}) and \eqref{eq:eta}
imply that $\eta_R$ is related to the normalized differential rate,
\be
\eta_R=
\frac{\displaystyle\pi m_R \Gamma_R\, \frac{D \, \Gamma (m_R^2)}{dm^2_{12}}}
{\displaystyle\int \frac{D \, \Gamma (m_{12}^2)}{dm^2_{12}} \,d m^2_{12}}
=  \pi m_R \Gamma_R\,
\frac{\displaystyle
\int |A(m_R, m_{23})|^2 dm_{23}^2}
 {\displaystyle \int  |A(m_{12}, m_{23})|^2 dm_{12}^2 \, dm_{23}^2}.
\label{eq: eta A}
\en
With the help of the following identity \footnote{This follows from the formula:
$\lim_{\epsilon\to 0}{\epsilon\over \epsilon^2+x^2}=\pi\delta(x)$.}
\be
\lim_{\Gamma_R\to 0} \frac{m_R \Gamma_R/\pi}{(m^2_{12}-m_R^2)^2+m^2_R\Gamma^2_R}
=\delta(m^2_{12}-m^2_R),
\label{eq: identity}
\en
one can readily verify that $\eta_R$ given in the above equation approaches unity in the narrow width limit, reproducing the well-known result of Eq.~\eqref{eq:NWA}.

\subsection{General case}

Although Eqs.~(\ref{eq: Gamma B}) and (\ref{eq: eta A}) are derived for the case of a scalar resonance, they can be generalized to a more generic case, where the resonance particle has spin $J$.
Instead of Eq.~(\ref{eq: A scalar}), the general amplitude has the following expression:
\be \label{eq: A generic}
A(m_{12}, m_{23})
=
{\cal M}(m_{12}, m_{12})R_J(m_{12}) {\cal T}_J(m_{12},m_{23}),
\en
where ${\cal M}(m_{12}, m_{12})$ is a regular function containing the information of $B\to R(m_{12})P_3$ weak decay and $R(m_{12})\to P_1P_2$ strong decay, $R_J$ describes the line shape of the resonance and ${\cal T}_J$ encodes the angular dependence.
Resonant contributions are commonly depicted by the relativistic Breit-Wigner (BW) line shape,
\be
R_J^{\rm BW}(m_{12})=\frac{1}{(m^2_{12}-m_R^2)+i m_R \Gamma_R(m_{12})}.
\label{eq: RJ BW}
\en
In general, the mass-dependent width is expressed as
\be \label{eq:GammaR}
\Gamma_R(m_{12})=\Gamma_R^0\left( {q\over q_0}\right)^{2J+1}
{m_R\over m_{12}} {X^2_J(q)\over X^2_J(q_0)},
\en
where $q=|\vec{p}_1|=|\vec{p}_2|$ is the center-of-mass (c.m.) momentum in the rest frame of the resonance $R$, $q_0$ is the value of $q$ when $m_{12}$ is equal to the pole mass $m_R$, and $X_J$ is a Blatt-Weisskopf barrier factor given by
\be \label{eq:XJ}
X_0(z)=1, \qquad X_1(z)=\sqrt{1\over (z\,r_{\rm BW})^2+1}, \qquad
X_2(z)=\sqrt{1\over (z\,r_{\rm BW})^4+3(z\,r_{\rm BW})^2+9},
\en
with $r_{{\rm BW}}\approx 4.0\,{\rm GeV}^{-1}$.  In Eq.~(\ref{eq:GammaR}), $\Gamma_R^0$ is the nominal total width of $R$ with $\Gamma_R^0=\Gamma_R(m_R)$. One advantage of using the energy-dependent decay width is that $\Gamma_R(m_{12})$ vanishes when $m_{12}$ is below the $m_1+m_2$ threshold (see the expression of $q$ in Eq.~(\ref{eq: q p3}) below).  Hence, the factor $q^{2L+1}$ with $L$ being the orbital angular momentum between $R$ and $P_3$ guarantees the correct threshold behavior. The rapid growth of this factor for angular momenta $>L$ is compensated at higher energies by the Blatt-Weisskopf barrier factors \cite{PDG}.

From Eqs.~(\ref{eq: TJ epsilon}), (\ref{eq: T'J}), (\ref{eq:Af2pi_bar}), (\ref{eq:Arhopi_bar}) and (\ref{eq:sigmapipi_1}) below, we find that the angular distribution term ${\cal T}_J$ in Eq.~(\ref{eq: A generic}) at the resonance is governed by the Legendre polynomial $P_J(\cos\theta)$, where $\theta$ is the angle between $\vec{p}_1$ and $\vec{p}_3$ measured in the rest frame of the resonance (see also~\cite{Asner:2003gh}).
Explicitly, we have
\be
P_0(\cos\theta)=1, \qquad P_1(\cos\theta)=\cos\theta, \qquad P_2(\cos\theta)={1\over 2}(-1+3\cos^2\theta),
\en
and
\be
{\cal T}_0(m_R, m_{23})=1, \qquad
{\cal T}_1(m_R, m_{23})\propto \cos\theta, \qquad
{\cal T}_2(m_R, m_{23})\propto 1-3\cos^2\theta.
\label{eq: TJ}
\en
Note that ${\cal T}_0(m_{12}, m_{23})=1$ throughout the entire phase space. This means that the strong and weak amplitudes can always be separated for the scalar case, as shown in Eq.~(\ref{eq: A scalar}).

Instead of Eq.~(\ref{eq: A at mR}), the general amplitude {\it at the resonance} takes the form
\be
i\sqrt{\pi m_R \Gamma^0_R} \, A(m_R,m_{23})
=\sum_\lambda {\cal M}_\lambda[B\to R(m_R) P_3] \frac{{\cal M}_\lambda[R(m_R)\to P_1 P_2]}{\sqrt{m_R \Gamma^0_R/\pi}},
\label{eq: A at mR generic}
\en
where $\lambda$ is the helicity of the resonance $R$.
Such a relation is expectable because there is a propagator of the resonance $R$ in the amplitude $A(m_{12}, m_{23})$ and its denominator reduces to $i m_R \Gamma^0_R$ on the mass shell of $m_{12}$ while its numerator reduces to a polarization sum of the polarization vectors, producing the above structure after contracted with the rest of the amplitude.

From Eq.~(\ref{eq: formular 2}) and Eq.~\eqref{eq: A at mR generic}, we have
\be
 \frac{D \, \Gamma(m^2_R)}{d m^2_{12}}
&=&\frac{1}{(32\pi^2)^2 \pi m_R (\Gamma^0_R)^2}
\int \bigg |\sum_\lambda {\cal M}_\lambda[B\to R(m_R) P_3] {\cal M}_\lambda[R(m_R)\to P_1 P_2]\bigg |^2
\frac{|\vec p_1|}{m_R^2} \frac{|\vec p_3|}{m_B^2} d\Omega_3
d\Omega_1,
\non\\
\label{eq: dGamma generic}
\en
where $|\vec p_1|$ and $\Omega_1$ are evaluated in the $R$ rest frame.
In this frame the sum over helicities in the amplitude can be replaced by the sum over spins. Consequently, ${\cal M}_\lambda[B\to R(m_R) P_3]$ and ${\cal M}_\lambda[R(m_R)\to P_1 P_2]$ are proportional  to $Y^*_{J\lambda}(\Omega_3)$ and $Y_{J\lambda}(\Omega_1)$, respectively.~\footnote{
For example, in the $J=1$ case and at the resonance, ${\cal M}_\lambda[B\to V(m_R) P_3]$ is proportional to
$p_B\cdot\epsilon^*(p_{12},\lambda)$, while ${\cal M}_\lambda[V(m_R)\to P_1 P_2]$  is proportional to
$\epsilon(p_{12},\lambda)\cdot(p_1-p_2)$.
See also Eq.~(\ref{eq: TJ epsilon}) below.
It can be easily seen that in the $V$ rest frame, these terms provide the  $Y^*_{1\lambda}(\Omega_3)$ and $Y_{1\lambda}(\Omega_1)$ factors, respectively.
}
As a cross check, we note that Eq.~(\ref{eq: TJ}) can be reproduced by using the well-known addition theorem of spherical harmonics, $(2J+1)P_J(\cos\theta)=4\pi\sum_\lambda Y^*_{J\lambda}(\Omega_3) Y_{J\lambda}(\Omega_1)$. Alternatively, we can start from Eq.~(\ref{eq: TJ}) and make use of the addition theorem to obtain the $\sum_\lambda Y^*_{J\lambda}(\Omega_3) Y_{J\lambda}(\Omega_1)$ factor.

We now see that %, from the orthogonality of the spherical harmonics,
the interference terms in Eq.~(\ref{eq: dGamma generic}) from different helicities (or spins) vanish after the angular integrations.  As a result, we obtain
\be
\pi m_R \Gamma_R \frac{D \, \Gamma(m^2_R)}{d m^2_{12}}
&=&\frac{1}{32\pi^2}\sum_\lambda \int \left|{\cal M}_\lambda[B\to R(m_R) P_3]\right|^2\frac{|\vec p_3|}{m_B^2} d\Omega_3
\non\\
&&\times
\bigg(\frac{1}{32\pi^2}\int \left|{\cal M}_\lambda[R(m_R)\to P_1 P_2]\right|^2\frac{|\vec p_1|}{m_R^2} d\Omega_1\bigg)/ \Gamma^0_R
\non\\
&=&\Gamma(B\to R P_3)\B(R\to P_1 P_2),
\label{eq: RP3 RPP generic}
\en
where we have made use of the fact that the branching fraction $\B(R\to P_1 P_2)$ is independent of the helicity (or spin) in the last step.
The above equation agrees with Eq.~(\ref{eq: Gamma B}), and consequently Eq.~(\ref{eq: eta A}) follows.

Eq.~(\ref{eq: eta A}) can be easily generalized to the case with identical particles in the final state. Let $P_2$ and $P_3$ be identical particles so that the decay amplitude reads ${\cal M}=A(m_{12},m_{23})+A(m_{13},m_{23})$, giving
\be
\eta_R=\pi m_R \Gamma_R\,
\frac{\displaystyle 2\int |A(m_R, m_{23})|^2 dm_{23}^2}
{\displaystyle \int  |A(m_{12}, m_{23})+A(m_{13}, m_{23})|^2 dm_{12}^2 \, dm_{23}^2}.
\label{eq: eta A identical particle}
\en

Furthermore, from Eqs.~(\ref{eq: identity}) and (\ref{eq: A generic}), we see that in the narrow width limit, the amplitude squared takes the form
\be
|A(m_{12}, m_{23})|^2_{\Gamma^0_R\to 0}=\pi m_R\Gamma^0_R\delta(m^2_{12}-m_R^2) |A(m_R, m_{23})|^2.
\en
Substituting this into Eq.~(\ref{eq: eta A}), we obtain $\eta_R= 1$ in the limit of zero width, hence reproducing the well known result in Eq.~\eqref{eq:NWA}.

%In Eq.~(\ref{eq: A generic}), the ${\cal M}(m_{12},m_{23})$ terms contain both weak and strong dynamics,
%while the $R_J  {\cal T}_J$ factor is basically known.
In this work, we will consider $A(m_{12}, m_{23})$ using the experimental parameterization (EXPP) and the QCDF calculation and compute $\eta^{\rm EXPP}_R$ and $\eta^{\rm QCDF}_R$, respectively.
In the latter case, we shall see that in the narrow width limit, the weak interaction part of the amplitude does reduce to the QCDF amplitude of the $B\to R P_3$ decay.
We will also show explicitly the validity of the factorization relation in the zero width limit for several selected examples of three-body decays involving tensor, vector and scalar mediating resonances.

\subsection{$\eta_R$ and the normalized differential rate}

As suggested by Eq.~(\ref{eq: eta A}), $\eta_R$ can be expressed in terms of the normalized differential rate,
\be
\eta_R=\pi m_R \Gamma_R \frac{d\tilde\Gamma (m_R^2)}{dm^2_{12}}
=\frac{1}{2}\pi\Gamma_R \frac{d\tilde\Gamma (m_R)}{dm_{12}},
\label{eq: eta dGamma tilde}
\en
where we have defined
\be
\frac{d\tilde\Gamma (m_{12}^2)}{dm^2_{12}}
\equiv
{\frac{D \, \Gamma (m_{12}^2)}{dm^2_{12}}}
\bigg/
{\int \frac{D \, \Gamma (m_{12}^2)}{dm^2_{12}} dm_{12}^2}.
\label{eq: normalized dGamma}
\en
Hence $\eta_R$ is determined by the value of the normalized differential rate at the resonance.
It should be noted that as the normalized differential rate is always positive and normalized to 1 after integration,
the value of $d\tilde \Gamma(m_R) /dm_{12}$ is anticorrelated with $d\tilde \Gamma(m_{12}) /dm_{12}$ elsewhere.
Hence, it is the shape of the (normalized) differential rate that matters in the determination of $\eta_R$.

The above point can be made more precise. When $\Gamma_R/m_R\ll 1$, we expect that the normalized differential rate around the resonance is reasonably well described as
\be
\frac{d\tilde\Gamma(m^2_{12})}{dm^2_{12}}\bigg|_{m^2_{12}\simeq m^2_R}\simeq \frac{m^2_R \Gamma^2_R}{(m^2_{12}-m_R^2)^2+m^2_R\Gamma^2_R} \frac{d\tilde\Gamma(m^2_R)}{dm^2_{12}}.
\en
It is straightforward to show that as a result, Eq.~(\ref{eq: eta dGamma tilde}) can be approximated by
\be
\eta_R\simeq
\frac{\pi}{2\tan^{-1}2}
\int_{(m_R-\Gamma_R)^2}^{(m_R+\Gamma_R)^2} \frac{d\tilde\Gamma(m^2_{12})}{dm^2_{12}} dm^2_{12},
\en
or, equivalently,
\be
\eta_R\simeq
\frac{\pi}{2\tan^{-1}2}
\int_{m_R-\Gamma_R}^{m_R+\Gamma_R} \frac{d\tilde\Gamma(m_{12})}{dm_{12}} dm_{12}
=\frac{\pi}{2\tan^{-1}2}\bigg(1
-\int_{\rm elsewhere} \frac{d\tilde\Gamma (m_{12})}{dm_{12}} dm_{12}\bigg).
\label{eq: anticorrelation}
\en
It becomes clear that $\eta_R$ represents the fraction of rates around the resonance and is anticorrelated with the fraction of rates off the resonance.

The EXPP and the QCDF approaches may have different shapes in the differential rates, resulting in different $\eta_R$'s, {\it i.e.}, $\eta^{\rm EXPP}_R\neq\eta^{\rm QCDF}_R$ in general.
The two-body rate reported by experiments should be corrected using $\eta_R=\eta^{\rm EXPP}_R$ in Eq.~\eqref{eq:BRofRP}, as the data are extracted using the experimental parameterization.
On the other hand, the experimental parameterization on normalized differential rates should be compared with the theoretical predictions from QCDF calculation as the latter takes into account the energy dependence of weak interaction amplitudes. As we shall show in Sec. V.A,
the usual experimental parameterization ignores the momentum dependence in weak dynamics and would lead to incorrect extraction of quasi-two-body decay rates in the case of broad resonances, as contrasted with the estimates using the QCDF approach.

\subsection{Formula of $\eta_R$ in the case of the Gounaris-Sakurai line shape}

A popular choice for describing the broad $\rho(770)$ resonance is the Gounaris-Sakurai (GS) model~\cite{Gounaris:1968mw}.  It was employed by both BaBar~\cite{BaBarpipipi} and LHCb~\cite{Aaij:3pi_1,Aaij:3pi_2} Collaborations in their analyses of the $\rho(770)$ resonance in the $B^-\to \pi^+\pi^-\pi^-$ decay.
The GS line shape for $\rho(770)$ is given by
\be
T_\rho^{\rm GS}(s)={ 1+D \, \Gamma_\rho^0/m_\rho \over s-m^2_{\rho}-f(s)+im_{\rho}\Gamma_{\rho}(s)},
\label{eq: T GS}
\en
where
\be
\Gamma_{\rho}(s)=\Gamma_{\rho}^0\left( {q\over q_0}\right)^3
{m_{\rho}\over \sqrt{s}} {X^2_1(q)\over X^2_1(q_0)},
\en
the Blatt-Weisskopf barrier factor is given in Eq.~(\ref{eq:XJ}), $\Gamma_\rho^0$ is the nominal total $\rho$ width with $\Gamma_\rho^0=\Gamma_\rho(m_\rho^2)$. The quantities $q$ and $q_0$ are already introduced before in Sec.~II.A.
In this model, the real part of the pion-pion scattering amplitude with an intermediate $\rho$ exchange calculated from the dispersion relation is taken into account by the $f(s)$ term in the propagator of $T_\rho^{\rm GS}(s)$. Unitarity far from the pole mass is thus ensured.
Explicitly,
\be \label{eq:f(s)}
f(s)=\Gamma_\rho^0{m_\rho^2\over q_0^3}\left[ q^2[h(\sqrt{s})-h(m_\rho)]+(m_\rho^2-s)q_0^2\left.{dh\over ds}\right\vert_{m_\rho}\right],
\en
and
\be
h(s)={2\over \pi}{q\over \sqrt{s}}\ln\left( {\sqrt{s}+2q\over 2m_\pi}\right), \qquad
\left.{dh\over ds}\right\vert_{m_\rho}=h(m_\rho)\left[ {1\over 8q_0^2}-{1\over 2m_\rho^2}\right]+{1\over 2\pi m_\rho^2}.
\en
The constant parameter $D$ is given by
\be
D={3\over \pi}\,{m_\pi^2\over q_0^2}\ln \left( {m_\rho+2q_0\over 2m_\pi} \right)+{m_\rho\over 2\pi q_0}-{m_\pi^2 m_\rho\over \pi q^3_0}.
\en

The $( 1+D \, \Gamma_\rho^0/m_\rho)$ factor in Eq.~(\ref{eq: T GS}) will modify the relation in Eq.~(\ref{eq: A at mR generic}) into
\be
i\frac{\sqrt{\pi m_R \Gamma^0_R}}{( 1+D \, \Gamma_\rho^0/m_\rho)} \, A^{\rm GS}(m_\rho,m_{23})
=\sum_\lambda {\cal M}_\lambda[B\to \rho(m_R) P_3] \frac{{\cal M}_\lambda[\rho(m_R)\to P_1 P_2]}{\sqrt{m_R \Gamma^0_R/\pi}}
\label{eq: A at mR GS}
\en
instead.
It can be easily seen that Eqs.~(\ref{eq: Gamma B}), (\ref{eq: eta A}) and (\ref{eq: eta dGamma tilde}) all need to be corrected by the factor of $1/(1+D \, \Gamma_\rho^0/m_\rho)^2$ accordingly. More explicitly, Eqs.~(\ref{eq: eta A}) and (\ref{eq: eta dGamma tilde}) should be replaced by
\be
\eta^{\rm GS}_\rho
=
\frac{\pi m_\rho \Gamma_\rho}{(1+D \, \Gamma_\rho^0/m_\rho)^2}\,
\frac{\displaystyle \int |A(m_\rho, m_{23})|^2 dm_{23}^2}
{\displaystyle \int  |A(m_{12}, m_{23})|^2 dm_{12}^2 \, dm_{23}^2}
\label{eq: eta A GS}
\en
and
\be
\eta^{\rm GS}_\rho=\frac{\pi m_\rho \Gamma^0_\rho}{(1+D \, \Gamma_\rho^0/m_\rho)^2} \frac{d\tilde\Gamma (m_\rho^2)}{dm^2_{12}}
=\frac{\pi\Gamma^0_\rho}{2(1+D \, \Gamma_\rho^0/m_\rho)^2} \frac{d\tilde\Gamma (m_\rho)}{dm_{12}},
\label{eq: eta dGamma tilde GS}
\en
respectively.

\subsection{Formula of $\eta_R$ in the case of the $\sigma/f_0(500)$ resonance}

As stressed in~\cite{Pelaez:2015qba}, the scalar resonance $\sigma/f_0(500)$ is very broad and cannot be described by the usual Breit-Wigner line shape. The partial wave amplitude does not resemble a Breit-Wigner shape with a clear peak and a simultaneous steep rise in the phase. The mass and width of the $\sigma$ resonance are identified from the associated pole position $\sqrt{s_\sigma}$ of the partial wave amplitude in the second Riemann sheet as $\sqrt{s_\sigma}=m_\sigma-i\Gamma_\sigma/2$~\cite{Pelaez:2015qba}. Hence, we shall follow the LHCb Collaboration~\cite{Aaij:3pi_2} to use a simple pole description
\be \label{eq:T sigma}
T_\sigma(s)={1\over s-s_\sigma}={1\over s-m_\sigma^2+\Gamma_\sigma^2(s)/4+im_\sigma\Gamma_\sigma(s)},
\en
with
\be
\Gamma_{\sigma}(s)=\Gamma_{\sigma}^0\left( {q\over q_0}\right)
{m_{\sigma}\over \sqrt{s}},
\en
and
$\Gamma_\sigma(m_\sigma^2)=\Gamma_\sigma^0$.

The factor of $1/[(\Gamma_\sigma^0)^2/4+i m_\sigma\Gamma^0_\sigma]=(i m_\sigma\Gamma^0_\sigma)^{-1}(1-i \Gamma^0_\sigma/4 m_\sigma)^{-1}$
in Eq.~(\ref{eq:T sigma}) at the resonance will modify the relation in Eq.~(\ref{eq: A at mR generic}) into
\be
\bigg(1-i \frac{\Gamma^0_\sigma}{4 m_\sigma}\bigg)i\sqrt{\pi m_R \Gamma^0_R} \, A(m_\sigma,m_{23})
=\sum_\lambda {\cal M}_\lambda[B\to \sigma(m_R) P_3] \frac{{\cal M}_\lambda[\sigma(m_R)\to P_1 P_2]}{\sqrt{m_R \Gamma^0_R/\pi}}
\label{eq: A at mR sigma}
\en
instead.
It can be easily seen that Eqs.~(\ref{eq: Gamma B}), (\ref{eq: eta A}) and (\ref{eq: eta dGamma tilde}) all need to be corrected by the factor of $r_\sigma\equiv[1+ (\Gamma^0_\sigma/4 m_\sigma)^2]$ accordingly. More explicitly, Eqs.~(\ref{eq: eta A}) and (\ref{eq: eta dGamma tilde}) should be replaced by
\be
\eta_\sigma
= \pi r_\sigma  m_\sigma \Gamma_\sigma\,
\frac{\displaystyle \int |A(m_\sigma, m_{23})|^2 dm_{23}^2}
{\displaystyle \int  |A(m_{12}, m_{23})|^2 dm_{12}^2 \, dm_{23}^2}
\label{eq: eta A sigma}
\en
and
\be
\eta_\sigma=\pi r_\sigma m_\sigma \Gamma^0_\sigma\frac{d\tilde\Gamma (m_\sigma^2)}{dm^2_{12}}
=\frac{\pi r_\sigma \Gamma^0_\sigma}{2} \frac{d\tilde\Gamma (m_\sigma)}{dm_{12}},
\label{eq: eta dGamma tilde sigma}
\en
respectively.

\section{Differential rates and $\eta^{\rm EXPP}_R$ using the experimental parameterization }

The following parameterization of the decay amplitude is widely used in the experimental studies of $B\to R P_3\to P_1 P_2 P_3$ decays (see, for example,~\cite{BaBar:Kmpippim}):
\be \label{eq:cF}
A(m_{12}, m_{23})
=c F(m_{12}, m_{23})
=c R_J(m_{12})\times X_J(p_3)\times X_J(p_1)\times T_J(m_{12}, m_{23}),
\en
where $R_J$  describes the line shape of the resonance introduced before in Eq.~(\ref{eq: RJ BW}), $X_J$ is the Blatt-Weisskopf barrier form factor as defined in Eq.~(\ref{eq:XJ}) with both $p_1$ and $p_3$ evaluated in the $R(m_{12})$ rest frame,
$T_J(m_{12}, m_{23})$ is an angular distribution term given by~\cite{Asner:2003gh},
\be
T_0(m_{12}, m_{23})&=&1,
\non\\
T_1(m_{12}, m_{23})&=&m^2_{23}-m^2_{13}+\frac{(m^2_B-m^2_3) (m_1^2-m_2^2)}{m^2_{12}},
\non\\
T_2(m_{12}, m_{23})
&=&\bigg(m^2_{23}-m^2_{13}+\frac{(m^2_B-m^2_3) (m_1^2-m_2^2)}{m^2_{12}}\bigg)^2
\non\\
&&-\frac{1}{3}\bigg(m^2_{12}-2m^2_B-2m_3^2+\frac{(m^2_B-m^2_3)^2}{m^2_{12}}\bigg)
\non\\
&&\qquad\times \bigg(m^2_{12}-2m^2_1-2m^2_2+\frac{(m^2_1-m^2_2)^2}{m^2_{12}}\bigg),
\label{eq: TJ trans}
\en
and $c$ is an unknown complex coefficient to be fitted to the data.
Basically the information of weak decay amplitude is included in $c$.  However, it is assumed to be a constant and have no dependence on the energy or momentum of the decay products.

The quantities $\Gamma(B\to R P_3)\B(R\to P_1 P_2)$ and $\eta^{\rm EXPP}_R$ can be obtained by using Eqs.~(\ref{eq: Gamma B}) and (\ref{eq: eta A}) as
\be
\Gamma(B\to R P_3)\B(R\to P_1 P_2)
&=&\frac{|c|^2}{8\pi^2 m_R \Gamma_R }\frac{1}{32 m_B^3}
\int_{(m_{23}^2)_{\rm min.}(m_R)}^{(m_{23}^2)_{\rm max.}(m_R)}
(|X_J(p_3)|^2 |X_J(p_1)|^2)_{m_{12}\to m_R}
\non\\
&&\qquad\qquad\qquad\qquad\qquad\qquad\qquad
\times |T_J(m_R, m_{23})|^2
 dm_{23}^2,
 \label{eq: Gamma B BaBar}
\en
and
\be
\eta^{\rm EXPP}_R={\pi\over m_R\Gamma_R}\,
\frac
{\displaystyle \int_{(m_{23}^2)_{\rm min.}(m_R)}^{(m_{23}^2)_{\rm max.}(m_R)}
 (|X_J(p_3)|^2 |X_J(p_1)|^2)_{m_{12}\to m_R}
\times |T_J(m_R, m_{23})|^2 dm_{23}^2}
{\displaystyle \int  |R_J(m_{12})\times X_J(p_3)\times X_J(p_1)\times T_J(m_{12}, m_{23}))|^2
dm_{12}^2 \, dm_{23}^2}.
 \label{eq: eta F BaBar}
\en
Note that being a constant, the factor $c$ in $A(m_{12}, m_{23})$ is canceled out between the numerator and the denominator in $\eta^{\rm EXPP}_R$.
One can readily verify that $\eta^{\rm EXPP}_R$ approaches unity in the narrow width limit by virtue of Eq.~(\ref{eq: identity}).

We can express $\eta^{\rm EXPP}_R$ in terms of the normalized differential rate,
\be
\eta^{\rm EXPP}_R
=\pi m_R \Gamma_R \frac{d\tilde\Gamma (m_R^2)}{dm^2_{12}}
=\frac{1}{2}\pi\Gamma_R \frac{d\tilde\Gamma (m_R)}{dm_{12}},
\label{eq: eta dGamma tilde BaBar}
\en
with
\be
\frac{d\tilde\Gamma (m_{12}^2)}{dm^2_{12}}=
\frac
{|R_J(m_{12})|^2
\int
 |X_J(p_3)\times X_J(p_1)
\times T_J(m_R, m_{23})|^2 dm_{23}^2}
{\displaystyle \int  |R_J(m_{12})\times X_J(p_3)\times X_J(p_1)\times T_J(m_{12}, m_{23})|^2
dm_{12}^2 \, dm_{23}^2}.
\label{eq: dGamma tilde BaBar}
\en
In the case that $P_2$ and $P_3$ are identical particles, we shall use Eq.~(\ref{eq: eta A identical particle}) to obtain $\eta^{\rm EXPP}_R$, giving
\be
\eta^{\rm EXPP}_R=\frac
{\displaystyle 2\pi\int_{(m_{23}^2)_{\rm min.}(m_R)}^{(m_{23}^2)_{\rm max.}(m_R)}
 (|X_J(p_3)|^2 |X_J(p_1)|^2)_{m_{12}\to m_R}
\times |T_J(m_R, m_{23})|^2 dm_{23}^2}
{\displaystyle m_R \Gamma_R \int  |R_J(m_{12})\times X_J(p_3)\times X_J(p_1)\times T_J(m_{12}, m_{23}))+(2\leftrightarrow 3)|^2
dm_{12}^2 \, dm_{23}^2}.
 \label{eq: eta F BaBar identical}
\en
Note that when the Gounaris-Sakurai line shape is used in place of $R_J$, we should use Eqs.~(\ref{eq: eta A GS})
and (\ref{eq: eta dGamma tilde GS}), instead of Eq.~(\ref{eq: eta dGamma tilde BaBar}), while Eqs.~(\ref{eq: eta F BaBar}) and (\ref{eq: dGamma tilde BaBar}) are still valid.
For the case of the $\sigma$ resonance, we should use Eqs.~(\ref{eq: eta A sigma})
and (\ref{eq: eta dGamma tilde sigma}).

In the case of narrow width, it is legitimate to use a complex constant $c$ to represent the weak dynamics.
As noted previously, it is the shape of the entire normalized differential rate that matters in determining $\eta_R$.
Hence, in the case of finite-width, the momentum dependence in the weak amplitude will play some role.

There are some subtleties in the angular terms.
Note that Eq.~(\ref{eq: TJ trans}) was obtained with transversality conditions, $p^\mu_{12}\epsilon_\mu=0$ and $p^\mu_{12}p^\nu_{12}\epsilon_{\mu\nu}=0$, enforced for $J=1, 2,$~\cite{Asner:2003gh}
\be
T_1(m_{12}, m_{23})&=&\sum_\lambda (p_B+p_3)_\mu \epsilon^{*\mu}(p_{12},\lambda) \epsilon^\nu(p_{12},\lambda) (p_1-p_2)_\nu
\non\\
&=& -2\vec{p}_1\cdot\vec{p}_3=-2q |\vec{p}_3|\cos\theta,
\non\\
T_2(m_{12}, m_{23})&=&\sum_\lambda (p_B+p_3)_\mu (p_B+p_3)_\nu
\epsilon^{*\mu\nu}(p_{12},\lambda)\epsilon^{\alpha\beta} (p_{12},\lambda)(p_1-p_2)_\alpha(p_1-p_2)_\beta
\non\\
&=& {4\over 3}\left[3(\vec{p}_1\cdot \vec{p}_3)^2-(|\vec{p}_1||\vec{p}_3|)^2\right]
={4\over 3}q^2 |\vec{p}_3|^2(3\cos^2\theta-1),
\label{eq: TJ epsilon}
\en
where $\epsilon^\mu$ and $\epsilon^{\mu\nu}$ are the polarization vector and tensor, respectively, and
\be
q = |\vec{p}_1|=|\vec{p}_2|&=&{\sqrt{[m_{12}^2-(m_1+m_2)^2][m_{12}^2-(m_1-m_2)^2]}\over 2m_{12} }, \non \\
|\vec{p}_3| &=& \left({(m_B^2-m_3^2-m_{12}^2)^2\over 4m_{12}^2} - m_3^2\right)^{1/2},
\label{eq: q p3}
\en
with $q= |\vec{p}_{1,2}|$ and $|\vec p_3|$ being the momenta of $P_{1,2}$ and $P_3$ in the $R(m_{12})$ rest frame, respectively.~\footnote{Note that $|\vec p_3|$ is related to $\tilde p_c$ through the relation $\tilde p_c=(m_{12}/m_B)|\vec{p}_3|$, where $\tilde p_c$ is the c.m. momentum of $P_3$ or $R(m_{12})$ in the $B$ rest frame. This relation can be easily verified using the conservation of momentum.}
Note that in Eq.~(\ref{eq: TJ epsilon}), the factor contracted with $(p_B+p_3)$ comes from the $B\to R(m_{12}) P$ weak decay amplitude, while the one contracted with $(p_1-p_2)$ comes from the $R(m_{12})\to P_1 P_2$ strong decay amplitude. To obtain the $\cos\theta$ dependence, it is useful to recall $\sum_\lambda\epsilon^*_{\mu}(p_{12},\lambda) \epsilon_\nu(p_{12},\lambda)=g^i_\mu g^j_\nu \delta_{ij}$ in the rest fame of $R(m_{12})$.

Alternatively, using the standard expressions of vector and tensor propagators, which are contracted with the $B\to R(m_{12}) P$ and the $R(m_{12})\to P_1 P_2$ parts, we expect the angular terms to take the following forms,
\be
T'_1(m_{12}, m_{23})&=&m^2_{23}-m^2_{13}+\frac{(m^2_B-m^2_3) (m_1^2-m_2^2)}{m^2_R},
\non\\
T'_2(m_{12}, m_{23})
&=&\bigg(m^2_{23}-m^2_{13}+\frac{(m^2_B-m^2_3) (m_1^2-m_2^2)}{m^2_R}\bigg)^2
\non\\
&&-\frac{1}{3}\bigg(m^2_{12}-2m^2_B-2m_3^2+\frac{(m^2_B-m^2_3)^2}{m^2_R}\bigg)
\non\\
&&\qquad\times \bigg(m^2_{12}-2m^2_1-2m^2_2+\frac{(m^2_1-m^2_2)^2}{m^2_R}\bigg).
\label{eq: T'J}
\en
The transversality condition, however, is not imposed on the above equations as the denominators become $m_R^2$ instead of $m^2_{12}$.
In general, these $T'_J$ cannot be expressed as Eq.~(\ref{eq: TJ epsilon}) except on the mass shell of $p_{12}$, where these two angular terms coincide, {\it i.e.}, $T'_J(m_R, m_{23})=T_J(m_R, m_{23})$.
In the case of a vector resonance, except for modes with the intermediate resonance decaying to daughters of different masses, these two angular terms are identical throughout the entire phase space.
We will also consider the case where the transversality condition is not imposed.

\section{Analysis in the QCD factorization approach}

In this section we will evaluate the decay amplitudes of $B\to RP_3$ and $B\to RP_3\to P_1P_2P_3$ within the framework of QCD factorization~\cite{BBNS,BN}. For the latter, its general amplitude has the expression
\be \label{eq:general three-bodyA}
A(B\to RP_3\to P_1P_2P_3) &\equiv& A(m_{12},m_{23})   \\
&=& g^{R\to P_1P_2}F(s_{12},m_R)\tilde A(B\to R(m_{12})P_3)R_J(m_{12}) {\cal T}_J(m_{12},m_{23}), \non
\en
where $g^{R\to P_1P_2}$ is the strong coupling constant associated with the strong decay $R(m_{12})\to P_1P_2$,  $F(s,m_R)$ is a form factor to be introduced later (see Eq.~(\ref{eq:FF for coupling}) below),
$R_J$ is the resonance line shape, and ${\cal T}_J$ is the angular distribution function. In this work, we find
\be
{\cal T}_0=1, \qquad {\cal T}_1=2q\cos\theta, \qquad {\cal T}_2={q^2\over\sqrt{6}}(1-3\cos^2\theta),
\en
where $\theta$ is the angle between $\vec{p}_1$ and $\vec{p}_3$ measured in the rest frame of the resonance and $q$ is given before in Eq.~(\ref{eq: q p3}). In Eq.~(\ref{eq:general three-bodyA}), the weak decay amplitude   $\tilde A(B\to R(m_{12})P_3)$  will be reduced to the QCDF amplitude
$A(B\to R(m_R)P_3)$ when $m_{12}\to m_R$.

Taking the relativistic Breit-Wigner line shape Eq.~(\ref{eq: RJ BW}), it follows from Eqs.~(\ref{eq: A at mR}) and (\ref{eq: formular 2}) that
\be
\frac{D \, \Gamma(m^2_R)}{d m^2_{12}}
&=&\frac{1}{(2\pi)^5}\frac{1}{32 m_R m_B^2}
\int |A(m_R,m_{23})|^2  |\vec p_1| |\vec p_3| \, d\Omega_1\, d\Omega_3, \non \\
&=& {1\over \pi m_R \Gamma_R^2}\bigg(\frac{1}{32\pi^2}\int |A(B\to R(m_R) P_3)|^2\frac{|\vec p_3|}{m_B^2} d\Omega_3\bigg)
\bigg(\frac{1}{32\pi^2}\int |g^{R\to P_1P_2}{\cal T}_J|^2\frac{|\vec{p}_1|}{m_R^2} d\Omega_1\bigg),
\non\\
&=&  {1\over \pi m_R\Gamma_R} \Gamma(B\to R P_3)\B(R\to P_1 P_2).
\label{eq: RP3 RPP}
\en
Indeed,  it is straightforward to show that the partial rate of $R(m_R)\to P_1P_2$ given by
\be
\Gamma(R\to P_1P_2)=\frac{1}{32\pi^2}\int |g^{R\to P_1P_2}{\cal T}_J|^2\frac{q_0}{m_R^2} d\Omega_1
\en
has the following expressions (see also Eq.~(2.39) of~\cite{Cheng:2020ipp})
\be \label{eq:partialwidth_1}
 \Gamma_{S\to P_1P_2}={q_0\over 8\pi m_S^2}g_{S\to P_1P_2}^2,\quad
 \Gamma_{V\to P_1P_2}={q_0^3\over 6\pi m_V^2}g_{V\to P_1P_2}^2,\quad
 \Gamma_{T\to P_1P_2}={q_0^5\over 60\pi m_T^2}g_{T\to P_1P_2}^2,
\en
for different types of resonances.
Therefore, the decay rate of $B\to R(m_R)P_3$ can be related to the differential rate of $B\to RP_3\to P_1P_2P_3$ at the resonance. This means that $\eta_R$ can be obtained from the normalized differential rate as shown in Eq.~(\ref{eq: eta A}) or (\ref{eq: eta dGamma tilde}).

Most of the input parameters employed in this section such as decay constants, form factors, CKM matrix elements can be found in Appendix A of \cite{Cheng:2020ipp}.

%%%%%%%%%%%%%%%%%%%%%
\subsection{Tensor resonances}
%%%%%%%%%%%%%%%%%%%%%

We begin with the tensor resonances and consider the three-body decay processes: $B^-\to f_2(1270)\pi^-\to\pi^+\pi^-\pi^-$ and $B^-\to \ov K_2^{*0}(1430)\pi^-\to K^-\pi^+\pi^-$.
Since the decay widths of $f_2(1270)$ and $K_2^*(1430)$ are around 187 and 109 MeV, respectively, it is na{\"i}vely expected that the deviation of $\eta_{f_2}$ from unity is larger than that of $\eta_{K_2^*}$ in both QCDF and EXPP schemes. We shall see below that this is not respected in the QCDF scheme and barely holds in the EXPP scheme.

\subsubsection{$f_2(1270)$}

\noindent \underline{$B^-\to f_2(1270)\pi^-$ decay in QCDF}
\vskip 0.5 cm

Consider the process $B^-\to f_2(1270)\pi^-\to\pi^+\pi^-\pi^-$. In QCDF, the amplitude of the quasi-two-body decay $B^-\to f_2(1270)\pi^-$ is given by~\cite{Cheng:TP}
\be \label{eq:Af2pi}
A(B^-\to f_2(1270)\pi^-) &=&
  \frac{G_F}{2}\sum_{p=u,c}\lambda_p^{(d)}   \non \\
  &&\times \Bigg\{ \left[a_1 \delta_{pu}+a^p_4+a_{10}^p-(a^p_6+a^p_8) r_\chi^\pi +\beta_2^p \delta_{pu}+\beta_3^p+\beta^p_{\rm 3,EW}\right]_{f_2\pi}  X^{({B} f_2,\pi)} \non \\
   &&~~~~~+  \Big[a_2\delta_{pu}+2(a_3^p+a_5^p)+a_4^p+r_\chi^{f_2}a_6^p+{1\over 2}(a_7^p+a_9^p)-{1\over 2}(a_{10}^p+r_\chi^{f_2}a_8^p)   \non \\
   &&~~~~~+ \beta_2^p \delta_{pu}+\beta_3^p+\beta^p_{\rm 3,EW}\Big]_{\pi f_2} X^{({B} \pi, f_2)}
    \Bigg\},
\en
where $\lambda_p^{(d)}\equiv V_{pb} V^*_{pd}$, and
\be \label{eq:Xf2}
X^{({B} f_2,\pi)} = 2f_{\pi}A_0^{ B f_2}(m_\pi^2){m_{f_2}\over m_B}\epsilon^{*\mu\nu}(0)p_{B\mu}p_{B\nu}, \quad
%= 2\sqrt{2\over 3}\,f_{\pi}{m_B\over m_{f_2}}p_c^2A_0^{ B f_2}(m_{\pi}^2), \
X^{({B} \pi, f_2)}= 2  f_{f_2}  m_B\,p_cF_1^{B\pi}(m_{f_2}^2),
\en
with $p_c$ being the c.m. momentum of either $f_2$ or $\pi^-$ in the $B$ rest frame. The chiral factors $r_\chi^\pi$ and $r_\chi^{f_2}$ in Eq.~(\ref{eq:Af2pi}) are given by
\be
r_\chi^\pi(\mu)={2m_\pi^2\over m_b(\mu)(m_u+m_d)(\mu)}, \qquad r_\chi^{f_2}(\mu) = \frac{2m_{f_2}}{m_b(\mu)}\,\frac{f_{f_2}^\perp(\mu)}{f_{f_2}}.
\en
For the definition of the scale-dependent decay constants $f_{f_2}$ and $f_{f_2}^\perp$, see, for example, Ref.~\cite{Cheng:TP}.
The coefficients $\beta_i^p$ describe weak annihilation contributions to the decay. The order of the arguments in the $a_i^p(M_1M_2)$ and $\beta_i^p(M_1M_2)$ coefficients  is dictated by the subscript $M_1M_2$ given in Eq.~(\ref{eq:Af2pi}).

In Eq.~(\ref{eq:Xf2}), $X^{({B} f_2,\pi)}$ is factorizable and given by $\la\pi|J^\mu|0\ra\la f_2|J'_\mu|B\ra$, while $X^{({B} \pi, f_2)}$ is a nonfactorizable amplitude as the factorizable one $\la f_2 |J^{\mu}|0\ra\la \pi^-|J'_{\mu}|B^- \ra$ vanishes owing to the fact that the tensor meson cannot be produced through the $V-A$ current.  Nevertheless, beyond the factorization approximation, contributions proportional to the decay constant $f_{f_2}$ can be produced from vertex, penguin and spectator-scattering corrections~\cite{Cheng:TP}. Therefore, when the strong coupling $\alpha_s$ is turned off, the nonfactorizable contributions vanish accordingly.

The factorizable amplitude $X^{({B} f_2,\pi)}$ can be further simplified by working in the $B$ rest frame  and assuming that $f_2$ ($\pi$) moves along the $-z$ ($z$) axis~\cite{Cheng:TP}. In this case, $p_B^\mu=(m_B, 0,0,0)$ and $\epsilon^{*\mu\nu}(0)=\sqrt{2/3}\,\epsilon^{*\mu}(0)\epsilon^{*\nu}(0)$ with $\epsilon^{*\mu}(0)=(p_c,0,0,E_{f_2})^\mu/m_{f_2}$ and, consequently,
\be \label{eq:Xf2-2}
X^{({B} f_2,\pi)}= 2\sqrt{2\over 3}\,f_{\pi}{m_B\over m_{f_2}}p_c^2A_0^{ B f_2}(m_{\pi}^2).
\en

\vskip 0.3 cm
\noindent \underline{Three-body decay $B^-\to f_2(1270)\pi^-\to \pi^+\pi^-\pi^-$}
\vskip 0.3 cm

As shown in~\cite{Cheng:2020ipp},  the decay amplitude $\A_{f_2(1270)}\equiv A(B^-\to \pi^-f_2(1270)\to \pi^-(p_1)\pi^+(p_2)\pi^-(p_3))$ evaluated in the factorization approach
\footnote{The study of charmless three-body $B$ decays in the factorization approach can be found in~\cite{CCS:nonres,Cheng:2013dua,Cheng:2016shb,Cheng:2020ipp} and references therein.
}
arises from the matrix element $\la \pi^+(p_2)\pi^-(p_3)|(\bar ub)|B^-\ra^{f_2}\la \pi^-(p_1)|(\bar du)|0\ra$,  where $(\bar q_1q_2)\equiv \bar q_1\gamma_\mu(1-\gamma_5)q_2$ and the superscript $f_2$ denotes the contribution from the $f_2$ resonance to the matrix element  $\la \pi^+(p_2)\pi^-(p_3)|(\bar ub)|B^-\ra$.
We shall use the relativistic Breit-Wigner line shape to describe the distribution of $f_2(1270)$:
\be
T_{f_2}^{\rm BW}(s)={1\over s-m^2_{f_2}+im_{f_2}\Gamma_{f_2}(s)},
\en
with
\be
\Gamma_{f_2}(s)=\Gamma_{f_2}^0\left( {q\over q_0}\right)^5
{m_{f_2}\over \sqrt{s}} {X_2^2(q)\over X_2^2(q_0)},
\en
where the quantities $q$, $q_0$, $X_2$ and $\Gamma_T^0$ are already introduced before in Eq.~(\ref{eq:GammaR}).
One advantage of using the energy-dependent decay width is that $\Gamma_{f_2}(s)$ will vanish when $s$ goes below the $2\pi$ threshold.

Consequently,
\be
&&\la \pi^+(p_2)\pi^-(p_3)|(\bar ub)|B^-\ra^{f_2}\la \pi^-(p_1)|(\bar du)|0\ra  \non \\
&=&
\la \pi^+(p_2)\pi^-(p_3)|f_2\ra \,T_{f_2}^{\rm BW}(s_{23}) \la f_2|(\bar ub)|B^-\ra\la \pi^-(p_1)|(\bar du)|0\ra  \non \\
&=& \sum_\lambda \vp^{*{\mu\nu}}(\lambda)p_{2\mu} p_{3\nu}\,g^{f_2\to \pi^+\pi^-} \,T_{f_2}^{\rm BW}(s_{23}) {2m_{f_2}\over m_B}f_\pi A_0^{Bf_2}(m_\pi^2)\vp_{\alpha\beta}(\lambda)p_B^\alpha p_1^\beta
\non \\
&=& {2 m_{f_2}\over m_B}g^{f_2\to \pi^+\pi^-}f_\pi A_0^{Bf_2}(m_\pi^2) \,T_{f_2}^{\rm BW}(s_{23})\left[{1\over 3}(|\vec{p}_1||\vec{p}_2|)^2-(\vec{p}_1\cdot\vec{p}_2)^2\right],
\en
where we have followed~\cite{Wang:2010ni} for the definition of the $B\to T$ transition form factors \footnote{The $B\to T$ transition form factors defined in~\cite{Wang:2010ni} and~\cite{Cheng:TP} differ  by a factor of $i$. We shall use the former as they are consistent with the normalization of $B\to S$ transition given in~\cite{CCH}.}
and employed the relation~\cite{Dedonder:2010fg,Asner:2003gh}
\be \label{eq:tensorangular}
\sum_\lambda \vp^{*{\mu\nu}}(\lambda)\vp_{\alpha\beta}(\lambda)p_{2\mu} p_{3\nu} p_B^\alpha p_1^\beta={1\over 3}(|\vec{p}_1||\vec{p}_2|)^2-(\vec{p}_1\cdot\vec{p}_2)^2,
\en
with
\be \label{eq:3momentum}
|\vec{p}_1|=\left({(m_B^2-m_\pi^2-s_{23})^2\over 4s_{23}} - m_\pi^2\right)^{1/2}, \quad |\vec{p}_2|=|\vec{p}_3|=q={1\over 2}\sqrt{s_{23}-4m_\pi^2}.
\en
Hence, factorization leads to
\be
\A_{f_2(1270)} &=& {1\over\sqrt{2}}{G_F\over \sqrt{2}}\sum_{p=u,c}\lambda_p^{(d)} \left[a_1 \delta_{pu}+a^p_4+a_{10}^p-(a^p_6+a^p_8) r_\chi^\pi\right]    \\
&&\times {2 m_{f_2}\over m_B}g^{f_2\to \pi^+\pi^-}f_\pi\, A_0^{Bf_2}(m_\pi^2)T_{f_2}^{\rm BW}(s_{23}){1\over 3}|\vec{p}_1|^2|\vec{p}_2|^2(1-3\cos^2\theta)+(s_{23}\leftrightarrow s_{12}), \non
\en
where the identical particle effect due to the two identical $\pi^-$ has been taken into account.
Comparing with Eq.~(\ref{eq:Af2pi}), we see that the nonfactorizable contribution characterized by $X^{({B} \pi, f_2)}$ and the weak annihilation described by $\beta^p$ terms are absent in the na{\"i}ve factorization approach. We shall use the QCDF expression for $B^-\to f_2(1270)\pi^-$ and write
\be \label{eq:Af2pi_a}
\A_{f_2(1270)} =g^{f_2\to \pi^+\pi^-}\,T_{f_2}^{\rm BW}(s_{23}){q^2\over\sqrt{6}} (1-3\cos^2\theta)\tilde A(B^-\to f_2(m_{23})\pi^-) +(s_{23}\leftrightarrow s_{12}),
\en
with
\be \label{eq:Af2pi_bar}
\tilde A(B^-\to f_2(m_{23})\pi^-) &=&
  \frac{G_F}{2}\sum_{p=u,c}\lambda_p^{(d)}{m_{f_2}^2\over s_{23}}
  \Bigg\{ \left[a_1 \delta_{pu}+a^p_4+\cdots+\beta^p_{\rm 3,EW} \right]_{f_2\pi}  \tilde X^{({B} f_2,\pi)} \non \\
   &&+  \Big[a_2\delta_{pu}+2(a_3^p+a_5^p)+\cdots+\beta^p_{\rm 3,EW} \Big]_{\pi f_2} \tilde X^{({B} \pi, f_2)}  \Bigg\},
\en
where
\be \label{}
\tilde X^{({B} f_2,\pi)}= 2\sqrt{2\over 3}\,f_{\pi}{m_B\over m_{f_2}}\tilde p_c^2A_0^{ B f_2}(m_{\pi}^2), \qquad
\tilde X^{({B} \pi, f_2)}&=& 2  f_{f_2} m_B\,\tilde p_cF_1^{B\pi}(s_{23}),
\en
and
\be
\tilde p_c=\left({(m_B^2-m_\pi^2-s_{23})^2\over 4m_B^2} - m_\pi^2\right)^{1/2}.
\en
It is easily seen that $\tilde A(B^-\to f_2(m_{23})\pi^-)$ is reduced to the QCDF amplitude
$A(B^-\to f_2\pi^-)$ given in Eq.~(\ref{eq:Af2pi}) when $m_{23}\to m_{f_2}$.

Before proceeding, we would like to address an issue. The strong coupling constant $|g^{f_2(1270)\to\pi^+\pi^-}|=18.56\,{\rm GeV}^{-1}$ extracted from the measured $f_2(1270)$ width (see Eq.~(\ref{eq:Gamma_2body}) below) is for the physical $f_2(1270)$. When $f_2$ is off the mass shell, especially when $s_{23}$ is approaching the upper bound of $(m_B-m_\pi)^2$, it is necessary to account for the off-shell effect. For this purpose, we shall follow~\cite{Cheng:FSI} to introduce a form factor $F(s,m_R)$ parameterized as \footnote{Note that the form factor $F(t,m)$ used in~\cite{Cheng:FSI} is for the $t$-channel off-shell effect.}
\be \label{eq:FF for coupling}
F(s,m_R)=\left( {\Lambda^2+m_R^2 \over \Lambda^2+s}\right)^n,
\en
with the cutoff $\Lambda$ not far from the resonance,
\be
\Lambda=m_R+\beta\Lambda_{\rm QCD},
\en
where the parameter $\beta$ is expected to be of order unity. We shall use $n=1$, $\Lambda_{\rm QCD}=250$ MeV and $\beta=1.0\pm0.2$ in subsequent calculations.

Finally, the decay rate is given by
\be \label{eq:Gamma_f2}
&&\Gamma(B^-\to f_2\pi^-\to \pi^+\pi^-\pi^-) \non
\\
&&={1\over 2}\,{1\over(2\pi)^3 32 m_B^3}\int_{(m_\pi+m_\pi)^2}^{(m_B-m_\pi)^2}ds_{23}\int_{(s_{12})_{\rm min}}^{(s_{12})_{\rm max}}ds_{12}\, |\A_{f_2}|^2
\non \\
&&={1\over 2}\,{1\over(2\pi)^3 32 m_B^3}\int_{(m_\pi+m_\pi)^2}^{(m_B-m_\pi)^2}ds_{23}\int_{(s_{12})_{\rm min}}^{(s_{12})_{\rm max}}ds_{12}\Bigg\{  {|g^{f_2\to \pi^+\pi^-}|^2 F(s_{23},m_{f_2})^2\over (s_{23}-m^2_{f_2})^2+m_{f_2}^2\Gamma_{f_2}^2(s_{23})}
\non \\
&&~~~~\times {q^4\over 6}(1-3\cos^2\theta)^2\left|\tilde A(B^-\to f_2\pi^-)\right|^2 +(s_{23}\leftrightarrow s_{12}) + {\rm interference~terms} \Bigg\},
\en
where the factor of 1/2 accounts for the identical-particle effect.
Note that $\cos\theta$ can be expressed as in terms of of $s_{12}$ and $s_{23}$:
\be
\cos\theta=a(s_{23})s_{12}+b(s_{23}),
\en
with~\cite{Bediaga:2015}
\be
a(s) &=& {1\over (s-4m_\pi^2)^{1/2}\left({(m_B^2-m_\pi^2-s)^2\over 4s} - m_\pi^2\right)^{1/2}}, \non \\
b(s) &=& -{m_B^2+3m_\pi^2-s\over 2(s-4m_\pi^2)^{1/2}\left({(m_B^2-m_\pi^2-s)^2\over 4s} - m_\pi^2\right)^{1/2}}.
\en
It follows that $(s_{12})_{\rm min}=-(1+b)/a$ and $(s_{12})_{\rm max}=(1-b)/a$. It is straightforward to show that
\be
\int_{(s_{12})_{\rm min}}^{(s_{12})_{\rm max}}ds_{12}(1-3\cos^2\theta)^2={8\over 5a}={16\over 5}{m_B\over\sqrt{s_{23}}}q\tilde p_c.
\en

In the narrow width limit, we have
\be
{m_{f_2}\Gamma_{f_2}(s)\over (s-m^2_{f_2})^2+m_{f_2}^2\Gamma_{f_2}^2(s)} \xlongrightarrow[]{\; \Gamma_{f_2}\to 0 \;}\pi\delta(s-m_{f_2}^2).
\en
Under the NWA, $|g^{f_2\to\pi^+\pi^-}|^2/\Gamma_{f_2}$ is finite as it is proportional to the branching fraction $\B(f_2\to \pi^+\pi^-)$. Due to the Dirac $\delta$-function in the above equation,
we have $s_{23}\to m_{f_2}^2$ in the zero width limit. As a result, $\tilde p_c\to p_c$, $q\to q_0$, $\tilde X^{({B} f_2,\pi)}\to X^{({B} f_2,\pi)}$,
$\tilde X^{({B} \pi, f_2)}\to X^{({B} \pi, f_2)}$ and $\tilde A(B^-\to f_2\pi^-)\to A(B^-\to f_2\pi^-)$. Likewise, the second term in Eq.~(\ref{eq:Gamma_f2}) with the replacement $s_{23}\leftrightarrow s_{12}$ has a similar expression. However, the interference term vanishes in the NWA due to different $\delta$-functions.
Using
\be \label{eq:Gamma_2body}
 \Gamma_{f_2\to \pi^+\pi^-} = {q_0^5\over 60\pi m_{f_2}^2}\left|g^{f_2\to \pi^+\pi^-}\right|^2, \qquad
 \Gamma_{B^-\to f_2\pi^-} = {p_c\over 8\pi m_B^2}|A(B^-\to f_2\pi^-)|^2,
\en
we are led to the desired factorization relation:
\be \label{eq:factorization_f2}
\Gamma(B^-\to f_2\pi^-\to \pi^+\pi^-\pi^-) \xlongrightarrow[]{\; \Gamma_{f_2}\to 0 \;}
 \Gamma(B^-\to f_2\pi^-) \B(f_2\to \pi^+\pi^-).
\en

\vskip 0.3 cm
\noindent \underline{Numerical results}
\vskip 0.3 cm

To compute $B^-\to f_2\pi^-$ and the three-body decay $B^-\to f_2\pi^-\to \pi^+\pi^-\pi^-$, we need to know the values of the flavor operators $a_i^p(M_1, M_2)$. In the QCDF approach, the flavor operators have the expressions~\cite{BBNS,BN}
 \be \label{eq:ai}
  a_i^{p}(M_1, M_2) =
 \left(c_i+{c_{i\pm1}\over N_c}\right)N_i(M_2)
  + {c_{i\pm1}\over N_c}\,{C_F\alpha_s\over
 4\pi}\Big[V_i(M_2)+{4\pi^2\over N_c}H_i(M_1M_2)\Big]+P_i^{p}(M_2),
 \en
where $i=1,\cdots,10$,  the upper (lower) sign is for odd (even) $i$, $c_i$ are the Wilson coefficients,
$C_F=(N_c^2-1)/(2N_c)$ with $N_c=3$, $M_2$ is the emitted meson,
and $M_1$ shares the same spectator quark with the $B$ meson. The detailed expressions for the
vertex corrections $V_i(M_2)$, hard spectator interactions
$H_i(M_1M_2)$  and  penguin contractions $P_i^p(M_2)$ for $M_1M_2=TP$ and $PT$ can be found in~\cite{Cheng:TP}. Note that the parameters $N_i(M)$ in Eq.~(\ref{eq:ai}) vanish if $M$ is a tensor meson; otherwise, it is equal to one.
Therefore, the coefficient $a_2(\pi f_2)$ appearing in Eq.~(\ref{eq:Af2pi}) vanishes when the strong coupling $\alpha_s$ is turned off.
We see from Table~\ref{tab:aiTP} that $a_i^p(f_2 P)$ and $a_i^p(Pf_2)$ can be quite different.

It is known that power corrections in QCDF always involve troublesome endpoint divergences. For example,
the annihilation amplitude has endpoint divergences even at twist-2
level, and the hard spectator scattering diagram at twist-3 order is power
suppressed and possesses soft and collinear divergences arising from the soft
spectator quark. Since the treatment of endpoint divergences is model-dependent,  we shall follow~\cite{BBNS} to model the endpoint
divergence $X\equiv\int^1_0 dx/\bar x$ in the annihilation and hard spectator
scattering diagrams as
\be \label{eq:XA}
 X_A=\ln\left({m_B\over \Lambda_h}\right)(1+\rho_A e^{i\phi_A}), \qquad
 X_H=\ln\left({m_B\over \Lambda_h}\right)(1+\rho_H e^{i\phi_H}),
\en
with $\Lambda_h$ being a typical hadronic scale of 0.5 GeV.
In this work we use
\be
\rho_A^{TP}=\rho_A^{PT}=0.7\,, \qquad \phi_A^{TP}=\phi_A^{PT}=-30^\circ,
\en
leading to
\be
&& \beta_2^p(f_2\pi)=0.023-0.010i, \qquad~~ (\beta_3^p+\beta_{\rm 3,EW}^p)(f_2\pi)=-0.047+ 0.053i,  \non \\
&& \beta_2^p(\pi f_2)=-0.033+0.018i, \qquad (\beta_3^p+\beta_{\rm 3,EW}^p)(\pi f_2)=-0.050+ 0.047i,
\en
for both $p=u$ and $c$.

Following~\cite{Cheng:TP}, we obtain the branching fraction and \CP asymmetry for $B^-\to f_2(1270)\pi^-$ as
\be \label{eq:2bodyf2:QCDF}
\B(B^-\to f_2(1270)\pi^-)_{\rm QCDF} &=& (2.65^{+1.29}_{-1.22})\times 10^{-6}, \non
\\
A_{C\!P}(B^-\to f_2(1270)\pi^-)_{\rm QCDF} &=& (46.7^{+32.6}_{-62.5})\%,
\en
where the decay constants $f_{f_2}=102\pm6$ MeV and $f_{f_2}^\perp=117\pm25$ MeV  both at $\mu=1$ GeV~\cite{Cheng:tensor}, the form factors $A_0^{Bf_2(1270)}(0)=0.13\pm0.02$, derived from large energy effective theory (see Table II of~\cite{Cheng:TP}), and
\be
F_1^{B\pi}(q^2)={0.26\pm0.03\over 1-{q^2\over m_{B^*}^2}}\left( 1+{0.64{q^2\over m_{B^*}^2}\over 1-0.40 {q^2\over m_{B}^2}}\right)
\en
have been used.
The theoretical errors correspond to the uncertainties due to the variation of Gegenbauer moments, decay constants, quark masses, form factors, the $\lambda_B$ parameter for the $B$ meson wave function and the power-correction parameters $\rho_{A,H}$, $\phi_{A,H}$ (see~\cite{Cheng:TP} for details), all added in quadrature.
In the narrow width limit, we find the central values
\be
\B(B^-\to f_2(1270)\pi^-\to \pi^+\pi^-\pi^-)_{\Gamma_{f_2}\to 0} &=& 1.485\times 10^{-6}, \non \\
A_{C\!P}(B^-\to f_2(1270)\pi^-\to \pi^+\pi^-\pi^-)_{\Gamma_{f_2}\to 0} &=& 46.23\%.
\en
Since $\B(f_2(1270)\to\pi^+\pi^-)=(0.842^{+0.029}_{-0.009})\times {2\over 3}$, it is easily seen that the factorization relation Eq.~(\ref{eq:factorization_f2}) is indeed numerically valid in the narrow width limit.

\begin{table}[t]
\caption{Numerical values of the flavor operators $a_i^p(M_1M_2)$ for $M_1M_2=f_2(1270)\pi$ and $\pi f_2(1270)$ at the scale $\mu=\ov m_b(\ov m_b)=4.18$ GeV. }
\label{tab:aiTP}
\begin{center}
\begin{tabular}{ l c c | l r r} \hline \hline
 $a_i^p$ & ~~$f_2\pi$~~ & ~~~$\pi f_2$~~~ & ~~$a_i^p$~~  & $f_2\pi$~~~~~~~~ & $\pi f_2$~~~~~~~~ \\
\hline
 $a_1$ & ~~~$1.011+0.014i$~~~ & ~~$-0.035+0.014i$~~ & ~~$a_6^c$ & $-0.053-0.005i$ & $(6.3+1.6i)10^{-3}$  \\
 $a_2$ & $0.123-0.080i$ & $0.133-0.078i$ &  ~~$a_7$ & $(-0.2+3.4i)10^{-5}$ & $(9.5-3.4i)10^{-5}$ \\
 $a_3$ & $0.0014+0.0027i$ & $-0.006+0.003i$ & ~~$a_8^u$ & $(3.6-1.0i)10^{-4}$ & $(-2.1+0.1i)10^{-5}$  \\
 $a_4^u$ & $-0.027-0.014i$ & $0.0064-0.0016i$ & ~~$a_8^c$ & $(3.4-0.5i)10^{-4}$  &  $(3.3+1.0i)10^{-5}$\\
 $a_4^c$ & $-0.032-0.006i$ & $0.0091+0.0064i$ & ~~$a_9$  & $(-9.1-0.1i)10^{-3}$ & $(3.0-1.2i)10^{-4}$\\
 $a_5$ & $0.0009-0.0031i$ & $-0.008+0.003i$ & ~~$a_{10}^u$ & $(-8.2+6.2i)10^{-4}$  & $(-9.6+7.0i)10^{-4}$ \\
 $a_6^u$ & $-0.050-0.014i$ & $-(3.52+0.02i)10^{-3}$  & ~~$a_{10}^c$ & $(-8.5+6.7i)10^{-4}$ & $(-9.4+7.5i)10^{-4}$ \\
\hline \hline
\end{tabular}
\end{center}
\end{table}

%====================================================================
\begin{figure}[t]
\begin{center}
\includegraphics[width=0.7\textwidth]{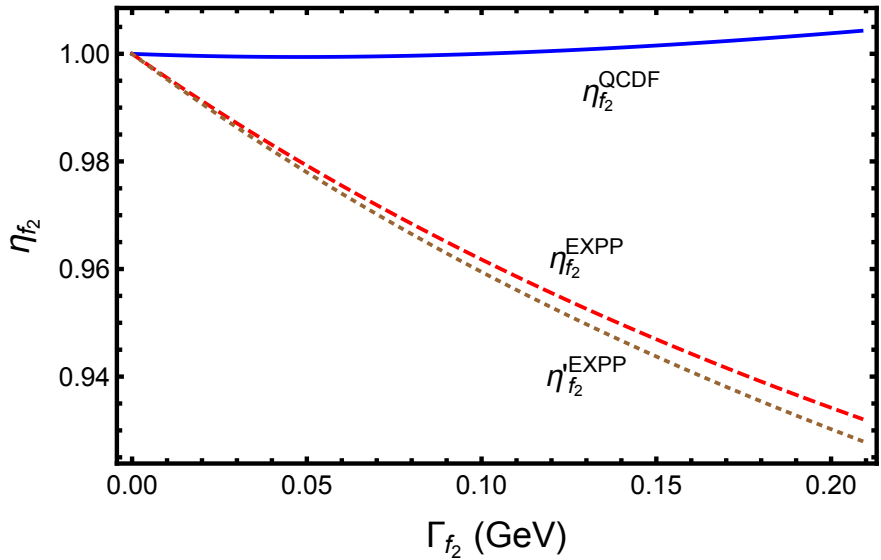}
\vspace{0.1cm}
\caption{The parameter $\eta_{f_2}$ as a function of the $f_2(1270)$ width, where the solid curve is derived from the QCDF calculation and the dashed (dotted) curve from the experimental parameterization (EXPP) with (without) the transversality condition imposed.
}
\label{fig:eta_f2}
\end{center}
\end{figure}
%=====================================================================

For the finite-width $\Gamma_{f_2}^0=186.7^{+2.2}_{-2.5}$ MeV~\cite{PDG}, we find
\be \label{eq:3bdoyf2_QCDF}
\B(B^-\to f_2(1270)\pi^-\to \pi^+\pi^-\pi^-) &=& (1.48^{+0.42}_{-0.37})\times 10^{-6}
~~~\left[ (1.52^{+0.43}_{-0.38})\times 10^{-6} \right], \non \\
A_{C\!P}(B^-\to f_2(1270)\pi^-\to \pi^+\pi^-\pi^-) &=& (44.56^{+0.41}_{-0.39})\%
~~~\left[ (47.20^{+0.45}_{-0.43})\% \right],
\en
where the values in square parentheses are obtained with the form factor $F(s,m_{f_2})$ being set as unity.
They are in agreement with the recent LHCb measurements~\cite{Aaij:3pi_1,Aaij:3pi_2}
\be
\B(B^-\to f_2(1270)\pi^-\to \pi^+\pi^-\pi^-)_{\rm LHCb} &=& (1.37\pm0.26)\times 10^{-6}, \non \\
A_{C\!P}(B^-\to f_2(1270)\pi^-\to \pi^+\pi^-\pi^-)_{\rm LHCb} &=& (46.8\pm7.7)\%,
\en
and consistent with the earlier BaBar measurements~\cite{BaBarpipipi}
\be
\B(B^-\to f_2(1270)\pi^-\to \pi^+\pi^-\pi^-)_{\rm BaBar} &=& (0.9\pm0.2^{+0.3}_{-0.1})\times 10^{-6}, \non \\
A_{C\!P}(B^-\to f_2(1270)\pi^-\to \pi^+\pi^-\pi^-)_{\rm BaBar} &=& (41\pm25^{+18}_{-15})\%.
\en
Notice that a large \CP asymmetry in the $f_2(1270)$ component was firmly established by the LHCb Collaboration.

We are now in the position to compute the parameter $\eta_{f_2(1270)}$ defined in Eq.~(\ref{eq:eta})
\be
\eta_{f_2}= \frac{\Gamma(B^-\to f_2(1270)\pi^-)\B(f_2(1270)\to \pi^+\pi^-)}{\Gamma(B^-\to f_2(1270)\pi^-\to \pi^+\pi^-\pi^-)}.
\en
From Eqs.~(\ref{eq:2bodyf2:QCDF}) and (\ref{eq:3bdoyf2_QCDF})   we find
\be
\eta_{f_2(1270)}^{\rm QCDF}=1.003^{+0.001}_{-0.002} \quad~~ (0.9743\pm0.0003).
\en
Since the theoretical uncertainties in the numerator and denominator essentially cancel out, the errors on $\eta_{f_2}$ mainly arise from the uncertainties in $\beta$ and the $f_2$ width. As discussed in Sec. II, $\eta_R$ can be expressed in terms of the normalized differential rate.
In general, the calculation done in this way is simpler.
From Eqs.~(\ref{eq: eta A identical particle}) and (\ref{eq:Af2pi_a}) we obtain the same result for $\eta_{f_2}^{\rm QCDF}$.
The dependence of the parameter $\eta_{f_2}$ on the width $\Gamma_{f_2}$ is plotted as the solid blue curve in Fig.~\ref{fig:eta_f2}.
It is somewhat surprising that the deviation of $\eta_{f_2}^{\rm QCDF}$ from unity is very tiny, even though $\Gamma_{f_2}/m_{f_2}$ is about $0.146$\,.

The parameter $\eta_{f_2}^{\rm EXPP}$ is calculated
using Eq.~(\ref{eq: eta A identical particle}) together with the experimental parameterization, Eq.~(\ref{eq:cF}) for $A(m_{12},m_{23})$. Its dependence on the $f_2(1270)$ width is depicted by the dashed red curve in Fig.~\ref{fig:eta_f2}.
At the resonance, we obtain
\be
\eta_{f_2(1270)}^{\rm EXPP}=0.937^{+0.006}_{-0.005}\,.
\en
We see that the the physical with $\Gamma_{f_2}^0=186.7^{+2.2}_{-2.5}$ MeV, the results in the QCDF and EXPP schemes differ by about 7\%.

\subsubsection{$K_2^*(1430)$}
We next turn to the $B^-\to \ov K_2^{*0}(1430)\pi^-\to K^-\pi^+\pi^-$ decay. The QCDF amplitude of the quasi-two-body $B^-\to \ov K_2^{*0}(1430)\pi^-$ decay is given by~\cite{Cheng:TP}
\be \label{eq:AK2pi}
A(B^-\to \ov K_2^{*0}\pi^-) &=&
  \frac{G_F}{\sqrt{2}}\sum_{p=u,c}\lambda_p^{(s)}
    \Big[a_4^p+r_\chi^{K_2^*}a_6^p-{1\over 2}(a_{10}^p+r_\chi^{K_2^*}a_8)   \non \\
   &&~~~~+ \beta_2^p \delta_{pu}+\beta_3^p+\beta^p_{\rm 3,EW}\Big]_{\pi K_2^*}  X^{({B} \pi,\bar K_2^*)},
\en
with  $\lambda_p^{(s)}\equiv V_{pb} V^*_{ps}$ and
\be \label{eq:XK2}
X^{({B} \pi, K_2^*)}= 2  f_{K_2^*}  m_B\,p_cF_1^{B\pi}(m_{K_2^*}^2).
\en
Note that this decay proceeds  only through nonfactorizable diagrams.

Analogous to the $f_2(1270)$ resonance, the decay amplitude $\A_{K_2^*(1430)}\equiv A(B^-\to \ov K_2^{*0}(1430)\pi^-\to K^-(p_1)\pi^+(p_2)\pi^-(p_3)$ reads (see the second term of Eq.~(\ref{eq:Af2pi_a}))
\be
\A_{K_2^*(1430)} =g^{\ov K_2^{*0}\to K^-\pi^+}\,F(s_{12},m_{K_2^*})T_{K_2^*}^{\rm BW}(s_{12}){q^2\over\sqrt{6}} (1-3\cos^2\theta)\tilde A(B^-\to \ov K_2^{*0}\pi^-),
\en
with
\be
q={\sqrt{[s_{12}-(m_K+m_\pi)^2][s_{12}-(m_K-m_\pi)^2]}\over 2\sqrt{s_{12}} }
\en
and
\be \label{eq:Af2pi_bar}
\tilde A(B^-\to \ov K_2^{*0}\pi^-) &=&
  \frac{G_F}{\sqrt{2}}\sum_{p=u,c}\lambda_p^{(s)} {m_{K_2^*}^2\over s_{12}}
  \Big[a_4^p+\cdots+\beta^p_{\rm 3,EW}\Big]_{\pi K_2^*}  \tilde X^{({B} \pi,\bar K_2^*)},
\en
where $\tilde X^{({B} \pi,\bar K_2^*)}$ has the same expression as $X^{({B} \pi,\bar K_2^*)}$ except for a replacement of $p_c$ by $\tilde p_c$ and $F_1^{B\pi}(m^2_{K_2^*})$ by $F_1^{B\pi}(s_{12})$.
Following the previous case, it is straightforward to show that the factorization relation
\be \label{eq:factorization_K2}
\Gamma(B^-\to \ov K_2^{*0}\pi^-\to K^-\pi^+\pi^-) \xlongrightarrow[]{\; \Gamma_{K_2^*}\to 0 \;}
 \Gamma(B^-\to \ov K_2^{*0}\pi^-) \B(\ov K_2^{*0}\to K^-\pi^+),
\en
holds in the NWA.

%====================================================================
\begin{figure}[t]
\begin{center}
\includegraphics[width=0.7\textwidth]{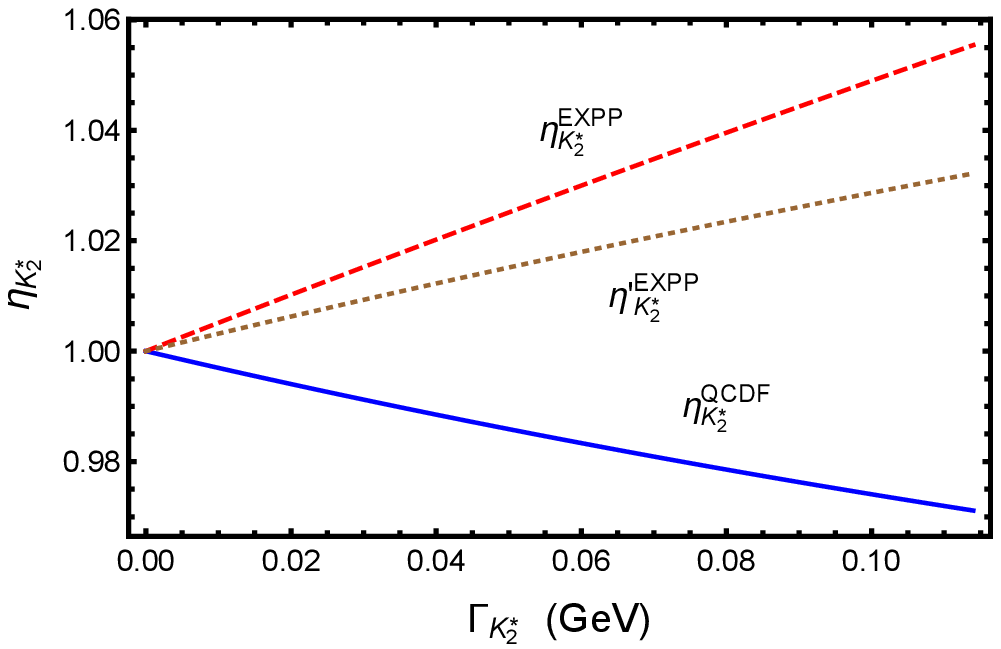}
\vspace{0.1cm}
\caption{Same as Fig. \ref{fig:eta_f2} for the $\ov K_2^*(1430)$  mediating the  $B^-\to K^-\pi^+\pi^-$ decay.}
\label{fig:eta_K2}
\end{center}
\end{figure}
%=====================================================================

In QCDF, we obtain
\be
\B(B^-\to \ov K_2^{*0}(1430)\pi^-)_{\rm QCDF} &=& (2.60^{+9.07}_{-2.53})\times 10^{-6}, \non
\\
A_{CP}(B^-\to \ov K_2^{*0}(1430)\pi^-)_{\rm QCDF} &=& (1.72^{+2.12}_{-1.95})\%,
\en
where the decay constants $f_{K_2^*}=118\pm5$ MeV and $f_{K_2^*}^\perp=77\pm14$ MeV  at $\mu=1$ GeV~\cite{Cheng:tensor}, and the penguin annihilation effects
\be
\beta_2^p(\pi K_2^*)=0.017+0.006i, \qquad (\beta_3^p+\beta^p_{\rm 3,EW})(\pi K_2^*)=-0.027+0.022i
\en
have been used. In the narrow width limit, we find that
\be
\B(B^-\to \ov K_2^{*0}(1430)\pi^-\to K^-\pi^+\pi^-)_{\Gamma_{K^*_2}\to 0} &=& 0.864\times 10^{-6}.
\en
Since $\B(K_2^{*0}(1430)\to\pi^+\pi^-)=(0.499\pm0.012)\times {2\over 3}$~\cite{PDG}, it is seen that the factorization relation Eq.~(\ref{eq:factorization_K2}) is numerically satisfied.

With the finite-width $\Gamma_{K_2^{*0}}^0=109\pm5$ MeV, we obtain \footnote{Contrary to the phase space integration in Eq.~(\ref{eq:Gamma_f2}) for $B^-\to f_2\pi^-\to \pi+\pi^-\pi^-$, here one should integrate over $s_{12}$ first and then $s_{23}$ owing to a pole structure in $T^{\rm BW}(s_{12})$ at $s_{12}=m_{K_2^*}^2$.}
\be \label{eq:BR&CP_K2}
\B(B^-\to \ov K_2^{*0}(1432)\pi^-\to K^-\pi^+\pi^-) &=& (0.89^{+0.22}_{-0.19})\times 10^{-6}, \non \\
A_{CP}(B^-\to \ov K_2^{*0}(1432)\pi^-\to K^-\pi^+\pi^-) &=& (1.711\pm0.002)\%,
\en
and
\be
\eta_{K_2^*}^{\rm QCDF}=0.972\pm0.001~~~(0.715\pm0.009)\,.
\en
As for the $\eta_{K_2^*}$ parameter in the experimental parameterization, we need to consider
two possibilities for the angular distribution function: $T_2$ in Eq.~(\ref{eq: TJ trans}) imposed with the transversality condition and  $T'_2$ in Eq.~(\ref{eq: T'J}) without the transversality condition. We thus find
\be
\eta_{K_2^*}^{\rm EXPP}=1.053\pm0.002, \qquad \eta_{K_2^*}^{\prime\,\rm EXPP}=1.031\pm0.001\,.
\en
Therefore, the transversality condition has little impact on the determination of $\eta_R$.
The dependence of $\eta_{K_2^*}$ in QCDF and in experimental parameterization is shown in Fig.~\ref{fig:eta_K2}.
Experimentally, the BaBar measurement~\cite{BaBar:Kmpippim} yields
\be
\B(B^-\to \ov K_2^{*0}(1430)\pi^-\to K^-\pi^+\pi^-)_{\rm expt} = (1.85^{+0.73}_{-0.50})\times 10^{-6}.
\en
Our result of Eq.~(\ref{eq:BR&CP_K2}) for the branching fraction is consistent with experiment within uncertainties.

Comparing $\eta_{K_2^*}$'s with $\eta_{f_2}$'s, it is clear that the proximity of $\eta_{f_2}^{\rm QCDF}$ to unity in QCDF is unexpected, while the deviation of $\eta_R$ from unity in the EXPP scenario is barely consistent with the expectation from the ratio of $\Gamma_R/m_R$ for $R=f_2(1270)$ and $K_2^*(1430)$.

\subsection{Vector mesons}
We take the processes $B^-\to \rho(770)\pi^-\to\pi^+\pi^-\pi^-$ and $B^-\to \ov K^{*0}(892)\pi^-\to K^-\pi^+\pi^-$ as examples to illustrate the width effects associated with the vector mesons. \footnote{For an early discussion on the decay $B\to \rho\pi\to 3\pi$, see \cite{Gardner}.}
It is known that $\rho(770)$ is much broader than $K^*(892)$. Therefore, it is expected that the former is subject to a larger width effect.

\subsubsection{$\rho(770) \pi^-$}

\noindent \underline{$B^-\to\rho^0(770) \pi^-\to \pi^+\pi^-\pi^-$ decay in QCDF}
\vskip 0.3 cm

The decay amplitude of the quasi-two-body decay $B^-\to\rho^0\pi^-$ in QCDF reads~\cite{BN}
\be \label{eq:Amprhopi}
A(B^-\to\rho^0\pi^-) &=&
  \frac{G_F}{2}\sum_{p=u,c}\lambda_p^{(d)}\Bigg\{  \Big[\delta_{pu}(a_2-\beta_2)-a_4^p-r_\chi^\rho a_6^p+{3\over 2}(a_7^p+a_9^p)+{1\over 2}(a_{10}^p+r_\chi^\rho a_8^p)  \non \\
 &&\quad
-\beta_3^p-\beta^p_{\rm 3,EW}\Big]_{\pi\rho} X^{(B^-\pi,\rho)}
+  \Big[\delta_{pu}(a_1+\beta_2)+a_4^p-r_\chi^\pi a_6^p+a_{10}^p-r_\chi^\pi a_8^p \non \\
&&\quad
+\beta_3^p+\beta^p_{\rm 3,EW}\Big]_{\rho\pi}X^{(B^-\rho,\pi)}\Bigg\},
\en
with the chiral factor
\be \label{eq:rchirho}
r_\chi^\rho(\mu) = \frac{2m_\rho}{m_b(\mu)}\,\frac{f_\rho^\perp(\mu)}{f_\rho} \,,
\en
and the factorizable matrix elements
\be
X^{(B^-\pi,\rho)} =2f_\rho m_B p_c F_1^{B\pi}(m_\rho^2), \qquad
X^{(B^-\rho,\pi)} =2f_\pi m_B p_c A_0^{B\rho}(m_\pi^2),
\en
where we have followed~\cite{BSW} for the  definitions of $B\to P$ and $B\to V$ transition form factors.

The so-called Gounaris-Sakurai model~\cite{Gounaris:1968mw} is a  popular approach for describing the broad $\rho(770)$ resonance. The line shape is introduced in Eq.~(\ref{eq: T GS}).
Note that the GS line shape for $\rho(770)$ was employed by both BaBar~\cite{BaBarpipipi} and LHCb~\cite{Aaij:3pi_1,Aaij:3pi_2} in their analysis of the $\rho(770)$ resonance in the $B^-\to \pi^+\pi^-\pi^-$ decay.

For the three-body decay amplitude
$\A_{\rho(770)}\equiv A(\B^-\to\rho^0(770) \pi^-\to \pi^-(p_1)\pi^+(p_2)\pi^-(p_3))$, factorization leads to the expression~\cite{Cheng:2020ipp}
\be \label{eq:rhoamp}
\A_{\rho(770)} &=& -{G_F\over 2}\sum_{p=u,c}\lambda_p^{(d)}g^{\rho\to \pi^+\pi^-} \,F(s_{23},m_{\rho})T_\rho^{\rm GS}(s_{23})(s_{12}-s_{13}) \non \\
&&\times
\Bigg\{ f_\pi\Big[
m_{\rho}A_0^{B\rho}(m_\pi^2)
+  {1\over 2}\left(m_B-m_{\rho}-{m_B^2-s_{23}\over m_B+m_{\rho}}\right)A_2^{B\rho}(m_\pi^2)
\Big] \non \\
&&~~~~\times
\left[\delta_{pu}(a_1+\beta_2)+a_4^p-r_\chi^\pi a_6^p+a_{10}^p-r_\chi^\pi a_8^p  +\beta_3^p+\beta^p_{\rm 3,EW}\right]_{\rho\pi}
+ m_{\rho} f_{\rho} F_1^{B\pi}(s_{23})  \non \\
&&~~~~\times
\bigg[\delta_{pu}(a_2-\beta_2)-a_4^p-r_\chi^\rho a_6^p+{3\over 2}(a_7^p+a_9^p)+{1\over 2}(a_{10}^p+r_\chi^\rho a_8^p)
-\beta_3^p-\beta^p_{\rm 3,EW}\bigg]_{\pi\rho}\Bigg\} \non
\\
&& +~ (s_{23}\leftrightarrow s_{12}).
\en
Penguin annihilation terms characterized by $\beta_2$, $\beta_3$ and $\beta_{\rm 3,EW}$, which are absent in na{\"i}ve factorization, are included here. Note that
\be \label{eq:s&theta}
s_{12}-s_{13}=-4\vec{p}_1\cdot\vec{p}_2=4\vec{p}_1\cdot\vec{p}_3=4|\vec{p}_1||\vec{p}_3|\cos\theta
\en
in the rest frame of $\pi^+(p_2)$ and $\pi^-(p_3)$ with the expressions of $|\vec{p}_i|$ ($i=1,2,3$) given in Eq.~(\ref{eq:3momentum}). Then we can write
\be \label{eq:Arho}
\A_{\rho(770)} =-g^{\rho\to \pi^+\pi^-}\,F(s_{23},m_{\rho})T_{\rho}^{\rm GS}(s_{23})2q\cos\theta\, \tilde A(B^-\to \rho\pi^-) +(s_{23}\leftrightarrow s_{12}),
\en
with $q$ already introduced in Eq.~(\ref{eq:3momentum}),
where
\be \label{eq:Arhopi_bar}
\tilde A(B^-\to \rho\pi^-) &=&
  \frac{G_F}{2}\sum_{p=u,c}\lambda_p^{(d)}
  \Bigg\{ \left[\delta_{pu}(a_1+\beta_2)+a^p_4+\cdots \right]_{\rho\pi}  \tilde X^{({B}^- \rho,\pi)} \non \\
   &&~~~ +  \Big[\delta_{pu}(a_2-\beta_2)-a_4^p+\cdots \Big]_{\pi\rho} \tilde X^{({B}^- \pi, \rho)}  \Bigg\},
\en
with
\be
\tilde X^{(B^-\pi,\rho)} &=& 2f_\rho m_B \tilde p_c F_1^{B\pi}(s_{23}), \non \\
\tilde X^{(B^-\rho,\pi)} &=& 2 f_\pi m_B\tilde p_c\Big[
A_0^{B\rho}(m_\pi^2)
+  {1\over 2m_\rho}\left(m_B-m_{\rho}-{m_B^2-s_{23}\over m_B+m_{\rho}}\right)A_2^{B\rho}(m_\pi^2)\Big].
\en

The decay rate is given by
\be
&&\Gamma(B^-\to \rho\pi^-\to \pi^+\pi^-\pi^-) \non
\\
&&= {1\over 2}\,{1\over(2\pi)^3 32 m_B^3}\int ds_{23}\,ds_{12} \Bigg\{ {|g^{\rho\to \pi^+\pi^-}|^2 F(s_{23},m_{\rho})^2(1+D \, \Gamma_\rho^0/m_\rho)^2 \over (s_{23}-m^2_{f_2}-f(s_{23}))^2+m_{\rho}^2\Gamma_{\rho}^2(s_{23})}
\non \\
&&~~~~
\times 4q^2\cos^2\theta|\tilde A(B^-\to \rho\pi^-)|^2 +(s_{23}\leftrightarrow s_{12})
+{\rm interference}\Bigg\}.
\en
One can integrate out the angular distribution part by noting that
\be
\int_{(s_{12})_{\rm min}}^{(s_{12})_{\rm max}}ds_{12}\cos^2\theta={2\over 3a}={4\over 3}{m_B\over\sqrt{s_{23}}}q\tilde p_c.
\en
In the narrow width limit,
\be
{m_{\rho}\Gamma_{\rho}(s)(1+D \, \Gamma_\rho^0/m_\rho)^2 \over (s-m^2_\rho-f(s))^2+m_{\rho}^2\Gamma_{\rho}^2(s)} \xlongrightarrow[]{\; \Gamma_{\rho}\to 0 \;}\pi\delta(s-m_{\rho}^2-f(s)).
\en
We see from Eq.~(\ref{eq:f(s)}) that $f(s)$ vanishes when $s\to m_\rho^2$.
Hence, the $\delta$-function implies $s\to m_{\rho}^2$ in the zero width limit. As a result, $\tilde p_c\to p_c$, $q\to q_0$, and $\tilde A(B^-\to \rho\pi^-)\to A(B^-\to \rho\pi^-)$.
We then obtain the desired factorization relation
\be \label{eq:factorization}
\Gamma(B^-\to \rho\pi^-\to \pi^+\pi^-\pi^-) \xlongrightarrow[]{\; \Gamma_{\rho}\to 0 \;}
 \Gamma(B^-\to \rho\pi^-) \B(\rho\to \pi^+\pi^-),
\en
where use of the relations
\be
 \Gamma_{\rho\to \pi^+\pi^-} = {q_0^3\over 6\pi m_{\rho}^2}g_{\rho\to \pi^+\pi^-}^2, \qquad
 \Gamma_{B^-\to \rho\pi^-} = {p_c\over 8\pi m_B^2}|A(B^-\to \rho\pi^-)|^2,
\en
has been made.

\begin{table}[t]
\caption{Numerical values of the flavor operators $a_i^p(M_1M_2)$ for $M_1M_2=\rho(770)\pi$ and $\pi \rho(770)$ at the scale $\mu=\ov m_b(\ov m_b)=4.18$ GeV. }
\label{tab:aiRhoP}
\begin{center}
\begin{tabular}{ l c c | l r r} \hline \hline
 $a_i^p$ & ~~$\rho\pi$~~ & ~~~$\pi \rho$~~~ & ~~$a_i^p$~~  & $\rho\pi$~~~~~~~~ & $\pi \rho$~~~~~~~~ \\
\hline
 $a_1$ & ~~~$1.007+0.108i$~~~ & ~~$1.000+0.095i$~~ & ~~$a_6^c$ & $-0.045-0.005i$ & $-0.013-0.006i$  \\
 $a_2$ & $0.135-0.379i$ & $0.158-0.340i$ &  ~~$a_7$ & $(-0.13+2.9i)10^{-4}$ & $(-0.3+2.6i)10^{-4}$ \\
 $a_3$ & $0.0008+0.0183i$ & $-0.0004+0.016i$ & ~~$a_8^u$ & $(5.2-1.0i)10^{-4}$ & $(-8.9-8.5i)10^{-5}$  \\
 $a_4^u$ & $-0.026-0.022i$ & $-0.026-0.021i$ & ~~$a_8^c$ & $(5.0-0.5i)10^{-4}$  &  $(-10.7-3.7i)10^{-5}$\\
 $a_4^c$ & $-0.030-0.013i$ & $-0.031-0.012i$ & ~~$a_9$  & $(-9.1-0.9i)10^{-3}$ & $(-9.0-0.8i)10^{-3}$\\
 $a_5$ & $0.0018-0.0247i$ & $0.004-0.022i$ & ~~$a_{10}^u$ & $(-0.9+3.3i)10^{-3}$  & $(-1.1+2.9i)10^{-3}$ \\
 $a_6^u$ & $-0.042-0.014i$ & $-0.010-0.015i$  & ~~$a_{10}^c$ & $(-0.9+3.3i)10^{-3}$ & $(-1.2+3.0i)10^{-3}$ \\
\hline \hline
\end{tabular}
\end{center}
\end{table}

\vskip 0.3 cm
\noindent \underline{Numerical results}
\vskip 0.3 cm

To compute the flavor operators $a_i^p(\rho\pi)$ and $a_i^p(\pi\rho)$ in QCDF, we need to specify the parameters $\rho_{A,H}$ and $\phi_{A,H}$ for penguin annihilation and hard spectator scattering diagrams.
For $B\to V\!P$ decays, we use the superscripts `$i$' and `$f$'
\be \label{eq:XA}
 X_A^{i,f} = \ln\left({m_B\over \Lambda_h}\right)(1+\rho_A^{i,f} e^{i\phi_A^{i,f}}),
\en
to distinguish the gluon emission from the initial and final-state quarks, respectively.
We shall use
\be
(\rho_A^i,\rho_A^f)_{_{PV}}=(2.87^{+0.66}_{-1.95}, 0.91^{+0.12}_{-0.13}), \qquad
(\phi_A^i,\phi_A^f)_{_{PV}}=(-145^{+14}_{-21}, -37^{+10}_{-~9})^\circ,
\en
and the first order approximation of  $\rho_H\approx \rho_A^i$ and $\phi_H\approx \phi_A^i$
(see~\cite{Cheng:2020hyj} for details).
This leads to
\be
&& \beta_2^p(\rho\pi)=0.025+0.011i, \qquad~~ (\beta_3^p+\beta_{\rm 3,EW}^p)(\rho\pi)=0.034- 0.030i,  \non \\
&& \beta_2^p(\pi\rho)=-0.018-0.008i, \qquad (\beta_3^p+\beta_{\rm 3,EW}^p)(\pi\rho)=0.026- 0.021i,
\en
and the flavor operators $a_i^p(\rho\pi)$ and $a_i^p(\pi\rho)$ shown in Table~\ref{tab:aiRhoP}.

\begin{figure}[t]
\centering
 { \includegraphics[scale=0.7]{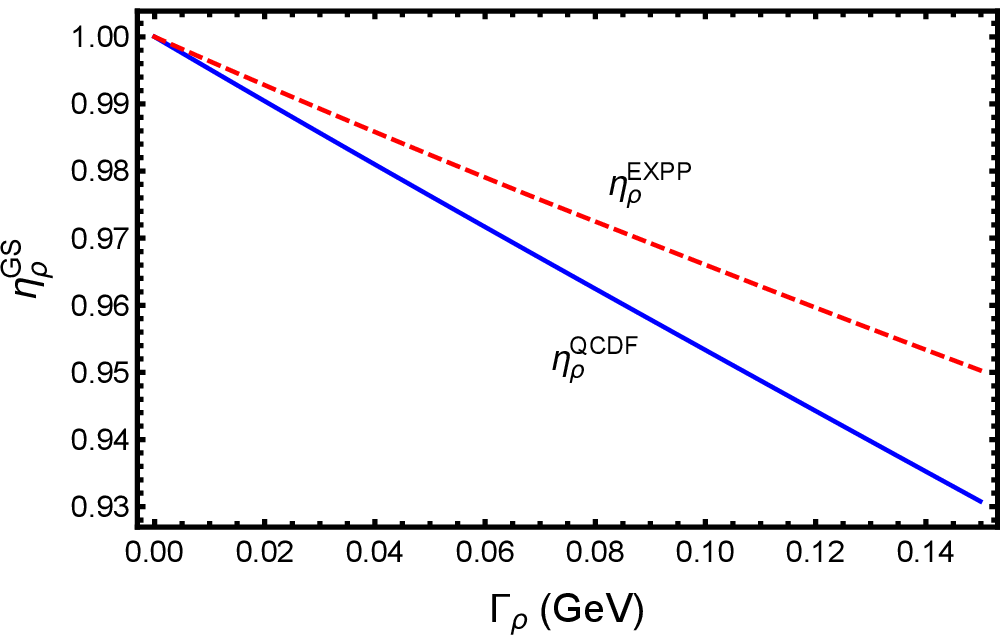} }\hskip 0.8cm
{  \includegraphics[scale=0.7]{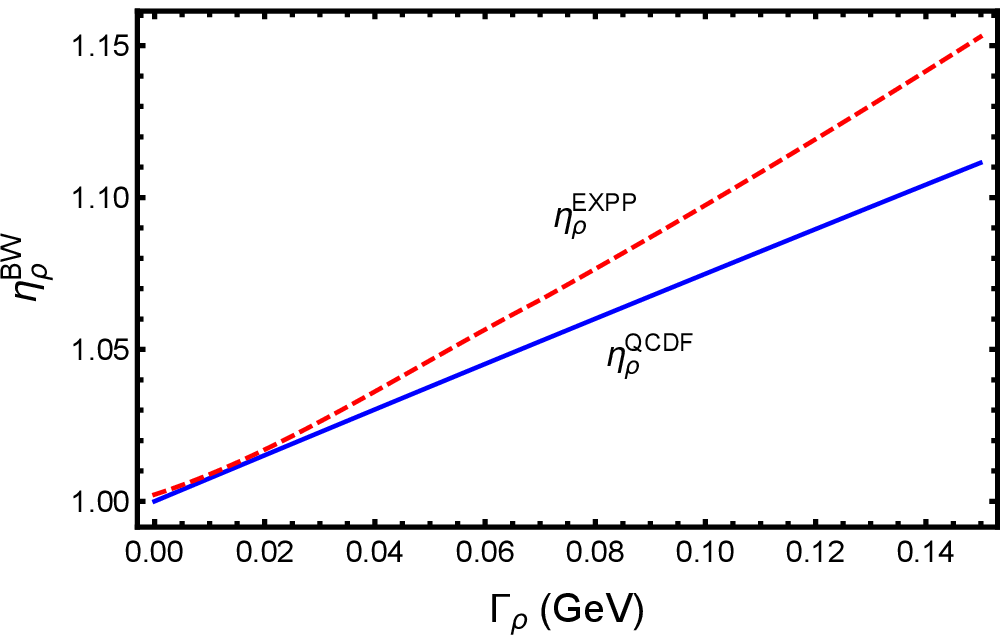} }
\vskip 0.2cm
\centering{(a) \hskip 7.0cm (b)}
\vspace{0.1cm}
\caption{Same as Fig.~\ref{fig:eta_f2} for the $\rho(770)$ resonance mediating the $B^-\to \pi^+\pi^-\pi^-$ decay  using (a) the Gounaris-Sakurai model and (b) the Breit-Wigner model to describe its line shape.
}
\label{fig:eta_rho}
\end{figure}

Following~\cite{CC:Bud}, we obtain in QCDF
\be
\B(B^-\to \rho(770)\pi^-)_{\rm QCDF} &=& (8.18^{+1.67}_{-0.81})\times 10^{-6}, \non \\ A_{CP}(B^-\to \rho(770)\pi^-)_{\rm QCDF} &=& (0.36^{+5.36}_{-4.54})\%,
\en
where use of the decay constants $f_\rho=216$ MeV and $f_\rho^\perp(\mu=1\,{\rm GeV})=165$ MeV~\cite{CC:Bud} has been made.
For the finite-width $\Gamma_\rho^0=149.1\pm0.8$ MeV, we find
\be \label{eq:BFCPrho}
\B(B^-\to \rho(770)\pi^-\to \pi^+\pi^-\pi^-) &=& (8.76^{+1.86}_{-1.68})\times 10^{-6}, \non \\
A_{CP}(B^-\to \rho(770)\pi^-\to \pi^+\pi^-\pi^-) &=& -(0.24^{+0.46}_{-0.54})\%,
\en
and
\be
\eta_{\rho \pi}^{\rm GS,QCDF}=0.931 ~~~(0.855)\,,
\en
with negligible uncertainties, where the value in parentheses is obtained with $F(s,m_{f_2})=1$.
The same results for $\eta_{\rho \pi}^{\rm GS,QCDF}$ can also be obtained using Eqs.~(\ref{eq: eta A GS}) and (\ref{eq:Arho}).
The deviation of $\eta_\rho^{\rm GS}$ from unity at 7\%  level is contrasted with the ratio $\Gamma_\rho/m_\rho=0.192$\,. For comparison, using the Breit-Wigner model to describe the $\rho$ line shape, we get
\be
\eta_{\rho\pi}^{\rm BW,QCDF}=1.111\pm0.001,~~~(1.033)\,.
\en
In the experimental parameterization scheme, we obtain
\be
\eta_{\rho\pi}^{\rm GS,EXPP}=0.950\,, \qquad \eta_{\rho\pi}^{\rm BW,EXPP}=1.152\pm0.001\,.
\en
The parameter $\eta_\rho$ as a function of the $\rho(770)$ width is shown in Fig.~\ref{fig:eta_rho} for both Gounaris-Sakurai and Breit-Wigner line shape models and for both QCDF and EXPP schemes.

%====================================================================
%=====================================================================

As shown in Eq.~(\ref{eq: eta dGamma tilde GS}), the expression of $\eta_\rho^{\rm GS}$ is the same as that of $\eta_\rho^{\rm BW}$ except for an additional $r^2\equiv (1+D \, \Gamma_\rho^0/m_\rho)^2$ factor in the denominator. This $r^2$ term accounts for the fact that $\eta_\rho^{\rm GS}<1<\eta_\rho^{\rm BW}$ in both QCDF and EXPP schemes. Since the Gounaris-Sakurai line shape was employed by both BaBar and LHCb Collaborations in their analyses of the $\rho$ resonance in $B^-\to \pi^+\pi^-\pi^-$ decay, the branching fraction of $B^-\to \rho\pi^-$ should be corrected using  $\eta_\rho^{\rm GS}$ rather than $\eta_\rho^{\rm BW}$.

From the measured branching fraction  $\B(B^-\to \rho(770)\pi^-\to \pi^+\pi^-\pi^-)=(8.44\pm0.87)\times 10^{-6}$ by LHCb~\cite{Aaij:3pi_1,Aaij:3pi_2} and $(8.1\pm0.7\pm1.2^{+0.4}_{-1.1})\times 10^{-6}$ by BaBar~\cite{BaBarpipipi}, we obtain the world average
\be
\B(B^-\to \rho(770)\pi^-\to \pi^+\pi^-\pi^-)_{\rm expt}=(8.36\pm0.77)\times 10^{-6}.
\en
It is worth emphasizing that the \CP asymmetry for the quasi-two-body decay $B^-\to\rho^0\pi^-$ has been found by LHCb to be consistent with zero in all three $S$-wave approaches.  For example, $\acp(\rho^0\pi^-)= (0.7\pm1.9)\%$ in the isobar model~\cite{Aaij:3pi_1,Aaij:3pi_2}. However,
previous theoretical
predictions all lead to a negative \CP asymmetry for $B^-\to \rho^0\pi^-$, ranging from $-7\%$ to $-45\%$ (see~\cite{Cheng:2020hyj} for a detailed discussion). The QCDF results for the branching fraction and \CP asymmetry presented in Eq.~(\ref{eq:BFCPrho}) agree with experiment.

\subsubsection{$\rho(770)K^-$}

The three-body decay amplitude
$\A_{\rho(770)K^-}\equiv A(\B^-\to K^- \rho(770)\to K^-(p_1)\pi^+(p_2)\pi^-(p_3))$ has the expression
\be
\label{}
\A_{\rho(770)K^-} &=& -{G_F\over 2}\sum_{p=u,c}\lambda_p^{(s)}g^{\rho\to \pi^+\pi^-} \,F(s_{23},m_\rho)T_\rho^{\rm GS}(s_{23})(s_{12}-s_{13}) \non \\
&&\times
\Bigg\{ f_K\Big[
m_{\rho}A_0^{B\rho}(m_K^2)
+  {1\over 2}\left(m_B-m_{\rho}-{m_B^2-s_{23}\over m_B+m_{\rho}}\right)A_2^{B\rho}(m_K^2)
\Big] \non \\
&&~~~~~~~\times
\left[\delta_{pu}(a_1+\beta_2)+a_4^p+a_{10}^p-r_\chi^K (a_6^p+a_8^p)  +\beta_3^p+\beta^p_{\rm 3,EW}\right]_{\rho K} \non \\
&&~~~~+ m_{\rho} f_{\rho} F_1^{BK}(s_{23})
\bigg[\delta_{pu}a_2+{3\over 2}(a_7^p+a_9^p)\bigg]_{K\rho}\Bigg\}, \non \\
&=&
-g^{\rho\to \pi^+\pi^-}F(s_{23},m_\rho)\,T_{\rho}^{\rm GS}(s_{23})\,2q\cos\theta\, \tilde A(B^-\to \rho K^-),
\en
where use of Eq.~(\ref{eq:s&theta}) has been made, and $\tilde A(B^-\to \rho K^-)$ has the same expression as the QCDF amplitude
for the quasi-two-body decay $B^-\to \rho K^-$~\cite{BN}
\begin{align}
\label{}
A(B^-\to \rho K^-) =&
  \frac{G_F}{2}\sum_{p=u,c}\lambda_p^{(s)}\Bigg\{
    2f_{\rho}m_Bp_c F_1^{BK}(m_\rho^2)
\bigg[\delta_{pu}a_2+{3\over 2}(a_7^p+a_9^p)\bigg]_{K\rho}  \\
&+ 2f_K m_B p_c A_0^{B\rho}(m_K^2)
\left[\delta_{pu}(a_1+\beta_2)+a_4^p+a_{10}^p-r_\chi^K (a_6^p+a_8^p)  +\beta_3^p+\beta^p_{\rm 3,EW}\right]_{\rho K}\Bigg\}, \non
\end{align}
except for a replacement of $p_cF_1^{BK}(m_\rho^2)$ by $\tilde p_c F_1^{BK}(s_{23})$ and $A_0^{B\rho}(m_K^2)$ by
\be
A_0^{B\rho}(m_K^2)
+  {1\over 2m_\rho}\left(m_B-m_{\rho}-{m_B^2-s_{23}\over m_B+m_{\rho}}\right)A_2^{B\rho}(m_K^2).
\en

%====================================================================
\begin{figure}[t]
\centering
 { \includegraphics[scale=0.7]{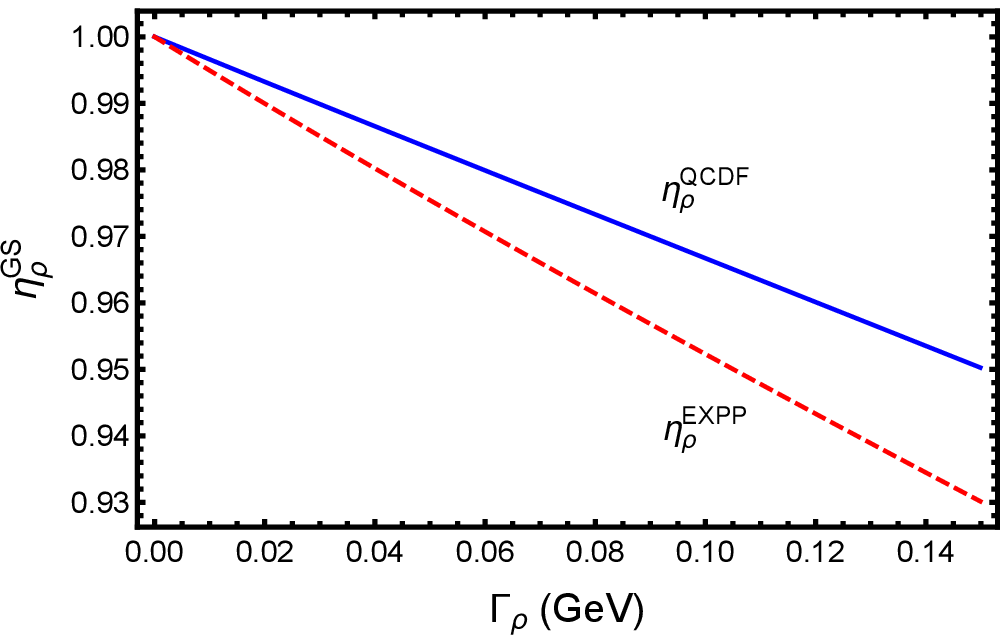} }\hskip 0.8cm
{  \includegraphics[scale=0.7]{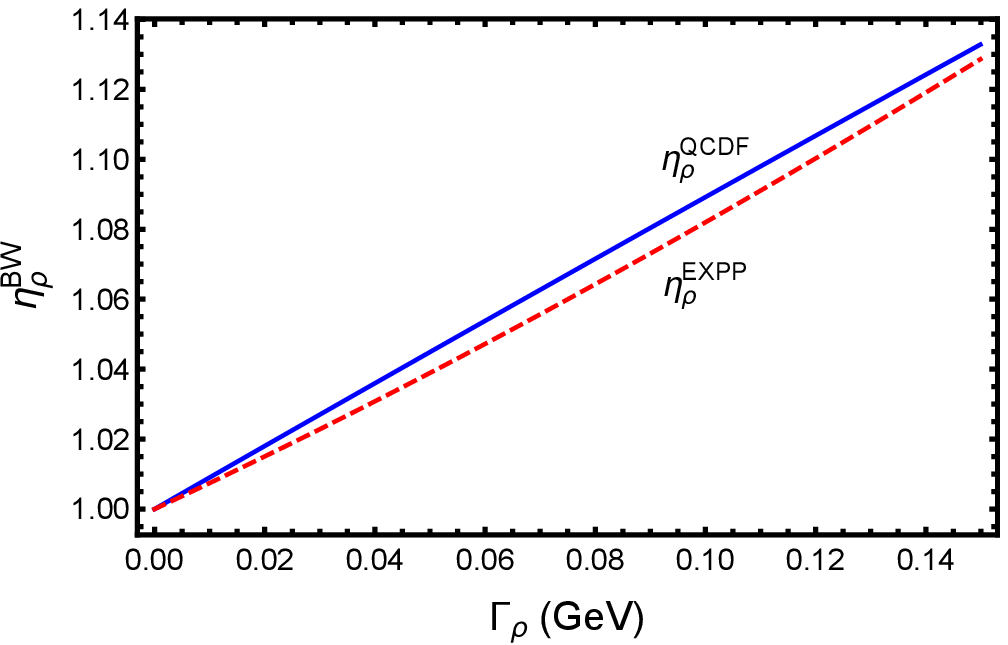} }
\vskip 0.2cm
\centering{(a) \hskip 7.0cm (b)}
\vspace{0.1cm}
\caption{Same as Fig.~\ref{fig:eta_rho} except that the $\rho(770)$ state is the resonance produced in the decay $B^-\to K^-\pi^+\pi^-$.
}
\label{fig:eta_rhoK}
\end{figure}
%=====================================================================

In QCDF, we obtain
\be
\B(B^-\to \rho(770)K^-)_{\rm QCDF} &=& (4.03^{+3.56}_{-1.67})\times 10^{-6}, \non \\ A_{C\!P}(B^-\to \rho(770)K^-)_{\rm QCDF} &=& (21.6^{+17.1}_{-16.6})\%.
\en
For the finite $\rho$ width, % $\Gamma_\rho^0=149.1\pm0.8$ MeV,
we find
\be \label{eq:BFCPrhoK}
\B(B^-\to K^-\rho(770)\to K^-\pi^+\pi^-) &=& (4.23^{+0.95}_{-0.84})\times 10^{-6}, \non \\
A_{C\!P}(B^-\to K^-\rho(770) \to K^-\pi^+\pi^-) &=& (20.5\pm0.8)\%,
\en
and
\be
\eta_{\rho K}^{\rm GS,QCDF}=0.951\pm0.003,~~~(0.899)\,.
\en
As a comparison, if the Breit-Wigner model is used to describe the $\rho$ line shape, we are led to have
\be
\eta_{\rho K}^{\rm BW,QCDF}=1.132\pm0.001, ~~~(1.086).
\en
In the experimental parameterization scheme, we obtain
\be
\eta_{\rho K}^{\rm GS,EXPP}=0.931, \qquad \eta_{\rho K}^{\rm BW,EXPP}=1.128\pm0.001\,.
\en
The dependence of $\eta_\rho$ as a function of the $\rho(770)$ width is shown in Fig.~\ref{fig:eta_rhoK}  for both the Gounaris-Sakurai and Breit-Wigner line shape models. It is evident that $\eta_{\rho\pi}$ and $\eta_{\rho K}$ are close to each other, as it should be.
Our predictions in Eq.~(\ref{eq:BFCPrhoK}) are consistent with the data:
\be \label{}
\B(B^-\to K^-\rho(770)\to K^-\pi^+\pi^-)_{\rm PDG} &=& (3.7\pm0.5)\times 10^{-6}, \non \\
A_{C\!P}(B^-\to K^-\rho(770)\to K^-\pi^+\pi^-)_{\rm PDG} &=& 0.37\pm0.10\,.
\en

\subsubsection{$K^*(892)$}

For the three-body decay amplitude
$\A_{K^*(892)}\equiv A(\B^-\to \ov K^{*0}(892) \pi^-\to K^-(p_1)\pi^+(p_2)\pi^-(p_3))$, factorization leads to the expression
\begin{align}
\label{}
\A_{K^*(892)} =& -{G_F\over \sqrt{2}}\sum_{p=u,c}\lambda_p^{(s)}g^{K^*\to K^-\pi^+} F(s_{12},m_{K^*})\,T_{K^*}^{\rm BW}(s_{12})\left[s_{13}-s_{23}-{(m_B^2-m_\pi^2)(m_K^2-m_\pi^2)
\over s_{12}}\right]  \non \\
&~~\times
\left[a_4^p-{1\over 2}a_{10}^p+r_\chi^{K^*}(a_6^p-{1\over 2} a_8^p)+\delta_{pu}\beta_2^p  +\beta_3^p+\beta^p_{\rm 3,EW}\right]_{\pi K^*}
m_{K^*} f_{K^*} F_1^{B\pi}(s_{12}).
\end{align}
Since
\be \label{}
s_{13}-s_{23}-{(m_B^2-m_\pi^2)(m_K^2-m_\pi^2) \over s_{12}}=4\vec{p}_2\cdot\vec{p}_3=4|\vec{p}_2||\vec{p}_3|\cos\theta
\en
in the rest frame of $K^-(p_1)$ and $\pi^+(p_2)$, the three-body amplitude can be recast to
\be
\A_{K^*(892)} =-g^{\bar K^*\to K^-\pi^+}F(s_{12},m_{K^*})\,T_{K^*}^{\rm BW}(s_{12})\,2q\cos\theta\, \tilde A(B^-\to \ov K^{*0}(892)\pi^-),
\en
where $\tilde A(B^-\to \ov K^{*0}(892)\pi^-)$ has the same expression as the QCDF amplitude
for the quasi-two-body decay $B^-\to \ov K^{*0}(892)\pi^-$~\cite{BN}
\be \label{eq:AK2pi}
A(B^-\to \ov K^{*0}\pi^-) &=&
  \frac{G_F}{\sqrt{2}}\sum_{p=u,c}\lambda_p^{(s)}
    \Big[a_4^p-{1\over 2}a_{10}^p+r_\chi^{K^*}(a_6^p-{1\over 2} a_8^p)   \non \\
   &&~~~
   + \beta_2^p \delta_{pu}+\beta_3^p+\beta^p_{\rm 3,EW}\Big]_{\pi K^*}  2  f_{K^*}  m_B\,p_cF_1^{B\pi}(m_{K^*}^2),
\en
except for a replacement of $p_cF_1^{B\pi}(m_{K^*}^2)$ by $\tilde p_c F_1^{B\pi}(s_{12})$
It is then straightforward to show the factorization relation
\be \label{}
\Gamma(B^-\to \ov K^{*0}(892)\pi^-\to K^-\pi^+\pi^-) \xlongrightarrow[]{\; \Gamma_{K^*}\to 0 \;}
 \Gamma(B^-\to \ov K^{*0}(892)\pi^-) \B(\ov K^{*0}(892)\to \pi^+\pi^-)
 \non \\
\en
being valid in the narrow width limit.

%====================================================================
\begin{figure}[t]
\begin{center}
\includegraphics[width=0.7\textwidth]{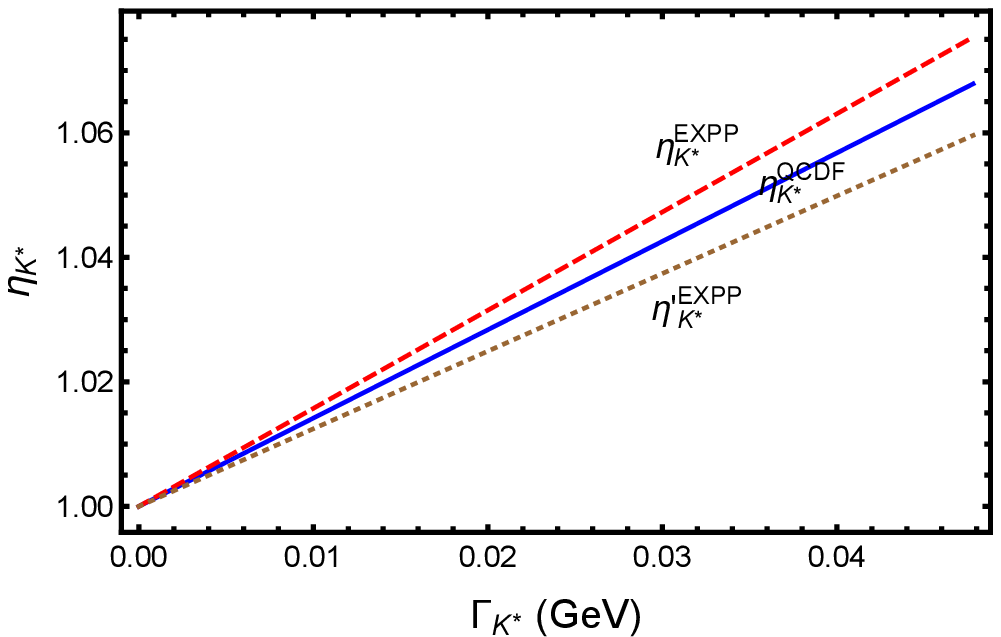}
\vspace{0.1cm}
\caption{Same as Fig.~\ref{fig:eta_f2} for the resonance $K^*(892)$ produced in the three-body decay $B^-\to K^-\pi^+\pi^-$.
}
\label{fig:eta_Kst}
\end{center}
\end{figure}
%=====================================================================

In QCDF, we obtain
\be
\B(B^-\to \ov K^{*0}(892)\pi^-) &=& (10.4^{+1.8}_{-1.7})\times 10^{-6}, \non \\ A_{C\!P}(B^-\to \ov K^{*0}(892)\pi^-) &=& (0.16^{+0.17}_{-0.14})\%,
\en
and
\be
\beta_2^p(\pi K^*)=0.017+0.006i, \qquad (\beta_3^p+\beta^p_{\rm 3,EW})(\pi K^*)=-0.027+0.022i.
\en
For the finite-width $\Gamma_{K^{*0}}^0=47.3\pm0.5$ MeV, we find
\be \label{eq:QCDF Kst}
\B(B^-\to \ov K^{*0}(892)\pi^-\to K^-\pi^+\pi^-) &=& (6.52^{+1.59}_{-1.42})\times 10^{-6}, \non \\
A_{C\!P}(B^-\to \ov K^{*0}(892)\pi^-\to K^-\pi^+\pi^-) &=& (0.166\pm0.002)\%,
\en
and
\be
\eta_{K^*}^{\rm BW,QCDF}=1.067\pm0.002, ~~~(0.9914\pm0.0001)\,.
\en
As for the $\eta_{K^*}$ parameter in the experimental parameterization, we obtain
\be
\eta_{K^*}^{\rm EXPP}=1.075\pm0.001, \qquad \eta_{K^*}^{\prime\,\rm EXPP}=1.059\pm0.001\,.
\en
The dependence of $\eta_{K^*}$ in QCDF and in experimental parameterization is shown in Fig.~\ref{fig:eta_Kst}.

The deviation of $\eta_{K^*}$ from unity is roughly consistent with the expectation from the ratio $\Gamma_{K^*}/m_{K^*}=0.053$. Experimentally, the average of BaBar~\cite{BaBar:Kmpippim} and Belle~\cite{Belle:Kmpippim} measurements yields
\be
\B(B^-\to \ov K^{*0}(892)\pi^-\to K^-\pi^+\pi^-)_{\rm expt} = (6.71\pm0.57)\times 10^{-6}.
\en
The result of the QCDF calculation of the branching fraction given in Eq.~(\ref{eq:QCDF Kst}) agrees with experimental data.

\subsection{Scalar resonances}

For examples of scalar intermediate states, we shall take the processes $B^-\to \sigma/f_0(500)\pi^-\to\pi^+\pi^-\pi^-$ and $B^-\to \ov K_0^{*}(1430)\pi^-\to K^-\pi^+\pi^-$ to illustrate their finite-width effects. Since $K_0^*(1430)$ and especially $\sigma$ are very broad, they are expected to exhibit large width effects. \footnote{The finite-width effect for $\sigma/f_0(500)$ had been considered in~\cite{Qi:2018lxy}. }

\subsubsection{$\sigma/f_0(500)$ }

In QCDF, the decay amplitude of $B^-\to\sigma\pi^-$ is given by (see Eq.~(A6) of ~\cite{CCY:SP}):
\begin{align}
\label{eq:sigmapi}
A(B^- \to \sigma \pi^- ) =&
 \frac{G_F}{\sqrt{2}}\sum_{p=u,c}\lambda_p^{(d)}
 \Bigg\{ \left[a_1 \delta_{pu}+a^p_4+a_{10}^p-(a^p_6+a^p_8) r_\chi^\pi \right]_{\sigma\pi} X^{(B\sigma,\pi)} \non \\
 &+
 \left[a_2\delta_{pu} +2(a_3^p+a_5^p)+{1\over 2}(a_7^p+a_9^p)+a_4^p-{1\over 2}a_{10}^p-(a_6^p-{1\over 2}a_8^p)\bar r^\sigma_\chi\right]_{\pi\sigma} X^{(B\pi,\sigma)}\non \\
 &- f_Bf_\pi\bar f_{\sigma}^u\bigg[\delta_{pu}b_2(\pi\sigma)+ b_3(\pi\sigma)
 + b_{\rm 3,EW}(\pi\sigma) +(\pi\sigma\to \sigma\pi) \bigg] \Bigg\},
\end{align}
where the factorizable matrix elements read
\be
X^{(B\sigma,\pi)}=-f_\pi F_0^{B\sigma^u}(m_\pi^2)(m_B^2-m_\sigma^2), \qquad
X^{(B\pi,\sigma)}=\bar f_\sigma^u F_0^{B\pi}(m_\sigma^2)(m_B^2-m_\pi^2),
\en
and $\bar r_\chi^{\sigma}(\mu)=2m_{\sigma}/m_b(\mu)$.
The superscript $u$ in the scalar decay constant $\bar f_\sigma^u$ and the form factor $F^{B\sigma^u}$ refers to the $u$ quark component of the $\sigma$ meson. The scale-dependent scalar decay constant is defined by $\la \sigma|\bar uu|0\ra=m_\sigma \bar f_\sigma^u$.  We follow~\cite{Cheng:2020hyj} to take $\bar f_\sigma^u=350$ MeV at $\mu=1$ GeV and $F_0^{B\sigma^u}(0)=0.25$, where the Clebsch-Gordon coefficient $1/\sqrt{2}$ is included in $\bar f_\sigma^u$ and $F_0^{B\sigma^u}$.

As discussed in Sec.~II.E, the $\sigma$ is too broad to be described by the usual Breit-Wigner line shape.
\footnote{Another issue with the Breit-Wigner line shape is that the Breit-Wigner mass and width agree with the pole parameters only if the resonance is narrow.}
We thuis follow the LHCb Collaboration~\cite{Aaij:3pi_2} to use the simple pole description
\footnote{In the analysis of $B^0\to \bar D^0 \pi^+\pi^-$ decays~\cite{Aaij:2015sqa}, LHCb has adopted the Bugg model~\cite{Bugg:2006gc} to describe the line shape of $\sigma/f_0(500)$.
However, the parameterization used in this model is rather complicated and the mass parameter $M\sim 1$ GeV is not directly related to the $\sigma$ pole mass. Hence, we shall follow~\cite{Aaij:3pi_2} to assume a simple pole model.
}
\be
T_\sigma(s)={1\over s-s_\sigma}={1\over s-m_\sigma^2+\Gamma_\sigma^2(s)/4+im_\sigma\Gamma_\sigma(s)},
\en
with $\sqrt{s_\sigma}=m_\sigma-i\Gamma_\sigma/2$ and
\be
\Gamma_{\sigma}(s)=\Gamma_{\sigma}^0\left( {q\over q_0}\right)
{m_{\sigma}\over \sqrt{s}}.
\en
Using the isobar description of the $\pi^+\pi^-$ $S$-wave to fit the $B^+\to\pi^+\pi^-\pi^+$ decay data, the LHCb Collaboration found~\cite{Aaij:3pi_2}
\be
\sqrt{s_\sigma}=(563\pm 10)-i(350\pm13)\,{\rm MeV},
\en
consistent with the PDG value of $\sqrt{s_\sigma}=(400-550)-i(200-350)\,{\rm MeV}$~\cite{PDG}.

With $\A_\sigma\equiv A(B^-\to\sigma\pi^-\to \pi^+\pi^-\pi^-)$, factorization leads to~\cite{Cheng:2020ipp}
\be \label{eq:sigmapipi_1}
\A_{\sigma}&=& {G_F\over 2}\sum_{p=u,c}\lambda_p^{(d)}
g^{\sigma\to \pi^+\pi^-} F(s_{23},m_\sigma)\,T_\sigma(s_{23})\Bigg\{ \tilde X^{(B\sigma,\pi)}\left[a_1 \delta_{pu}+a^p_4+a_{10}^p-(a^p_6+a^p_8) r_\chi^\pi\right]_{\sigma\pi}  \non \\
&&~~~~ + \tilde X^{(B\pi,\sigma)}\left[a_2\delta_{pu} +2(a_3^p+a_5^p)+{1\over 2}(a_7^p+a_9^p)+a_4^p-{1\over 2}a_{10}^p-(a_6^p-{1\over 2}a_8^p)\bar r^\sigma_\chi\right]_{\pi\sigma}\Bigg\}
\non \\
&&
+  (s_{23}\leftrightarrow s_{12}) \non \\
&=& g^{\sigma\to \pi^+\pi^-} F(s_{23},m_\sigma)\,T_\sigma(s_{23})\tilde A(B^-\to \sigma \pi^-)+ (s_{23}\leftrightarrow s_{12}),
\en
with
\be
\tilde X^{(B\sigma,\pi)}=-f_\pi (m_B^2-s_{23})F_0^{B\sigma^u}(m_\pi^2), \qquad
\tilde X^{(B\pi,\sigma)}=\bar f_\sigma^u (m_B^2-m_\pi^2)F_0^{B\pi}(s_{23}).
\en
Its decay rate reads
\begin{align}
\Gamma(B^-\to \sigma\pi^-\to \pi^+\pi^-\pi^-)
=& {1\over 2}\,{1\over(2\pi)^3 32 m_B^3}\int ds_{23}\,ds_{12} \Bigg\{ {|g^{\sigma\to \pi^+\pi^-}|^2 F(s_{23},m_\sigma)^2\over (s_{23}-m^2_{\sigma}+\Gamma_\sigma(s_{23})/4)^2+m_{\sigma}^2\Gamma_{\sigma}^2(s_{23})}
\non \\
&~~~\times |\tilde A(B^-\to \sigma\pi^-)|^2 +(s_{23}\leftrightarrow s_{12})+{\rm interference}
\Bigg\}.
\end{align}
Note that
\be
\int_{(s_{12})_{\rm min}}^{(s_{12})_{\rm max}}ds_{12}={2\over a}=4{m_B\over\sqrt{s_{23}}}q\,\tilde p_c.
\en
Applying the relations
\be
 \Gamma_{\sigma\to \pi^+\pi^-} = {q_0\over 8\pi m_{\sigma}^2}g_{\sigma\to \pi^+\pi^-}^2, \qquad
 \Gamma_{B^-\to \sigma\pi^-} = {p_c\over 8\pi m_B^2}|A(B^-\to \sigma\pi^-)|^2,
\en
we arrive at the desired factorization relation~\footnote{In the LHCb paper, the square of the pole position is defined by $\sqrt{s_\sigma}=m_\sigma-i\Gamma_\sigma$ rather than $m_\sigma-i\Gamma_\sigma/2$. In this case, the left-hand side of the factorization relation in Eq.~(\ref{eq:factorization_sigma}) should be multiplied by a factor of 2. }
\be \label{eq:factorization_sigma}
\Gamma(B^-\to \sigma\pi^-\to \pi^+\pi^-\pi^-) \xlongrightarrow[]{\; \Gamma_{\sigma}\to 0 \;}
 \Gamma(B^-\to \sigma\pi^-) \B(\sigma\to \pi^+\pi^-).
\en

%====================================================================
\begin{figure}[t]
\begin{center}
\includegraphics[width=0.7\textwidth]{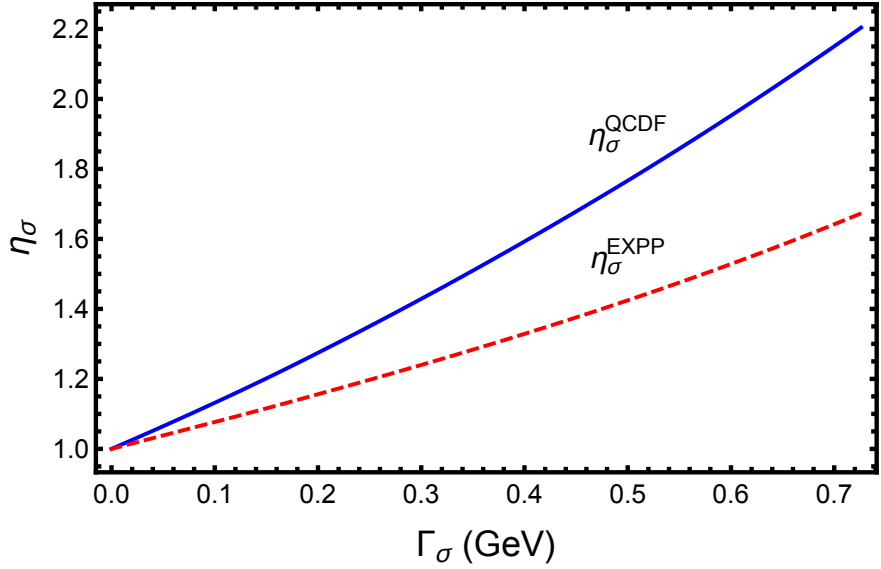}
\vspace{0.1cm}
\caption{The parameter $\eta_{\sigma}$ as a function of the $\sigma$ width, where the solid curve is derived from the QCDF calculation and the dashed curve from the experimental parameterization (EXPP).
}
\label{fig:eta_sigma}
\end{center}
\end{figure}
%=====================================================================

Using the input parameters given in~\cite{Cheng:2020ipp}, we obtain
\be
\B(B^-\to\sigma\pi^-)_{\rm QCDF} &=& (5.31^{+0.20+1.33+0.89}_{-0.19-1.18-1.33})\times 10^{-6}, \non \\
A_{C\!P}(B^-\to\sigma\pi^-)_{\rm QCDF} &=& (15.06^{+0.30+0.02+~8.34}_{-0.29-0.03-11.34})\%
\en
in QCDF.
For the finite-width $\Gamma_\sigma^0=700\pm 26$ MeV, we find
\be \label{eq:BRCP3pi}
\B(B^-\to \sigma\pi^-\to \pi^+\pi^-\pi^-) &=& (1.65^{+0.42}_{-0.37})\times 10^{-6}, \non \\
A_{C\!P}(B^-\to \sigma\pi^-\to \pi^+\pi^-\pi^-) &=& (14.7\pm0.1)\%,
\en
and
\be \label{eq:eta sigma}
\eta_{\sigma}^{\rm QCDF} &=& 2.15\pm0.05 ~~~~(1.629\pm0.025)\,, \non \\
\eta_{\sigma}^{\rm EXPP} &=& 1.64\pm0.03\,,
\en
where use of Eq.~(\ref{eq: eta dGamma tilde sigma}) has been made for the calculation of $\eta_{\sigma}^{\rm EXPP}$.
The dependence of $\eta_{\sigma}$ on the $\sigma$ width is shown in Fig.~\ref{fig:eta_sigma}.
Thus, the width correction is very large here. %for the intermediate $\sigma$ production.
In Sec.~V.B, we shall discuss its implications.

The LHCb measurement analyzed in the isobar model~\cite{Aaij:3pi_1,Aaij:3pi_2} yields
\be
\B(B^-\to \sigma\pi^-\to \pi^+\pi^-\pi^-)_{\rm expt} &=& (3.83\pm0.84)\times 10^{-6}, \non \\
A_{C\!P}(B^-\to \sigma\pi^-\to \pi^+\pi^-\pi^-)_{\rm expt} &=& (14.9^{+0.5}_{-0.6})\%.
\en
We see that while the calculated \CP asymmetry in Eq.~(\ref{eq:BRCP3pi}) based on QCDF is in excellent agreement with experiment, the predicted branching fraction is smaller than the measurement by a factor of about $2$.

\subsubsection{$K_0^*(1430)$}

For the three-body decay amplitude
$\A_{K_0^*(1430)}\equiv A(B^-\to \ov K_0^{*}(1430)^0 \pi^-\to K^-(p_1)\pi^+(p_2)\pi^-(p_3))$, factorization leads to the expression
\be \label{eq:K0stamp}
\A_{K_0^*(1430)} &=& {G_F\over \sqrt{2}}\sum_{p=u,c}\lambda_p^{(s)}g^{K_0^*\to K^-\pi^+} F(s_{12},m_{K_0^*})\,T_{K_0^*}^{\rm BW}(s_{12})
\Bigg[a_4^p-{1\over 2}a_{10}^p-r_\chi^{K_0^*}\Big({s_{12}\over m_{K_0^*}^2}\Big)\Big(a_6^p-{1\over 2} a_8^p\Big) \non \\
&&~~~+ \delta_{pu}\beta_2^p  +\beta_3^p+\beta^p_{\rm 3,EW}\Bigg]_{\pi K^*}
f_{\bar K_0^*} F_0^{B\pi}(s_{12})(m_B^2-m_\pi^2) \non \\
&=& g^{K_0^*\to K^-\pi^+} F(s_{12},m_{K_0^*})\,T_{K_0^*}^{\rm BW}(s_{12})\tilde A(B^-\to \ov K_0^{*}(1430)^0 \pi^-),
\en
where
\be \label{eq:chi_K0st}
  r^{K^*_0}_\chi(\mu)={2m_{K_0^*}^2\over
 m_b(\mu)(m_s(\mu)-m_q(\mu))},
\en
and the vector decay constant of $\ov K_0^*(1430)$ is related to the scalar one defined by
$\la \ov K_0^*|\bar sd|0\ra=m_{K_0^*}\bar f_{\bar K_0^*}$ via \footnote{
The decay constants of a scalar meson and its antiparticle are related by
$\bar f_{\bar S}=\bar f_S$ and $f_{\bar S}=-f_S$~\cite{Cheng:scalar}. Hence, the vector decay constants of  $K_0^*(1430)$ and $\ov K_0^*(1430)$ are of opposite signs. Using the QCD sum rule result for $\bar f_{\bar K_0^*}$ ~\cite{CCY:SP}, we obtain
$f_{\bar K_0^*(1430)}=36.4$ MeV.}
\be
f_{\bar K_0^*}={m_s(\mu)-m_q(\mu)\over m_{K^*_0}}\bar f_{\bar K_0^*}.
\en

In QCDF, the decay amplitude of $B^- \to \ov K^{*0}_0\pi^-$ reads~\cite{CCY:SP}
\be
A(B^- \to \ov K^{*0}_0\pi^- ) &=&
\frac{G_F}{\sqrt{2}}\sum_{p=u,c}\lambda_p^{(s)}
 \left[ a_4^p-r_\chi^{K^*_0}a_6^p
 -{1\over 2}(a_{10}^p-r_\chi^{K^*_0}a_8^p)+\delta_{pu}\beta_2^p+\beta_3^p+\beta_{\rm 3,EW}^p)\right]_{\pi K^*_0} \non \\
&&~~~~\times f_{\bar K_0^*}F_0^{B\pi}(m_{K_0^*}^2)(m_B^2-m_\pi^2).
\en
It is obvious that $\tilde A(B^-\to \ov K_0^{*}(1430)^0 \pi^-)$ has the same expression as $A(B^- \to \ov K^{*0}_0\pi^- )$ except that the chiral factor $r_\chi^{K_0^*}$ is multiplied  by  a factor of $s_{12}/m_{K_0^*}^2$ (see also~\cite{ElBennich:2009da}) and the form factor $F_0^{B\pi}(m^2_{K_0^*})$ is replaced by $F_0^{B\pi}(s_{12})$.
As before, we have the factorization relation
\be \label{eq:fact_K0st}
\Gamma(B^-\to \ov K_0^{*0}\pi^-\to K^-\pi^+\pi^-) \xlongrightarrow[]{\; \Gamma_{K_0^*}\to 0 \;}
 \Gamma(B^-\to \ov K_0^{*0}\pi^-) \B(\ov K_0^{*0}\to \pi^+\pi^-).
\en

\begin{figure}[t]
\includegraphics[width=0.7\textwidth]{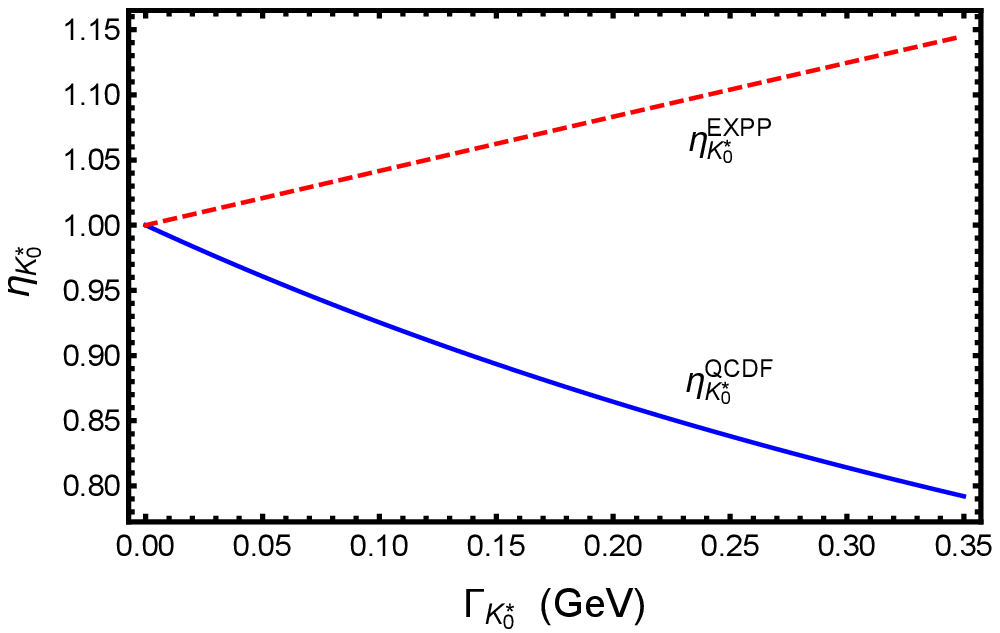}
\caption{Same as Fig.~\ref{fig:eta_sigma} for the resonance $\ov K_0^*(1430)$ produced in the three-body decay $B^-\to K^-\pi^+\pi^-$.
}
\label{fig:eta_K0st}
\end{figure}

Following~\cite{CCY:SP,Cheng:scalar}, we obtain
\be
\beta_2^p(\pi K_0^*)=-0.0969, \qquad (\beta_3^p+\beta^p_{\rm 3,EW})(\pi K_0^*)=-0.0323\,,
\en
and
\be \label{eq:QCDF_K0st}
\B(B^-\to \ov K^{*}_0(1430)^0\pi^-) &=& (13.6^{+39.9}_{-~9.3})\times 10^{-6}, \non \\ A_{C\!P}(B^-\to \ov K^{*}_0(1430)^0\pi^-) &=& (1.27^{+5.84}_{-4.75})\%.
\en
For the finite-width $\Gamma_{K_0^{*}(1430)}=270\pm80$ MeV, we find
\be \label{eq:QCDFK0st3body}
\B(B^-\to \ov K^{*}_0(1430)^0\pi^-\to K^-\pi^+\pi^-) &=& (10.2^{+3.0}_{-2.3})\times 10^{-6}, \non \\
A_{C\!P}(B^-\to \ov K^{*}_0(1430)^0\pi^-\to K^-\pi^+\pi^-) &=& (1.12\pm0.01)\%,
\en
and
\be
\eta_{K_0^*}^{\rm QCDF} &=& 0.83\pm0.04, ~~~(0.31^{+0.08}_{-0.05}), \non \\
\eta_{K_0^*}^{\rm EXPP} &=& 1.11\pm0.03.
\en
The dependence of $\eta_{K_0^*}$ on the $K_0^*(1430)$ width in  the Breit-Wigner model is shown in Fig.~\ref{fig:eta_K0st}. When off-shell effects on the strong coupling
$g^{\ov K^*_0\to K^-\pi^+}$ are turned off, $\eta_{K_0^*}^{\rm QCDF}$ is of order 0.30, rendering an extremely large deviation from unity, even much larger than $\eta_\sigma$.  Off-shell effects are particularly significant in this mode because the seemingly large QCDF enhancement in the large $s_{12}$ region is suppressed by the form factor $F(s_{12},m_{K^*_0}^2)$. As a consequence, $\eta_{K_0^*}^{\rm QCDF}$ becomes about 0.83.

%====================================================================
%=====================================================================

It has been argued that the Breit-Wigner parameterization is not appropriate for describing the broad
$K_0^*(1430)$ resonance.  LASS line shape is an alternative and popular description of the $K^*_0(1430)$ component proposed by the LASS Collaboration~\cite{LASS}.
In the analysis of three-body decays of $B$ mesons, BaBar and Belle often adopt different definitions for the $K_0^*(1430)$ resonance and nonresonant.
While Belle (see, {\it e.g.}, \cite{Belle:Kmpippim}) employed the relativistic Breit-Wigner model to describe the line shape of the $K_0^*(1430)$ resonance and an exponential parameterization for the nonresonant contribution, BaBar~\cite{BaBar:Kmpippim} used
the LASS parameterization to describe the elastic $K\pi$ $S$-wave and the $K_0^*(1430)$ resonance
by a single amplitude ~\cite{LASS}
\be
T_{K_0^*}^{\rm LASS}(s)={\sqrt{s}\over q\cot \delta_B-iq} -e^{2i\delta_B}
{ m_0\Gamma_0{m_0\over q_0}  \over s-m_0^2+im_0\Gamma_0{q\over q_0}{m_0\over \sqrt{s}} },
\en
with
\be
\cot\delta_B={1\over a q}+{1\over 2}rq,
\en
where $q$ is the c.m. momentum of $K^-$ and $\pi^+$ in the $K_0^*(1430)$ rest frame and ${q}_0$ is the value of $q$ when $s=m_{K_0^*}^2$.
The second term of $T_{K_0^*}^{\rm LASS}$ is similar to the relativistic Breit-Wigner function $T_{K_0^*}^{\rm BW}$ except for a phase factor $\delta_B$ introduced to retain unitarity. The first term is a slowly varying nonresonant component.

%%%%%%%%%%%%%%%%%%%%%%%%%
\begin{table}[t]
\caption{Branching fractions (in units of $10^{-6}$) of resonant and
nonresonant (NR) contributions to $B^-\to K^-\pi^+\pi^-$.  Note that the BaBar's branching fraction  $(2.4\pm0.5^{+1.3}_{-1.5})\times 10^{-6}$ given in Table II of~\cite{BaBar:Kmpippim} is for the phase-space nonresonant contribution to $B^-\to K^-\pi^+\pi^-$. }
\begin{center}
\begin{tabular}{l l l } \hline \hline \label{tab:Kpipi}
Decay mode~~ & BaBar~\cite{BaBar:Kmpippim} & Belle~\cite{Belle:Kmpippim}
\\ \hline
 $\overline K^{*0}_0(1430)\pi^-$~~ & $19.8\pm0.7\pm1.7^{+5.6}_{-0.9}\pm3.2$~~~~~~ &
$32.0\pm1.0\pm2.4^{+1.1}_{-1.9}$   \\
NR & $9.3\pm1.0\pm1.2^{+6.7}_{-0.4}\pm1.2$  &
$16.9\pm1.3\pm1.3^{+1.1}_{-0.9}$ \\
\hline \hline
\end{tabular}
\end{center}
\end{table}

The nonresonant branching fraction $(2.4\pm0.5^{+1.3}_{-1.5})\times 10^{-6}$ in $B^-\to K^-\pi^+\pi^-$ reported by BaBar~\cite{BaBar:Kmpippim} is much smaller than $(16.9\pm1.3^{+1.7}_{-1.6})\times 10^{-6}$ measured by Belle (see Table~\ref{tab:Kpipi}).
In the BaBar analysis, the nonresonant component of the Dalitz plot is modeled as a constant complex phase-space amplitude.
Since the first part of the LASS line shape is really nonresonant, it should be added to the phase-space nonresonant piece to get the total nonresonant contribution.  Indeed, by combining coherently the nonresonant part of the LASS parameterization and the phase-space nonresonant, BaBar found the total nonresonant branching fraction to be  $(9.3\pm1.0\pm1.2^{+6.8}_{-1.3})\times 10^{-6}$. Evidently, the BaBar result is now
consistent with Belle within errors. For the resonant contributions from $K_0^*(1430)$, the BaBar results were obtained from $(K\pi)_0^{*0}\pi^-$ by subtracting the elastic range term from the $K\pi$ $S$-wave~\cite{BaBar:Kmpippim}, namely, the Breit-Wigner component of the LASS parameterization.
\footnote{It should be stressed that the Breit-Wigner component of the LASS parameterization
does not lead to the factorization relation Eq.~(\ref{eq:fact_K0st}).}
Although both BaBar and Belle employed the Breit-Wigner model to describe the line shape of $K_0^*(1430)$, the discrepancy between BaBar and Belle for the $K_0^*\pi$ mode remains an issue to be resolved.

Note that our calculation of $\B(B^-\to \ov K^{*}_0(1430)^0\pi^-\to K^-\pi^+\pi^-)$ in Eq.~(\ref{eq:QCDFK0st3body}) based on QCDF is smaller by a factor of 2 (3) when compared to the BaBar (Belle) measurement. If we follow PDG~\cite{PDG} to apply the na{\"i}ve factorization relation (\ref{eq:fact}), we will obtain using Table~\ref{tab:Kpipi} the branching fraction of $B^-\to \ov K_0^*(1430)\pi^-$ to be $(32.0\pm1.2^{+10.8}_{-~6.0})\times 10^{-6}$ from BaBar
\footnote{Another BaBar measurement of $B^+\to K_0^{*0}\pi^+\to K_S^0\pi^0\pi^+$~\cite{Lees:2015uun} yields
$\B(B^+\to K_0^*(1430)\pi^+)_{\rm NWA}=(34.6\pm3.3\pm4.6)\times 10^{-6}$.}
and $(51.6\pm1.7^{+7.0}_{-7.5})\times 10^{-6}$ from Belle. Obviously, they are much larger than the QCDF prediction given in Eq.~(\ref{eq:QCDF_K0st}). Indeed, as pointed out before~\cite{CCY:SP,Cheng:scalar}, this has been a long-standing puzzle that for scalar resonances produced in $B$ decays, the QCDF predictions of $B^-\to \ov K_0^{*0}(1430)\pi^-$ and $\ov B^0\to K_0^{*-}(1430)\pi^+$ are in general too small compared to experiment by a factor of $2\sim 4$. Nevertheless, when the finite-width effect is taken into account, the PDG values of $\B(B^-\to \ov K_0^{*0}(1430)\pi^-)$ should be reduced by multiplying a factor of $\eta_{K_0^*}^{\rm QCDF}\simeq 0.83$ or further enhanced by a factor of $\eta_{K_0^*}^{\rm EXPP}\simeq 1.10$, depending on the scheme.

%====================================================================
\begin{figure}[tbp]
\centering
\subfigure[]{
  \includegraphics[width=6.5cm]{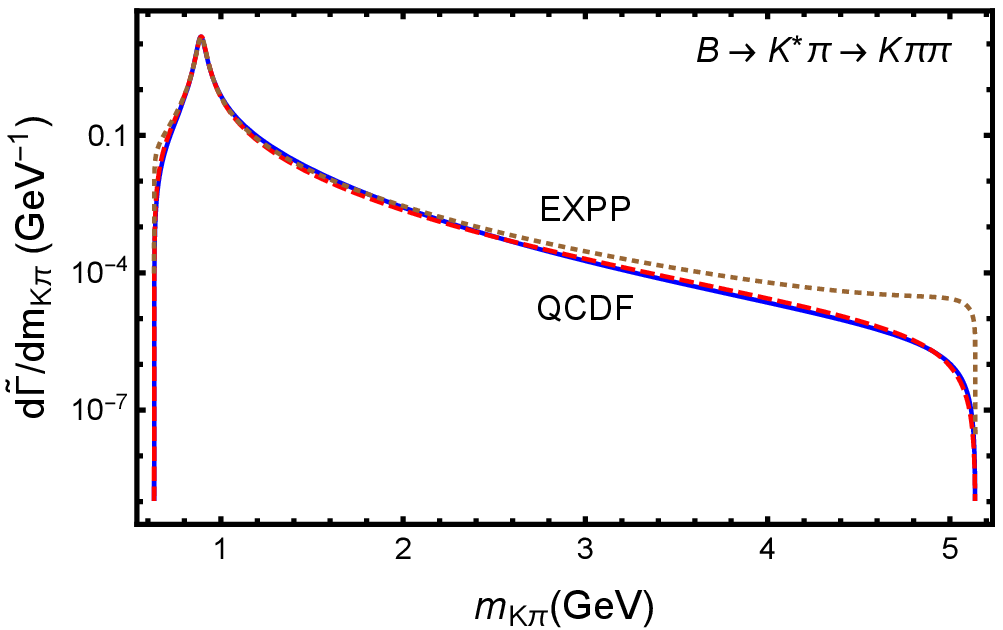}%{dGamma_Kst.pdf}
}
\hspace{0.5cm}
\subfigure[]{
  \includegraphics[width=6.5cm]{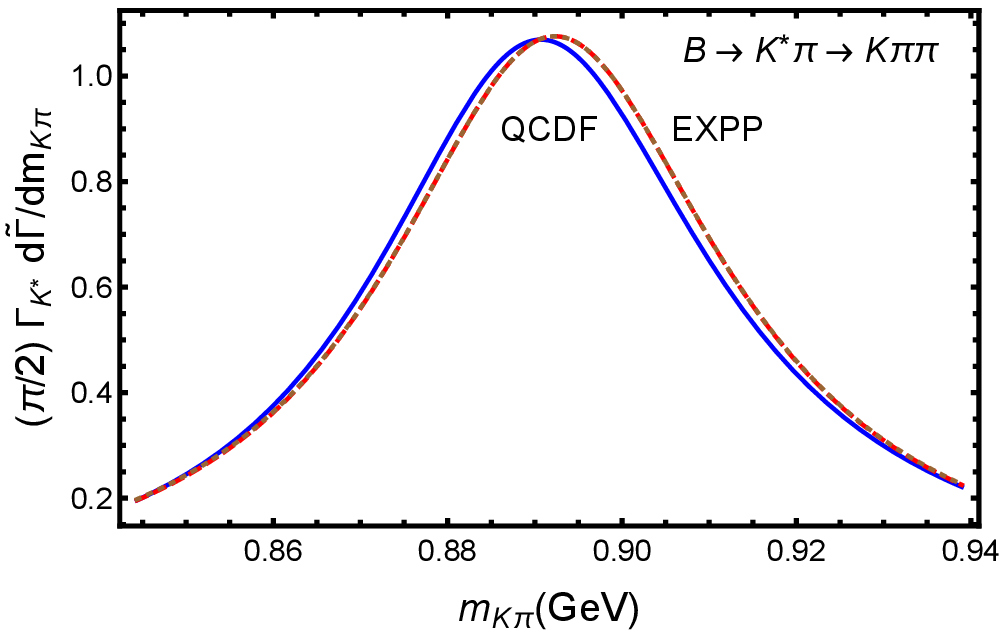}%{dGammaZ_Kst.pdf}
}
\subfigure[]{
  \includegraphics[width=6.5cm]{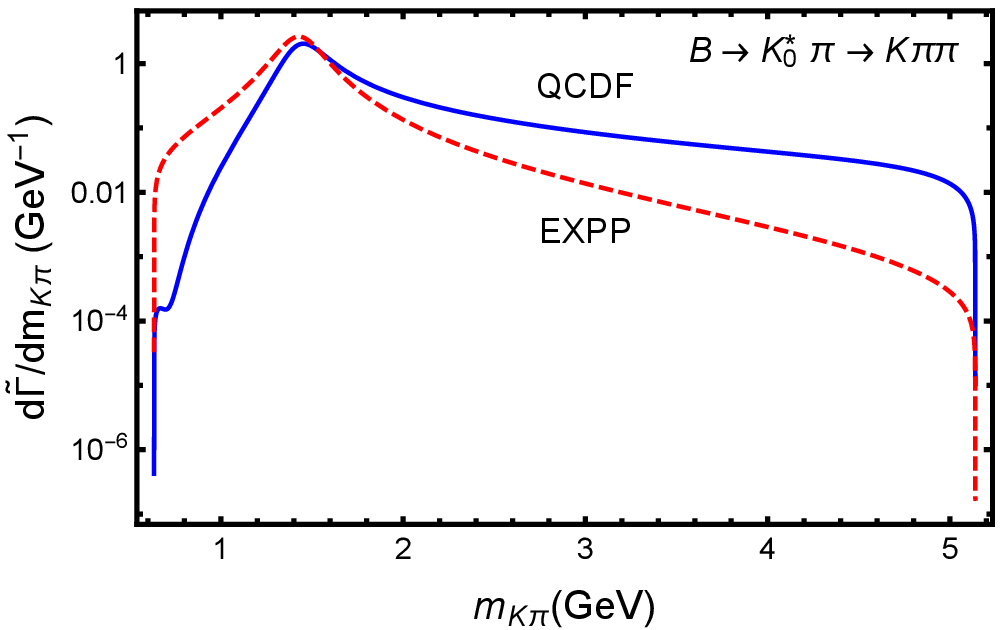}%{dGamma_K0st.pdf}
}
\hspace{0.5cm}
\subfigure[]{
  \includegraphics[width=6.5cm]{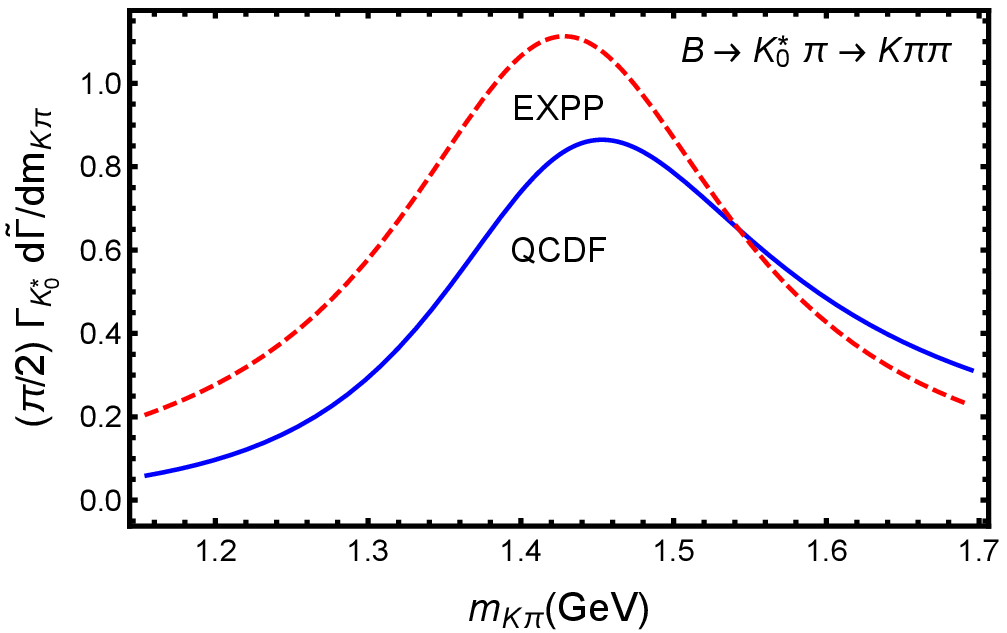}%{dGammaZ_K0st.pdf}
}
\subfigure[]{
  \includegraphics[width=6.5cm]{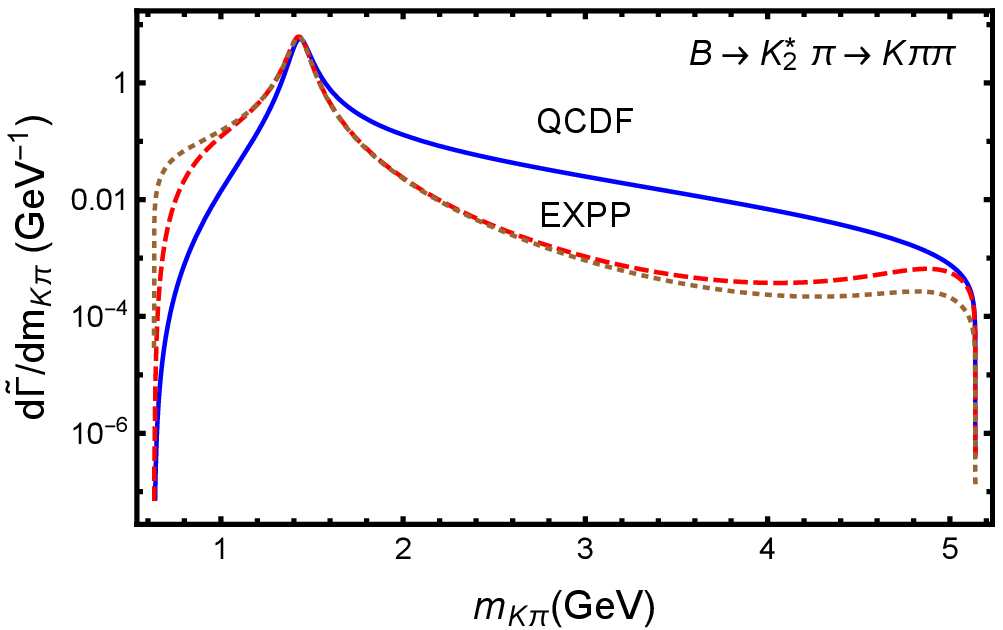}%{dGamma_K2.pdf}
}
\hspace{0.5cm}
\subfigure[]{
  \includegraphics[width=6.5cm]{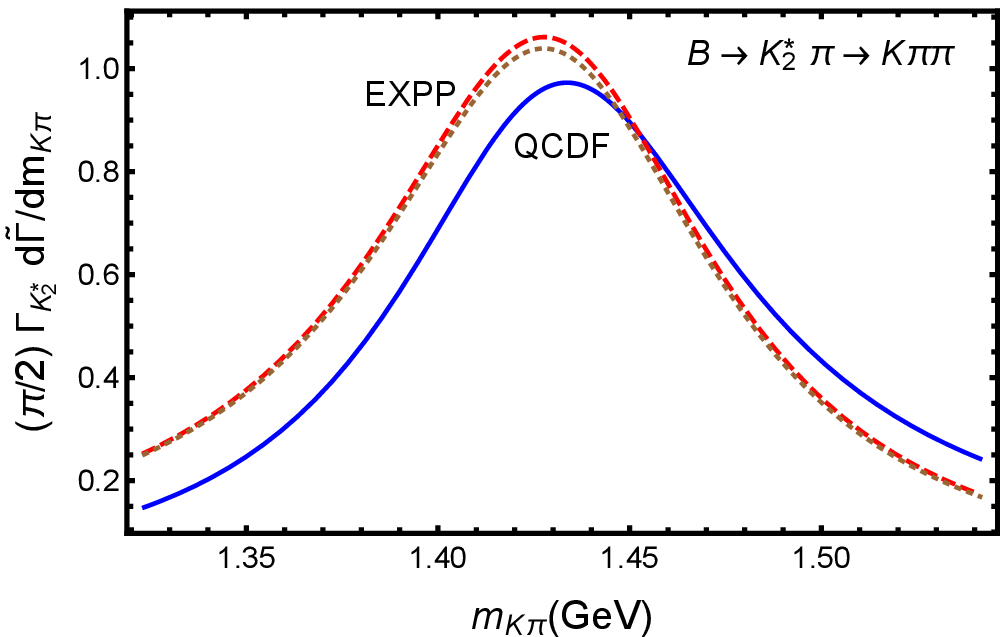}%{dGammaZ_K2.pdf}
}
\subfigure[]{
  \includegraphics[width=6.5cm]{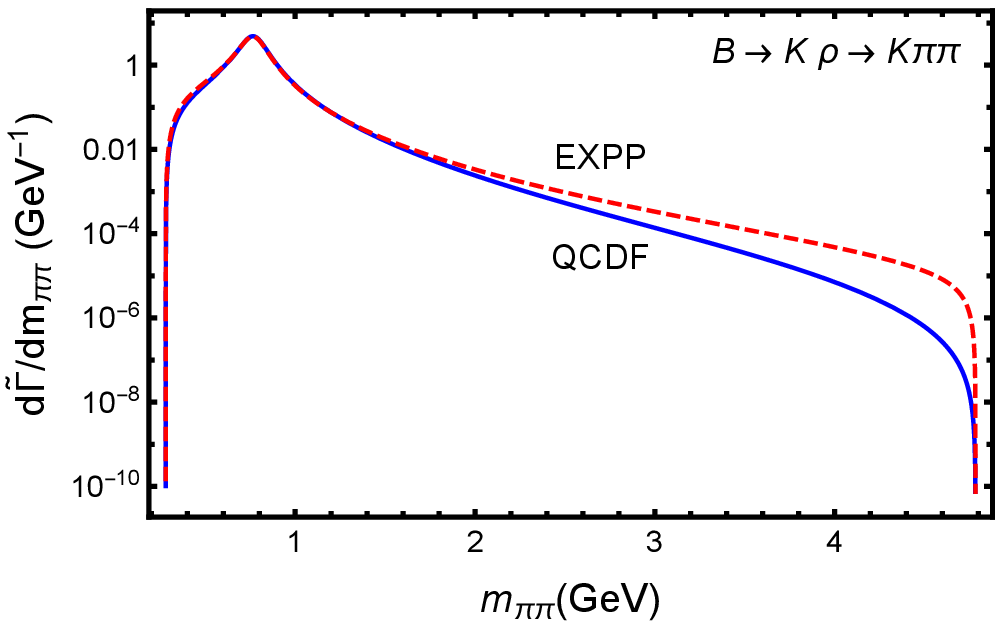}%{dGamma_Krho.pdf}
}
\hspace{0.5cm}
\subfigure[]{
  \includegraphics[width=6.5cm]{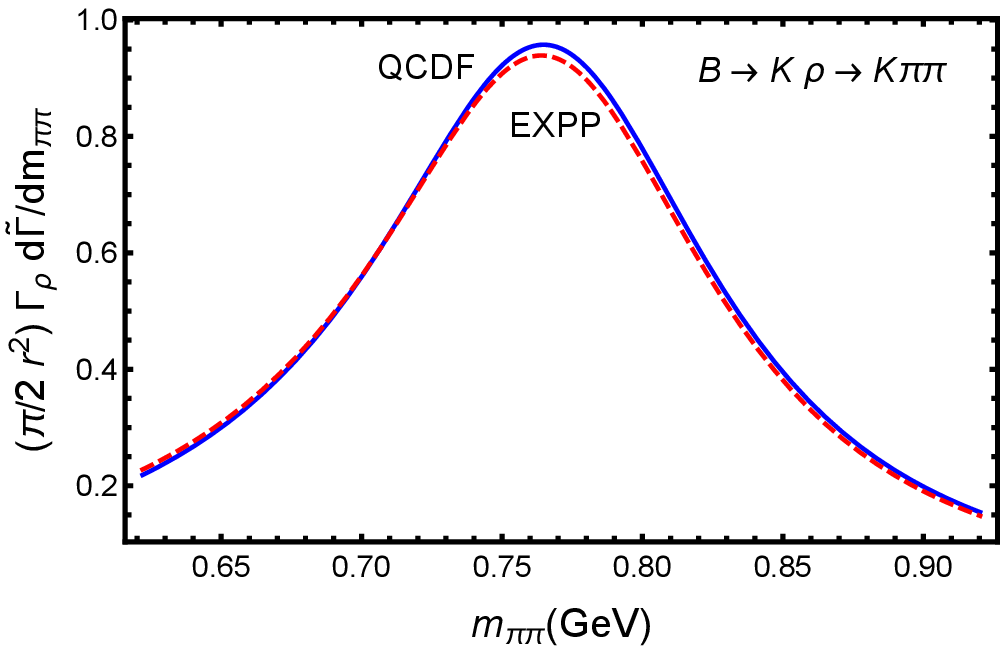}%{dGammaZ_Krho.pdf}
}
\caption{\small Left column: the normalized differential rates in $B^-\to R\pi^-\to K^-\pi^+\pi^-$ and
$B^-\to K^-\rho\to K^-\pi^+\pi^-$  decays.  Right column: plots scaled and blown-up in the resonance regions,
where the heights at the resonances equal $\eta_R$.  In plot (h), we use $r\equiv 1+D \, \Gamma_\rho^0/m_\rho$. The solid curves come from the QCDF calculation and the dashed (dotted) curves from the experimental parameterization with (without) the transversality condition imposed.
}
\label{fig: dGam}
\end{figure}
%=====================================================================

\section{Discussions}

\subsection{Finite-width and off-shell effects }

In Table~\ref{tab:eta}, we give a summary
of the $\eta_R$ parameters calculated using QCDF and the experimental parameterization for various resonances produced in the three-body $B$ decays. Since the strong coupling of $R(m_{12})\to P_1P_2$ will be suppressed by the form factor $F(s_{12},m_R)$ when $m_{12}$ is off shell from $m_R$ (see Eq.~(\ref{eq:FF for coupling})), this implies a suppression of the three-body decay rate in the presence of off-shell effects. Therefore, $\eta_R^{\rm QCDF}$ is always larger than $\bar\eta_R^{\rm QCDF}$, with the latter defined for $F(s,m_R)=1$.  We see from Table~\ref{tab:eta} that off-shell effects are small in vector meson productions, but prominent in the $K_2^*(1430)$, $\sigma/f_0(500)$ and $K_0^*(1430)$ resonances. Also, the parameters $\eta_R^{\rm QCDF}$ and $\eta_R^{\rm EXPP}$ are similar for vector mesons, but different for tensor and scalar resonances. To understand the origin of their differences, we need to study the differential decay rates.

\begin{table}[t]
\caption{A summary of the $\eta_R$ parameter for various resonances produced in the three-body $B$ decays. Off-shell effects on the strong coupling $g^{R\to h_1h_2}$ are taken into account in the determination of $\eta_R^{\rm QCDF}$ but not in $\bar\eta_R^{\rm QCDF}$. Uncertainties in $\eta_R$ are not specified whenever negligible.
}
\vskip 0.15cm
\label{tab:eta}
\footnotesize{
\begin{ruledtabular}
\begin{tabular}{ l l c c l l l}
 Resonance~~~ & ~$B^+\to Rh_3\to h_1h_2h_3$ ~~~ & ~$\Gamma_R$ (MeV)~\cite{PDG}~~ & $\Gamma_R/m_R$ & ~~~$\bar\eta_R^{\rm QCDF}$ & ~~~$\eta_R^{\rm QCDF}$ & ~~~$\eta^{\rm EXPP}_R$ \\
\hline
$f_2(1270)$ & $B^+\to f_2\pi^+\to \pi^+\pi^-\pi^+$ & ~$186.7^{+2.2}_{-2.5}$~~ & 0.146 & ~~0.974 &  ~~$1.003^{+0.001}_{-0.002}$ & ~~$0.937^{+0.006}_{-0.005}$ \\
$K_2^*(1430)$ & $B^+\to K^{*0}_2\pi^+\to K^+\pi^-\pi^+$ & ~$109\pm5$~~ & 0.076 & ~~$0.715\pm0.009$ & ~~$0.972\pm0.001$ & ~~$1.053\pm0.002$ \\
$\rho(770)$ & $B^+\to \rho^0\pi^+\to \pi^+\pi^-\pi^+$ & ~$149.1\pm0.8$~~ & 0.192 & ~~0.86 (GS) & ~~0.93 (GS) & ~~0.95 (GS)\\
 &  &  & & ~~1.03 (BW) & ~~1.11 (BW) & ~~1.15 (BW)\\
$\rho(770)$ & $B^+\to K^+\rho^0 \to K^+\pi^+\pi^-$ & ~$149.1\pm0.8$~~ & 0.192 & ~~0.90 (GS) & ~~0.95 (GS) & ~~0.93 (GS)\\
 &  &  & & ~~1.09 (BW) & ~~1.13 (BW) & ~~1.13 (BW)\\
$K^*(892)$ & $B^+\to K^{*0}\pi^+\to K^+\pi^-\pi^+$ & ~$47.3\pm0.5$~~ & 0.053 & ~~1.01 & ~~$1.067\pm0.002$ & ~~1.075 \\
$\sigma/f_0(500)$ & $B^+\to \sigma\pi^+\to \pi^+\pi^-\pi^+$ & ~$700\pm26$~\cite{Aaij:3pi_2}~~ & $\approx 1.24$ & ~~$1.63\pm0.03$ & ~~$2.15\pm0.05$ & ~~$1.64\pm0.03$     \\
$K_0^*(1430)$ & $B^+\to K^{*0}_0\pi^+\to K^+\pi^-\pi^+$ & ~$270\pm80$~~ & $\approx 0.19$ &~~$0.31^{+0.08}_{-0.05}$ & ~~$0.83\pm0.04$ & ~~$1.11\pm0.03$ \\
%\hline \hline
\end{tabular}
\end{ruledtabular} }
\end{table}

In Fig.~\ref{fig: dGam}, we show the normalized differential rates of the $B^-\to R\pi^-\to K^-\pi^+\pi^-$ and
$B^-\to K^- R\to K^-\pi^+\pi^-$ decays with $R=\overline K^{*0}(980), \overline K_0^{*0}(1430)$, $\overline K^0_2(1430)$ and $\rho^0$ respectively in the left plots.
The plots blown up in the resonance regions are also shown in the right plots.  Note that the figures on the right are scaled by a factor of $(\pi/2)\Gamma_R$ or $(\pi/2 r^2)\Gamma^0_\rho$ with $r\equiv (1+D \, \Gamma^0_\rho/m_\rho)$.
For the $B^-\to K^-\rho^0\to K^-\pi^+\pi^-$ decay, we only show the result using the Gounaris-Sakurai line shape, as this is employed by the experimental parameterization for the $\rho$ resonance.
The normalized differential rates obtained from the QCDF calculation and the experimental parameterization are shown in the plots.
For $R=\overline K^{*0}$ and $\overline K^0_2$, we also show the results using the experimental parameterization with or without enforcing the transversality condition (see Eqs.~(\ref{eq: TJ trans}) and (\ref{eq: T'J})). They are plotted in dashed and dotted curves, respectively.
Removing the transversality condition has mild effects on the normalized differential rates and little impacts on their values at the resonances.

As shown in Eqs.~(\ref{eq: eta dGamma tilde}) and (\ref{eq: eta dGamma tilde GS}), $\eta_R$ in these decays are given by
\be
\eta_R
=\frac{1}{2}\pi\Gamma_R \frac{d\tilde\Gamma (m_R)}{dm_{K\pi}},
\qquad
\eta^{\rm GS}_\rho
=\frac{\pi\Gamma^0_\rho}{2(1+D \, \Gamma_\rho^0/m_\rho)^2} \frac{d\tilde\Gamma (m_\rho)}{dm_{\pi\pi}}.
\label{eq: eta dGamma/dm}
\en
From the right plots in Fig.~\ref{fig: dGam}, one can read off the values of $\eta_R$ from the height of the curves at the resonances. The values agree with those shown in Table~\ref{tab:eta}.
Recall that for $\Gamma_R/m_R\ll 1$, we can approximate $\eta_R$ by the integration of the normalized differential rate around the resonance as shown in Eq.~(\ref{eq: anticorrelation}).
For example, for the $B^-\to R^0\pi^-\to K^-\pi^+\pi^-$ decays, $\eta_R$ can be approximately given by
\be
\eta_R
\simeq \frac{\pi}{2\tan^{-1}2}\int_{m_R-\Gamma_R}^{m_R+\Gamma_R} \frac{d\tilde\Gamma (m_{K\pi})}{dm_{K\pi}} dm_{K\pi}
=\frac{\pi}{2\tan^{-1}2}\bigg(1
-\int_{\rm elsewhere} \frac{d\tilde\Gamma (m_{K\pi})}{dm_{K\pi}} dm_{K\pi}\bigg).
\label{eq: anticorrelation1}
\en
Note that for the case of $\eta^{\rm GS}_\rho$ one needs to include the $1/(1+D \, \Gamma_\rho^0/m_\rho)^2$ factor.
Numerically, we find that this approximation works well for the decay modes considered in this section.
The above equation clearly shows that $\eta_R$ represents the fraction of rates around the resonance and it is anticorrelated with the fraction of rates off the resonance.

From the Figs.~\ref{fig: dGam}(a) and (g), we see that for $R=\overline K^{*0}$ and $\rho^0$ the normalized differential rates predicted by QCDF are very similar to those obtained by using the experimental parameterization,
while for $R=\overline K^{*0}_0$ and $\overline K^{*0}_2$ the QCDF results and experimental models are different.
Consequently, as shown in Figs.~\ref{fig: dGam}(b) and (h),  QCDF and the experimental model give similar values on $d\tilde\Gamma(m_{\overline K^*})/d m_{K\pi}$ and $d\tilde\Gamma(m_\rho)/d m_{\pi\pi}$, resulting in $\eta^{\rm QCDF}_R\simeq\eta^{\rm EXPP}_R$ for $R=\rho$ and $K^*$.
In contrast, as
shown in Figs.~\ref{fig: dGam}(d) and (f), the QCDF $d\tilde\Gamma(m_{\overline K^*_0})/d m_{K\pi}$ and $d\tilde\Gamma(m_{\overline K_2})/d m_{K\pi}$ are smaller than those from the experimental model, resulting in
$\eta^{\rm QCDF}_{K^*_0, K_2^*}<\eta^{\rm EXPP}_{K^*_0, K_2^*}$.

Using Eq.~(\ref{eq: anticorrelation1}), we can relate the smallness of $\eta^{\rm QCDF}_{K^*_0}$, comparing to $\eta^{\rm EXPP}_{K^*_0}$, to the fact that
the normalized differential rate obtained in the QCDF calculation is much larger than the one using the experimental parameterization in the off-resonance region, particularly in the large $m_{K\pi}$ region.
To verify the source of the enhancement, we note that, as shown in Eq.~(\ref{eq:K0stamp}), the $m_{K\pi}$ dependence in the QCDF amplitude is governed by the strong decay form factor, $F(m^2_{K\pi}, m_{K^*_0})$, the $B\to \pi$ form factor, $F^{B\pi}_0(m_{K\pi}^2)$, and a $m_{K\pi}^2$ factors sitting in front of the QCD penguin Wilson coefficient $(a^p_6-a^p_8/2)$ and related to the so-called chiral factor ($r^S_\chi$) in the two-body decay (see Eq.~(\ref{eq:chi_K0st})). The last two factors are responsible for the enhancement of the QCDF differential rate in the large $m_{K\pi}$ region.
As shown in Eq.~(\ref{eq:cF}) and the equations below it, these two factors are not included in the experimental parameterization for the scalar resonance.
As a result, QCDF and the experimental parameterization give different normalized differential rates and $\eta_R$ for this mode.

The momentum  dependence (such as $m_{K\pi}$) of weak dynamics is mode-dependent.
For example, in the above $B^-\to \overline K^*_0(1430)\pi^- \to K^-\pi^+\pi^-$ decay, we have a $m^2_{K\pi}$ factor from the chiral factor $r^S_\chi$, while the chiral factor $r^V_\chi$ in the $B^-\to \overline K^*(980)\pi^- \to K^-\pi^+\pi^-$ decay  does not provide the $m^2_{K\pi}$ factor (see Eq.~(\ref{eq:rchirho})).
Such a difference in the momentum dependence of weak dynamics has a visible effect on the shape of the normalized differential rates, as depicted in Figs.~\ref{fig: dGam}(a) and (c).

As shown in Eq.~(\ref{eq:cF}), the weak dynamics in the experimental parameterization is basically represented by a complex number, the coefficient $c$, which does not have any momentum dependence.
In the narrow width limit, the value of the normalized differential rate is highly dominated by its peak at the resonance, and the values of the normalized differential rate elsewhere cannot compete with it. Therefore, only $m_{K\pi(\pi\pi)}\simeq m_R$ matters and, consequently, it is legitimate to use a momentum-independent coefficient, namely $c$, to represent the weak dynamics.
However, in the case of a broad resonance, things are generally different. The peak at the resonance is no longer highly dominating, as its height is affected by the values of the normalized differential rate elsewhere. In this case, the momentum dependence of the weak dynamics cannot be ignored and, hence, using a momentum-independent coefficient to represent the weak dynamics is too na\"ive.

\subsection{Branching fractions of quasi-two-body decays}

For given experimental measurements of $\B(B^+\to R P_3\to P_1P_2P_3)$, we show in Table~\ref{tab:BF2body} various branching fractions of the quasi-two-body decays $B^+\to R P_3$. $\B(B^+\to R P_3)_{\rm NWA}$ denotes the branching fraction obtained from Eq.~(\ref{eq:NWA}) in the NWA. Our results of $\B(B^+\to R P_3)_{\rm NWA}$ for $B^+\to K_2^{*0}(1430)\pi^+$, $K^{*0}\pi^+$, and $K^+\rho^0$ modes agree with the PDG data~\cite{PDG}. For $B^+\to f_2(1270)\pi^+$, $\rho^0\pi^+$, and $\sigma\pi^+$ decays, we have included the new measurement of $B^+\to\pi^+\pi^+\pi^-$ performed by the LHCb Collaboration~\cite{Aaij:3pi_1,Aaij:3pi_2}. As for $\B(B^+\to K_0^{*0}\pi^+)_{\rm NWA}$, our value is different from $(39^{+6}_{-5})\times 10^{-6}$ given by PDG~\cite{PDG} as the contribution of $B^+\to K_0^{*0}\pi^+\to K_S^0\pi^0\pi^+$~\cite{Lees:2015uun} is included in the latter case.

When the resonance is sufficiently broad, it is necessary to take into account the finite-width effects characterized by the parameter $\eta_R$. In Table~\ref{tab:BF2body}, we have shown the corrections to $\B(B^+\to R P_3)_{\rm NWA}$ in both QCDF and EXPP schemes. Although the finite-width effects are generally small, they are significant in the $B^+\to \rho\pi^+$ decay and prominent in $B^+\to\sigma/f_0(500)\pi^+$ and $B^+\to K_0^{*0}(1430)\pi^+$. For example, the PDG value of $\B(B^+\to\rho\pi^+)=(8.3\pm1.2)\times 10^{-6}$~\cite{PDG} should be corrected to $(7.7\pm1.1)\times 10^{-6}$ in QCDF or $(7.9\pm1.1)\times 10^{-6}$ in EXPP.
The large width effects in the $\sigma/f_0(500)$ production imply that $B^-\to \sigma\pi^-$ has a large branching fraction of order $10^{-5}$. More precisely, the LHCb value of
$\B(B^+\to\sigma\pi^+)=(5.8\pm1.3)\times 10^{-6}$  should be corrected to $(12.4\pm2.7)\times 10^{-6}$ in QCDF or $(9.4\pm2.1)\times 10^{-6}$ in EXPP.

\begin{table}[t]
\caption{Branching fractions of quasi-two-body decays $B^+\to R P_3$ (in units of $10^{-6}$) derived from the measured
$B^+\to R P_3\to P_1P_2P_3$ rates. $\B(B^+\to R P_3)_{\rm NWA}$ denotes the branching fraction obtained from Eq.~(\ref{eq:NWA}) in the narrow width approximation.
}
\vskip 0.1cm
\label{tab:BF2body}
\scriptsize{
\begin{ruledtabular}
\begin{tabular}{ l c c c c}
Mode & $\B(B^+\to R P_3\to P_1P_2P_3)_{\rm expt}$  & $\B(B^+\to R P_3)_{\rm NWA}$ & $\eta_R^{\rm QCDF}\B(B^+\to R P_3)_{\rm NWA}$ & $\eta_R^{\rm EXPP}\B(B^+\to R P_3)_{\rm NWA}$  \\
\hline
$B^+\to f_2\pi^+\to \pi^+\pi^-\pi^+$
& $1.17\pm0.20$~\cite{Aaij:3pi_1,Aaij:3pi_2,BaBarpipipi}
& $2.08\pm0.36$
& $2.09\pm0.36$
& $1.95\pm0.33$ \\
$B^+\to K^{*0}_2\pi^+\to K^+\pi^-\pi^+$
& $1.85^{+0.73}_{-0.50}$~\cite{BaBar:Kmpippim,Belle:Kmpippim}
& $5.56^{+2.19}_{-1.50}$
& $5.40^{+2.13}_{-1.46}$
& $5.85^{+2.31}_{-1.58}$ \\
$B^+\to \rho^0\pi^+\to \pi^+\pi^-\pi^+$
& $8.36\pm0.77$~\cite{Aaij:3pi_1,Aaij:3pi_2,BaBarpipipi}
& $8.36\pm0.77$
& $7.78\pm0.72$ (GS)
& $7.95\pm0.73$ (GS) \\
&  &  & ~$9.28\pm0.86$ (BW) & \\
$B^+\to K^+\rho^0\to K^+\pi^+\pi^-$
& $3.7\pm0.5$~\cite{BaBar:Kmpippim,Belle:Kmpippim}
& $3.7\pm0.5$
& $3.5\pm0.5$ (GS)
& $3.4\pm0.5$ (GS) \\
&  &  &  ~$4.2\pm0.6$  (BW) & \\
$B^+\to K^{*0}\pi^+\to K^+\pi^-\pi^+$
& $6.71\pm0.57$~\cite{BaBar:Kmpippim,Belle:Kmpippim}
& $10.1\pm0.8$
& $10.7\pm0.9$
& $10.9\pm0.9$ \\
$B^+\to \sigma\pi^+\to \pi^+\pi^-\pi^+$
& $3.83\pm0.84$ ~\cite{Aaij:3pi_1,Aaij:3pi_2}
& $5.75\pm1.26$
& $12.36\pm2.71$
& $9.44\pm2.08$ \\
$B^+\to K^{*0}_0\pi^+\to K^+\pi^-\pi^+$
& $27.9^{+5.6}_{-4.3}$~\cite{BaBar:Kmpippim,Belle:Kmpippim}
& $45^{+9}_{-7}$
& $37^{+8}_{-6}$
& $50^{+10}_{-~8}$ \\
\end{tabular}
\end{ruledtabular}
}
\end{table}

\section{Conclusions }
For the branching fractions of the quasi-two-body decays  $\B(B\to RP_3)$ with $R$ being an intermediate resonant state, it is a common practice to apply the factorization relation, also known as the narrow width approximation (NWA), to extract them from the measured process $B\to RP_3\to P_1P_2P_3$.  However, such a treatment is valid only in the narrow width limit of the intermediate resonance, namely $\Gamma_R\to 0$.  In this work, we have studied the corrections to $\B(B\to RP_3)$ arising from the finite-width effects.  We consider the parameter $\eta_R$ which is the ratio of the three-body decay rate without and with the finite-width effects of the resonance.
Our main results are:

\begin{itemize}

\item We have presented a general framework for the parameter $\eta_R$ and shown that it can be expressed in terms of the normalized differential rate and is determined by its value evaluated at the resonance.
Since the value of the normalized differential rate at the resonance is anticorrelated with the normalized differential rate off the resonance, it is the shape of the normalized differential rate that matters in the determination of $\eta_R$.

\item In the experimental analysis of $B\to RP_3\to P_1P_2P_3$ decays, it is customary to parameterize the amplitude as $A(m_{12}, m_{23})=c\,F(m_{12}, m_{23})$, where the strong dynamics is described by the function $F$ parameterized in terms of the resonance line shape, the angular dependence and Blatt-Weisskopf barrier factors, while the information of weak interactions in encoded in the complex coefficients $c$. We evaluate $\eta_R$ in this experimentally motivated parameterization and in the theoretical framework of QCDF.

\item In QCDF calculations, we have verified the NWA relation both analytically and numerically for some charged $B$ decays involving tensor, vector and scalar resonances.  We have introduced a form factor $F(s_{12},m_R)$ for the strong coupling of $R(m_{12})\to P_1P_2$ when $m_{12}$ is away from $m_R$.   We find that off-shell effects are small in vector meson productions, but prominent in the $K_2^*(1430)$, $\sigma/f_0(500)$ and $K_0^*(1430)$ resonances.

\item In principle, the two-body rates reported by experiments should be corrected using $\eta_R=\eta^{\rm EXPP}_R$ in Eq.~\eqref{eq:BRofRP}, as the data are extracted using the experimental parameterization.  On the other hand, the experimental parameterization of the normalized differential rates should be compared with the theoretical predictions using QCDF calculations as the latter take into account the energy dependence of weak interaction amplitudes. In some cases, where $\eta^{\rm EXPP}_R$ are very different from $\eta^{\rm QCDF}_R$, we note that using an energy-independent coefficient $c$, in the experimental parameterization, to represent the weak dynamics is too na\"ive.  Moreover, systematic uncertainties in these experimental results after being corrected by $\eta_R^{\rm EXPP}$ are still underestimated.

\item We have compared between $\eta_R^{\rm QCDF}$ and $\eta_R^{\rm EXPP}$ for their width dependence in Figs.~\ref{fig:eta_f2}--\ref{fig:eta_K0st}. Numerical results are summarized in Table~\ref{tab:eta}. In general, the two quantities are similar for vector mesons but different for tensor and scalar mesons.  A study of the differential rates in Fig.~\ref{fig: dGam} enables us to understand the origin of their differences. For example, the similar normalized differential rates for $\rho$ and $K^*$ at and near the resonance account for $\eta^{\rm QCDF}_{\rho,K^*}\simeq\eta^{\rm EXPP}_{\rho,K^*}$. In contrast, the $m_{K\!\pi}^2$ dependence associated with the penguin Wilson coefficients $(a^p_6-a^p_8/2)$ in $B^-\to \ov K_0^*(1430)\pi^-\to K^-\pi^+\pi^+$ yields a large enhancement in the QCDF differential rate in the large $m_{K\!\pi}$ distribution, rendering $\eta^{\rm QCDF}_{K^*_0}<\eta^{\rm EXPP}_{K^*_0}$.

\item Finite-width corrections to $\B(B^+\to RP)_{\rm NWA}$, the branching fractions of quasi-two-body decays obtained in the NWA, are summarized in Table~\ref{tab:BF2body} for both QCDF and EXPP schemes. In general, finite-width effects are small, less than 10\%, but they are prominent in $B^+\to\sigma/f_0(500)\pi^+$ and $B^+\to K_0^{*0}(1430)\pi^+$ decays.

\item It is customary to use the Gounaris-Sakurai model to describe the line shape of the broad $\rho(770)$ resonance to ensure the unitarity far from the pole mass. If the relativistic Breit-Wigner model is employed instead, we find $\eta_\rho^{\rm BW}>1>\eta_\rho^{\rm GS}$ in both QCDF and EXPP schemes owing to the $(1+D \, \Gamma^0_\rho/m_\rho)$ term in the GS model. For example, in the presence of finite-width corrections, the PDG value of $\B(B^+\to\rho\pi^+)=(8.3\pm1.2)\times 10^{-6}$  should be corrected to $(7.7\pm1.1)\times 10^{-6}$ in QCDF and $(7.9\pm1.1)\times 10^{-6}$ in EXPP.

\item The $\sigma/f_0(500)$ scalar resonance is very broad, and its line shape cannot be described by the familiar Breit-Wigner model.  We have followed the LHCb Collaboration to use a simple pole model description.  We have found very large width effects: $\eta_\sigma^{\rm QCDF}\sim 2.15$ and $\eta_\sigma^{\rm EXPP}\sim 1.64$\,. Consequently, $B^-\to \sigma\pi^-$ has a large branching fraction of order $10^{-5}$.

\item We have employed the Breit-Wigner line shape to describe the production of $K_0^*(1430)$ in three-body $B$ decays and found large off-shell effects.  The smallness of $\eta^{\rm QCDF}_{K^*_0}$ relative to $\eta^{\rm EXPP}_{K^*_0}$ is ascribed to the fact that the normalized differential rate obtained in the QCDF calculation is much larger than that using the EXPP scheme in the off-resonance region.  The large discrepancy between QCDF estimate and experimental data of $\Gamma(B^-\to \ov K_0^{*0}\pi^-\to K^-\pi^+\pi^-)$ still remains an enigma.

\item In the approach of QCDF, the calculated \CP asymmetries of $B^-\to f_2(1270)\pi^-$, $B^-\to\sigma/f_0(500)\pi^-$ and $B^-\to K^-\rho^0$ agree with the experimental observations. The non-observation of \CP asymmetry in $B^-\to \rho(770)\pi^-$ can also be accommodated in QCDF.

\end{itemize}

\vskip 2.5cm \acknowledgments

This research was supported in part by the Ministry of Science and Technology of R.O.C. under Grant Nos. MOST-106-2112-M-033-004-MY3 and MOST-108-2112-M-002-005-MY3.

%%%%%%%%%%%%%%%%%%%%%%%%%%%%%%%%%%%%%%%%%%%%%%%%%%%%%%%%

\end{document}